\documentclass[a4paper,10pt]{article}
\pdfoutput=1
\usepackage[utf8]{inputenc}
\usepackage{amsmath}
\usepackage{amssymb}
\usepackage{bbm}
\usepackage{todonotes}
\usepackage{color,soul}
\usepackage[ngerman,british]{babel}
\usepackage{graphicx}
\usepackage{cite}
\usepackage{slashed}
\usepackage{xspace}
\usepackage{hyperref}

\allowdisplaybreaks

\evensidemargin \oddsidemargin
\newcommand{\xf}{\hat X_t}
\newcommand{\yf}{\hat Y_t}
\newcommand{\mf}{\hat \mu}
\newcommand{\ma}{\hat M_A}
\newcommand{\at}{\hat A_t}

\newcommand{\tb}{t_\beta}
\newcommand{\sbe}{s_\beta}
\newcommand{\cbe}{c_\beta}
\newcommand{\tbe}{t_\beta}
\newcommand{\sbb}{s_\beta^2}
\newcommand{\cbb}{c_\beta^2}
\newcommand{\tbb}{t_\beta^2}

\newcommand{\sae}{s_\alpha}
\newcommand{\cae}{c_\alpha}

\newcommand{\saa}{s_\alpha^2}
\newcommand{\caa}{c_\alpha^2}

\newcommand{\DR}{{\overline{\text{DR}}}}
\newcommand{\MS}{{\overline{\text{MS}}}}
\newcommand{\OS}{{\text{OS}}}
\newcommand{\Pdd}{\Phi_1^\dagger\Phi_1}
\newcommand{\Puu}{\Phi_2^\dagger\Phi_2}
\newcommand{\Pdu}{\Phi_1^\dagger\Phi_2}
\newcommand{\Pud}{\Phi_2^\dagger\Phi_1}
\newcommand{\li}{{\lambda_1}}
\newcommand{\lii}{{\lambda_2}}
\newcommand{\liii}{{\lambda_3}}
\newcommand{\liv}{{\lambda_4}}
\newcommand{\lv}{{\lambda_5}}
\newcommand{\lvi}{{\lambda_6}}
\newcommand{\lvii}{{\lambda_7}}

\newcommand{\hgiuu}{{\hat g_{1uu}}}
\newcommand{\hgidd}{{\hat g_{1dd}}}
\newcommand{\hgiiuu}{{\hat g_{2uu}}}
\newcommand{\hgiidd}{{\hat g_{2dd}}}
\newcommand{\hgiud}{{\hat g_{1ud}}}
\newcommand{\hgidu}{{\hat g_{1du}}}
\newcommand{\hgiiud}{{\hat g_{2ud}}}
\newcommand{\hgiidu}{{\hat g_{2du}}}
\newcommand{\tgiu}{{\tilde g_{1u}}}
\newcommand{\tgid}{{\tilde g_{1d}}}
\newcommand{\tgiiu}{{\tilde g_{2u}}}
\newcommand{\tgiid}{{\tilde g_{2d}}}
\newcommand{\htn}{{h_t}}
\newcommand{\htp}{{h_t^{\prime}}}
\newcommand{\gy}{{g^\prime}}

\newcommand{\cp}{\ensuremath{{\cal CP}}}

\newcommand{\SM}{\text{SM}}
\newcommand{\MSSM}{\text{MSSM}}
\newcommand{\THDM}{\text{THDM}}

\newcommand{\SUSY}{{\text{SUSY}}}

\newcommand{\tev}{\text{ TeV}}
\newcommand{\gev}{\text{ GeV}}

\newcommand{\CP}{$\mathcal{CP}$}
\newcommand{\msusy}{M_\SUSY}

\newcommand{\FH}{\mbox{{\tt FeynHiggs}}\xspace}

\newcommand{\MhEFT}{\mbox{{\tt MhEFT}}\xspace}

\newcommand{\order}[1]{\ensuremath{{\cal O}(#1)}}
\newcommand{\al}{\alpha}
\newcommand{\als}{\al_s}
\newcommand{\alt}{\al_t}
\newcommand{\alb}{\al_b}

\newcommand{\Fig}[1]{Fig.~\ref{#1}}
\newcommand{\Sec}[1]{Section~\ref{#1}}
\newcommand{\App}[1]{App.~\ref{#1}}
\newcommand{\Eq}[1]{Eq.~(\ref{#1})}
\newcommand{\Eqs}[2]{Eqs.~(\ref{#1}) and (\ref{#2})}
\newcommand{\Eqss}[2]{Eqs.~(\ref{#1})-(\ref{#2})}

\begin{document}

\thispagestyle{empty}
\def\thefootnote{\fnsymbol{footnote}}

\begin{flushright}
MPP-2018-83
\end{flushright}
\vspace{3em}
\begin{center}
{\Large\bf Precise prediction of the MSSM Higgs boson masses for low $\boldsymbol{M}_{\boldsymbol{A}}$}
\\
\vspace{3em}
{
Henning Bahl\footnote{email: hbahl@mpp.mpg.de},
Wolfgang Hollik\footnote{email: hollik@mpp.mpg.de}
}\\[2em]
{\sl Max-Planck Institut f\"ur Physik, F\"ohringer Ring 6, D-80805 M\"unchen, Germany}\
\def\thefootnote{\arabic{footnote}}
\setcounter{page}{0}
\setcounter{footnote}{0}
\end{center}
\vspace{2ex}
\begin{abstract}
{}
Precise predictions for Higgs boson masses in the Minimal Supersymmetric Standard Model can be obtained by combining fixed-order calculations with effective field theory (EFT) methods for the resummation of large logarithms in case of heavy superpartners.  This hybrid approach is implemented in the computer code \FH\ and has been applied in previous studies for calculating the mass of the lightest \CP-even Higgs boson for low, intermediate and high SUSY scales. In these works it was assumed that the non-standard Higgs bosons share a common mass scale with the supersymmetric squark particles, leaving the Standard Model as the low-energy EFT. In this article, we relax this restriction and report on the implemention of a Two-Higgs-Doublet Model (THDM) as effective theory below the SUSY scale into our hybrid approach. We explain in detail how our EFT calculation is consistently combined with the fixed-order calculation within the code \FH. In our numerical investigation we find effects on the mass of the lightest \CP-even Higgs boson $h$ of up to 9 GeV in scenarios with low $M_A$, low $\tan\beta$ and high SUSY scales, when compared with previous versions of \FH. Comparisons to other publicly available pure EFT codes with a THDM show good agreement. Effects on the mass of the second lightest \CP-even Higgs boson $H$ are found to be negligible in the phenomenologically interesting parameter regions where $H$ can be traded for $h$ as the experimentally observed Higgs particle.

\end{abstract}

\newpage
\tableofcontents
\newpage
\def\thefootnote{\arabic{footnote}}

%%%%%%%%%%%%%%%%%%%%%%%%%%%%%%%%%%%%%%%%%%%%%%%%%%%%%%%%%%%%%%%%%%%%%%%%%%%%%%
%%%%%%%%%%%%%%%%%%%%%%%%%%%%%%%%%%%%%%%%%%%%%%%%%%%%%%%%%%%%%%%%%%%%%%%%%%%%%%

\section{Introduction}

Precise measurements of the properties of the Higgs boson, discovered by the ATLAS and CMS collaborations at the CERN Large Hadron Collider~\cite{Aad:2012tfa,Chatrchyan:2012xdj} in 2012, are not only crucial for testing the Standard Model (SM) but also allow to constrain physics beyond the Standard Model. Supersymmetric extensions of the SM are theoretically well motivated, in particular the Minimial Supersymmetric Standard Model (MSSM) with quite specific predictions for Higgs bosons.

In the MSSM, the Higgs sector consists of two Higgs doublets, with vacuum expectation values (vevs) $v_1$ and $v_2$ which can be chosen real and  non-negative without loss of generality. After electroweak symmetry breaking, the two Higgs doublets accommodate five physical Higgs states: the light and heavy $\cp$-even $h$ and $H$ bosons, the $\cp$-odd $A$ boson, and the pair  $H^\pm$ of charged Higgs bosons. At the tree level, all Higgs boson masses are determined by two parameters, conventionally  chosen to be $\tan\beta = v_2/v_1$ and the mass of the $A$ boson, $M_A$. These tree-level relations, however, are  affected by large higher-order corrections resulting from the quantum effects of the MSSM.

Since no direct evidence for SUSY particles has been found so far, the range of  MSSM parameters can only be constrained indirectly. In addition to the classic set of precision observables ($Z$ and $W$ boson masses, effective electroweak mixing angle, \dots), the mass of the Higgs boson, determined by ATLAS and CMS \cite{Aad:2015zhl} to be $125.09\pm 0.24\gev$, can serve as an additional powerful precision observable. When interpreted as the mass of a light Higgs boson  within the MSSM spectrum, it is very sensitive especially to the parameters of the top-squark sector and can therefore be used to assess the SUSY scale. In the light of the high level of precision reached by the experiments, an accurate and reliable theoretical prediction is essential.

Therefore, a substantial amount of work has been dedicated to reduce the uncertainty of the theoretical prediction. Full one-loop corrections have been calculated diagrammatically in~\cite{Chankowski:1992er,Dabelstein:1994hb,Pierce:1996zz,Frank:2006yh}, dominant two-loop corrections in~\cite{Hempfling:1993qq,Casas:1994us,Carena:1995wu,Heinemeyer:1998kz,Heinemeyer:1998jw,Heinemeyer:1998np,Zhang:1998bm,Heinemeyer:1999be,Espinosa:1999zm,Espinosa:2000df,Degrassi:2001yf,Brignole:2001jy,Martin:2001vx,Brignole:2002bz,Martin:2002iu,Dedes:2003km,Martin:2002wn,Martin:2003it,Heinemeyer:2004xw,Martin:2004kr,Martin:2005eg,Heinemeyer:2007aq,Hollik:2014bua,Passehr:2017ufr} and partial three-loop corrections in~\cite{Martin:2007pg,Harlander:2008ju,Kant:2010tf}. Besides fixed-order calculations, also effective field theory (EFT) methods were applied (see~\cite{Giudice:2011cg,Draper:2013oza,Bagnaschi:2014rsa,Lee:2015uza,Vega:2015fna,Bagnaschi:2017xid}) as an alternative strategy.

The advantage of EFT methods is a resummation of logarithms which become large if the relevant scales are widely separated, like in the case of a high SUSY scale $\msusy$. Fixed-order calculations become unreliable for such wide scale separations, since the appearance of large logarithms can spoil the perturbative expansion. Fixed-order calculations, on the other hand, capture all terms with inverse powers of $\msusy$. Though suppressed in case of a high scale, they can become relevant for lower $\msusy$   and thus are needed for accurate predictions as well. These terms are missed in EFT calculations, at least when no higher-dimensional operators are taken into account (see \cite{Bagnaschi:2017xid} for a study including higher-dimensional operators).

In order to obtain results as accurate as possible for all SUSY scales, hybrid methods have been developed~\cite{Hahn:2013ria,Bahl:2016brp,Athron:2016fuq,Staub:2017jnp,Bahl:2017aev,Athron:2017fvs}. In~\cite{Hahn:2013ria,Bahl:2016brp,Bahl:2017aev} the strategy has been pursued to incorporate an EFT calculation on top of a  fixed-order calculation. Additional subtraction terms are introduced to avoid double counting of terms contained in both the EFT and the fixed-order calculation. The method has been implemented in the publicly available computer code \FH~\cite{Heinemeyer:1998yj,Heinemeyer:1998np,Hahn:2009zz,Degrassi:2002fi,Frank:2006yh}. So far, the EFT calculation in that approach was restricted to scenarios in which all non-SM particles share a common mass scale (with the only exception of possibly light electroweakinos and/or a light gluino),  leaving the SM as the low-energy EFT.

In this paper, we report on an improvement of this method by introducing a Two-Higgs-Doublet Model (THDM) as the effective theory below the SUSY scale, in replacement of the SM. This setup allows to cover the possibility of light non-standard Higgs bosons in the EFT calculation. Also scenarios with additional light electroweakinos are considered,  which are especially interesting in view of the increasingly tight constraints on colored SUSY particles from the LHC. We give a detailed  description of the steps needed to combine the THDM EFT calculation with the fixed-order calculation and illustrate the impact of the new hybrid version on the Higgs boson masses by numerical comparisons  with previous versions of \FH. An earlier pure EFT study \cite{Lee:2015uza} with an effective THDM found potentially large effects  originating from the resummation of logarithms of the SUSY scale over $M_A$, and observed significant differences with respect to \FH in specific parameter regions. We will clarify this situation by a detailed comparison and explain the current differences between \cite{Lee:2015uza} and our new THDM-improved hybrid calculation.

The outline of this paper is as follows: In \Sec{EFT_Sec}, we detail on the EFT calculation. Subsequently, we describe the consistent combination with the fixed-order part in \Sec{Combination_Section}. In \Sec{OtherCodes_Sec}, we compare our approach to that of other publicly available codes. This is followed by a numerical analysis in \Sec{NumResults_Sec}, with conclusions in \Sec{Conclusions_Sec}. The sections A to E of the Appendix provide additional technical information.

\newpage
%%%%%%%%%%%%%%%%%%%%%%%%%%%%%%%%%%%%%%%%%%%%%%%%%%%%%%%%%%%%%%%%%%%%%%%%%%%%%%
%%%%%%%%%%%%%%%%%%%%%%%%%%%%%%%%%%%%%%%%%%%%%%%%%%%%%%%%%%%%%%%%%%%%%%%%%%%%%%

\section{EFT calculation of Higgs-boson masses}\label{EFT_Sec}

If the SUSY particles are significantly heavier than all SM particles, they can be integrated out. In the simplest case when all non-standard particles occur at a common mass scale, the SUSY scale $\msusy$, the remaining EFT is the SM, with the Higgs self-coupling determined via matching conditions at $\msusy$. This self-coupling and all the other remaining SM  couplings are evolved from the SUSY scale down to the electroweak scale by means of renormalization group equations (RGEs); fixing the remaining SM couplings at the electroweak scale by matching to physical observables determines the input quantitites from which the SM Higgs-boson mass can be calculated. This approach has the advantage that large logarithmic contributions are resummed. On the other hand, terms suppressed by $\msusy$ are missed unless higher-dimensional operators are included in the EFT Lagrangian.

The assumption that all non-standard particles have a common mass scale is quite restrictive. For a better flexibility and wider applicabilty, more refined EFTs have to be considered. In our approach, we allow for several independent mass scales where each of them corresponds to the appearance of distinguished new phenomena. To be specific, we take into account five mass scales in our EFT calculation: the SM scale $M_t$, the non-standard Higgs-boson scale $M_A$, the electroweakino scale $M_\chi$, the gluino mass scale $M_{\tilde g}$ and the SUSY scale $\msusy$. We define the SUSY scale to be the mass scale of sfermions, which we assume to be approximately mass degenerate. Below $\msusy$, sleptons and  squarks are removed from the EFT; below $M_{\tilde g}$, we remove the gluino. The electroweakino scale $M_\chi$ is defined by
\begin{align}
M_\chi \sim M_1 , M_2 , \mu,
\end{align}
where $M_{1,2}$ are the soft-breaking electroweakino mass parameters and $\mu$ is the Higgsino mass parameter.
Below $M_\chi$, we remove the electroweakinos from the EFT.
$M_A$ marks the scale at which the heavy Higgs bosons are integrated out.

%%%%%%%%%%%%%%%%%%%%%%%%% F I G U R E %%%%%%%%%%%%%%%%%%%%%%%%%%%%%%%%%%%%%%%%
\begin{figure}\centering
\includegraphics[scale=2.5]{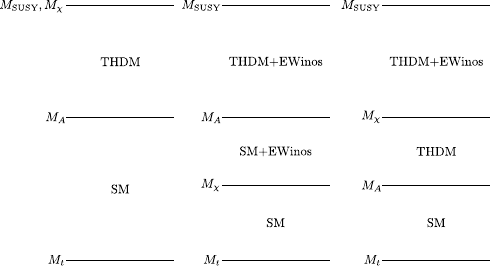}
\caption{EFT towers covered in this work (gluino threshold not shown).}
\label{FigHierarchies}
\end{figure}
%%%%%%%%%%%%%%%%%%%%%%%%% F I G U R E %%%%%%%%%%%%%%%%%%%%%%%%%%%%%%%%%%%%%%%%

We assume $M_A$ and $M_\chi$ to be smaller or equal to $\msusy$. Therefore, we have a set of eight different EFTs: the SM, the SM plus electroweakinos, the THDM and the THDM plus electroweakinos (plus the same with added gluino). This diversity leads to various different hierarchies, as illustrated in~\Fig{FigHierarchies}.

In our EFT calculation, we take into account full one-loop threshold corrections and full two-loop RGEs. This implies a full LL and NLL resummation. Additionally, we include \order{\als\alt} matching conditions for the Higgs self-couplings. \order{\alt^2} threshold corrections for matching the THDM to the MSSM are currently not known. Moreover, three-loop RGEs for the THDM are not yet available. Since the SM three-loop running is negligible, one may believe that this also holds for the three-loop THDM running~\cite{Lee:2015uza}. Nevertheless, the resummation of NNLL contributions is incomplete.

%%%%%%%%%%%%%%%%%%%%%%%%%%%%%%%%%%%%%%%%%%%%%%%%%%%%%%%%%%%%%%%%%%%%%%%%%%%%%%

\subsection{Relevant EFTs}

Here we give a brief overview of the various EFTs appearing in our calculation and specify our notation. We will not describe EFTs with gluino, since the presence of the gluino does not induce any effective couplings that are relevant at the order of the calculation presented in this paper. It, however, does alter the RGEs (see \cite{Bahl:2016brp}).

\subsubsection*{The Two-Higgs-Doublet Model}

Decoupling all sfermions, gauginos and Higgsinos from the full MSSM leads to a THDM as  the remaining effective theory below the SUSY scale. The THDM Higgs potential can be written as follows,
\begin{align}\label{HiggsPotential}
V_{\text{THDM}}(\Phi_1,\Phi_2) =& m_1^2\,\Pdd + m_2^2\,\Puu - m_{12}^2 (\Pdu + \Pud) + \frac{1}{2}\lambda_1 (\Pdd)^2 + \frac{1}{2}\lambda_2 (\Puu)^2 \nonumber\\
& + \lambda_3 (\Pdd)(\Puu) + \lambda_4 (\Pdu)(\Pud) + \frac{1}{2}\lambda_5 \left((\Pdu)^2 + (\Pud)^2\right) \nonumber\\
& + \lambda_6 (\Pdd)\left((\Pdu)+(\Pud)\right) + \lambda_7 (\Puu)\left((\Pdu)+(\Pud)\right),
\end{align}
where $\Phi_{1,2}$ denote  the two doublets of scalar fields. Since we consider only the real MSSM, all the coefficients can be chosen as real parameters. At the minimum of the potential each Higgs field $\Phi_i$ acquires a vacuum expectation value (vev),
\begin{align}
\langle\Phi_i\rangle = \begin{pmatrix}0 \\ v_i\end{pmatrix},
\quad
i = 1,2 .
\end{align}
Decomposing the Higgs fields into components according to
\begin{align}\label{HiggsFields}
\Phi_i =
\begin{pmatrix}
\phi_i^+ \\ v_i+\frac{1}{\sqrt{2}}(\phi_i + i\chi_i)
\end{pmatrix},
\end{align}
introducing the quantities
\begin{align}
v &= \sqrt{v_1^2+v_2^2}, \hspace{1cm} t_\beta \equiv\tan\beta = \frac{v_2}{v_1},
\end{align}
and expanding the potential around the minimum yields
the mass matrix of  the \CP-even neutral Higgs bosons,
\begin{align}
\mathcal{M}_{\phi\phi}^2 =
\begin{pmatrix}
m_{1}^2 & -m_{12}^2 \\
-m_{12}^2 & m_{2}^2
\end{pmatrix}
+ v^2
\begin{pmatrix}
a_{11} & a_{12} \\
a_{12} & a_{22}
\end{pmatrix},
\end{align}
with the entries
\begin{align}
a_{11} &= 3\li\cbb+(\liii+\liv+\lv)\sbb+6\lvi\sbe\cbe, \\
a_{12} &= 2(\liii+\liv+\lv)\sbe\cbe+6\lvi\cbb+6\lvii\sbb, \\
a_{22} &=  3\lii\sbb+(\liii+\liv+\lv)\cbb+6\lvii\sbe\cbe,
\end{align}
where the abbreviations
\begin{align}
s_{\gamma}\equiv\sin\gamma,\qquad c_{\gamma}\equiv\cos\gamma,\qquad t_{\gamma}\equiv\tan\gamma
\end{align}
for a generic  angle $\gamma$ have been introduced.

With the minimum conditions for the Higgs potential, $m_1^2$ and $m_2^2$ can be eliminated; the following relations for the masses of the \CP-odd neutral $A$ boson and of the charged $H^\pm$ bosons are obtained,
\begin{align}
 & m_A^2 = \frac{m_{12}^2}{\sbe\cbe}-v^2(2\lv+\lvi/\tbe+\lvii\tbe), \\
 & m_{H^\pm}  = m_A^2 + v^2(\lv-\liv) ,
\end{align}
and the \CP-even mass matrix $\mathcal{M}_{\phi\phi}^2$ can be cast into
the following form,
\begin{align}\label{MppEFT}
\mathcal{M}_{\phi\phi}^2 = m_A^2
\begin{pmatrix}
\sbb & -\sbe\cbe \\
-\sbe\cbe & \cbb
\end{pmatrix}
+ 2 v^2
\begin{pmatrix}
b_{11} & b_{12} \\
b_{12} & b_{22}
\end{pmatrix}
\end{align}
with
\begin{align}
b_{11} &= \li\cbb+2\lvi\sbe\cbe+\lv\sbb, \\
b_{12} &= (\liii+\liv)\sbe\cbe+\lvi\cbb+\lvii\sbb, \\
b_{22} &=  \lii\sbb+2\lvii\sbe\cbe+\lv\cbb.
\end{align}
The tree-level mass eigenstates $h$ and $H$ are obtained by a rotation,
\begin{align}\label{Alpha_Rot_Eq}
\begin{pmatrix} H \\ h \end{pmatrix} =
\begin{pmatrix}
c_\alpha & s_\alpha \\
-s_\alpha & c_\alpha
\end{pmatrix}
\begin{pmatrix} \phi_1 \\ \phi_2 \end{pmatrix},
\end{align}
with the angle $\alpha$ determined by
\begin{align}
s_{2\alpha} = \frac{2\mathcal{M}_{\phi_1\phi_2}^2}{\sqrt{\left(\mathcal{M}_{\phi_1\phi_1}^2-\mathcal{M}_{\phi_2\phi_2}^2\right)^2+4\left(\mathcal{M}_{\phi_1\phi_2}^2\right)^2}},
\hspace{1cm} -\frac{\pi}{2}< \alpha < \frac{\pi}{2}.
\end{align}
Often, it is useful to work in the Higgs basis instead of the $h,H$ mass eigenstate basis \cite{Gunion:2002zf}. It is obtained by rotating the original doublets $\Phi_{1,2}$ by the angle $\beta$,
\begin{align}\label{HiggsBasisRotation_Eq}
\begin{pmatrix} H_1 \\ H_2 \end{pmatrix} =
\begin{pmatrix}
c_\beta & s_\beta \\
-s_\beta & c_\beta
\end{pmatrix}
\begin{pmatrix} \Phi_1 \\ \Phi_2 \end{pmatrix}.
\end{align}
In this basis, only $H_1$ acquires a vev,
\begin{align}
\langle H_1\rangle = \begin{pmatrix} 0 \\ v \end{pmatrix} \text{ with }v \equiv \sqrt{v_1^2+v_2^2} ,
\end{align}
and the mass matrix (\ref{MppEFT}) is transformed into
\begin{align}\label{HiggsBasisMassMatrix_Eq}
\mathcal{M}_{HH}^2 = m_A^2
\begin{pmatrix}
0 & 0 \\
0 & 1
\end{pmatrix}
+ 2 v^2
\begin{pmatrix}
c_{11} & c_{12} \\
c_{12} & c_{22}
\end{pmatrix}
\end{align}
with
\begin{align}
\label{c11}
c_{11} &=
\li\cbe^4+\lii\sbe^4+2(\liii+\liv+\lv)\sbb\cbb+4\lvi\sbe\cbe^3+4\lvii\sbe^3\cbe, \\
c_{12} &= -\li\sbe\cbe^3+\lii\sbe^3\cbe+(\liii+\liv+\lv)\sbe\cbe c_{2\beta} - \lvi\cbb (3\sbb-\cbb) + \lvii\sbb(3\cbb-\sbb), \\
c_{22} &= (\li+\lii)\sbb\cbb-2(\liii+\liv)\sbb\cbb+\lv(\sbe^4+\cbe^4)-(\lvi-\lvii)s_{2\beta}c_{2\beta}.
\end{align}
To get from the Higgs basis to the mass eigenstate basis, we have to rotate by the angle $\alpha-\beta$.

We also need the Yukawa part of the effective THDM Lagrangian, which is given by
\begin{align}
\label{YukawaTHDM}
\mathcal{L}_{\text{Yuk}}(\Phi_1,\Phi_2) ={}& - \big[ h_t\, \bar t_R (-i\Phi_2^T\sigma_2)  Q_L
                               +h_t' \, \bar t_R (-i\Phi_1^T\sigma_2) Q_L + h.c. \big] ,
\end{align}
with the third-generation quark doublet $Q_L$  and the Pauli matrix $\sigma_2$. $\htn$ and $\htp$ are the effective top Yukawa couplings. All other Yukawa couplings are neglected in the EFT calculation; they are, however, fully captured through the diagrammatic calculation  at the one-loop level, in case of the bottom Yukawa coupling also at the two-loop level.

As already noted in \cite{Lee:2015uza}, the effective THDM with the Yukawa texture as given in \Eq{YukawaTHDM} is not a type~II model where only $\Phi_2$ couples to up-type quarks. Although the tree-level Yukawa sector of the MSSM is that of a THDM of type~II, loop corrections induce also a coupling of $\Phi_1$ to the top-quark, which enters through the matching procedure in the effective THDM. Differently from~\cite{Lee:2015uza}, we take this coupling  fully into account in all the affected RGEs and threshold corrections. Hence, we have to deal with 12 coupling constants, consisting of three gauge couplings, seven Higgs self-couplings, and two Yukawa couplings. We derived the RGEs for the considered THDM using the \texttt{Mathematica} package \texttt{SARAH}~\cite{Staub:2013tta}. The corresponding expressions are available from the authors upon request.

%%%%%%%%%%%%%%%%%%%%%%%%%%%%%%%%%

\subsubsection*{The Two Higgs-Doublet Model with Electroweakinos}

If in addition to the non-SM Higgs bosons also light electroweak gauginos and Higgsinos (EWinos) are present, the effective Lagrangian below the scale $\msusy$ is the one of the THDM described above, extended by extra mass and interaction terms
\begin{align}\label{LeffTHDM+EWinos}
\mathcal{L} = \, & \mathcal{L}_{\text{THDM}} \,
-\frac{1}{2}M_\chi \widetilde W \widetilde W -\frac{1}{2}M_\chi \widetilde B \widetilde B - M_\chi \,(i\widetilde{\mathcal{H}}_u^T\sigma_2)\widetilde{\mathcal{H}}_d \nonumber\\
& - \frac{1}{\sqrt{2}}H_u^\dagger\left(\hgiiuu\sigma_a\widetilde W^a+\hgiuu\widetilde B\right)\widetilde{\mathcal{H}}_u - \frac{1}{\sqrt{2}}H_d^\dagger\left(\hgiidd\sigma_a\widetilde W^a-\hgidd\widetilde B\right)\widetilde{\mathcal{H}}_d\nonumber\\
& - \frac{1}{\sqrt{2}}(i H_d^T \sigma_2)\left(\hgiidu\sigma_a\widetilde W^a+\hgidu\widetilde B\right)\widetilde{\mathcal{H}}_u - \frac{1}{\sqrt{2}}(-i H_u^T \sigma_2)\left(\hgiiud\sigma_a\widetilde W^a-\hgiud\widetilde B\right)\widetilde{\mathcal{H}}_d\nonumber\\
&+ h.c.
\end{align}
for the Bino field $\widetilde B$, the Wino fields $\widetilde W^a$, and the Higgsino fields $\widetilde{\mathcal{H}}_{u,d}$. The associated Higgs fields $H_{u,d}$ are related to the doublets $\Phi_{1,2}$ in (\ref{HiggsFields}) by
\begin{align}
H_u  = & \Phi_2 , \\
H_d  = & i \sigma_2 \Phi_1^* .
\end{align}
The coupling constants $\hat g_{1uu,1dd,1ud,1du,...}$ are effective Higgs-Higgsino-Gaugino couplings. The numeral in the subscript refers to the attached gauge symmetry (i.e.\ $U(1)$ or $SU(2)$), the first letter to the involved Higgs doublet, and the second letter to the involved Higgsino. Altogether, we now have 20  effective couplings in the game. Also the RGEs of the THDM+EWinos have been derived using \texttt{SARAH}. They are available from the authors upon request.

%%%%%%%%%%%%%%%%%%%%%%%%%%%%%%%%%

\subsubsection*{The Standard Model with Electroweakinos}

If we decouple all non-standard Higgs bosons, but keep light EWinos in the EFT, the effective Lagrangian simplifies to
\begin{align}\label{LeffSM+EWinos}
\mathcal{L} = \, & \mathcal{L}_{\text{SM}} \,
-\frac{1}{2}M_\chi \widetilde W \widetilde W -\frac{1}{2}M_\chi \widetilde B \widetilde B - M_\chi \,(i\widetilde{\mathcal{H}}_u^T\sigma_2)\widetilde{\mathcal{H}}_d - \frac{1}{\sqrt{2}}\Phi_\SM^\dagger\left(\tgiiu\sigma_a\widetilde W^a+\tgiu\widetilde B\right)\widetilde{\mathcal{H}}_u\nonumber\\
& -  \frac{1}{\sqrt{2}}(-i\Phi_\SM^T\sigma_2)\left(\tgiid\sigma_a\widetilde
  W^a-\tgid\widetilde B\right)\widetilde{\mathcal{H}}_d + h.c.
\end{align}
with $\Phi_\SM$ being the SM-like Higgs doublet,
\begin{align}
\Phi_{\text{SM}}=
\begin{pmatrix}
\phi^+ \\ v+\frac{1}{\sqrt{2}}(\phi + i\chi)
\end{pmatrix} .
\end{align}
The scalar potential in the SM part of the Lagrangian is given by
\begin{align}
V_{\text{SM}}(\Phi_{\text{SM}})=\frac{\lambda}{2}\left[(\Phi_{\text{SM}}^{\dagger} \Phi_{\text{SM}})-v^2\right]^2 .
\end{align}
 $\tilde g_{1u,1d,2u,2d}$ are the effective Higgs-Higgsino-Gaugino couplings of the SM+EWinos, in obvious notation. The number of couplings is reduced to 8. Two-loop RGEs can be found in \cite{Giudice:2011cg}. Below the electroweakino scale the effective model is eventually the pure SM.

%%%%%%%%%%%%%%%%%%%%%%%%%%%%%%%%%%%%%%%%%%%%%%%%%%%%%%%%%%%%%%%%%%%%%%%%%%%%%%

\subsection{Matching the EFTs}\label{MatchingProcedure_Sec}

After having specified the various EFTs, we describe how they are matched to each other. To derive the matching conditions, we have to compare physical amplitudes with external light particles computed in the EFT valid below the matching scale and in the full model (or the more complete EFT) valid above the matching scale. The difference between the physical amplitudes has to be absorbed by adapting the effective couplings in the particular EFT that is to be matched.

Terms contributing to the matching conditions arise from different vertex corrections and from different normalizations of the external fields. The part coming from the vertex corrections is obtained by calculating the vertex functions in the high-energy and the low-energy theory. The difference can then directly be absorbed into the effective coupling of the low-energy theory. At least at the one-loop level, at which we mostly work, this procedure is straightforward. Therefore, we will not go into more details.

If all external fields are non-mixed mass eigenstates, the external leg corrections are just given the corresponding LSZ factors, the wave-function renormalization. The difference between the LSZ factors in the high-energy and the low-energy theory has again to be absorbed by the low-energy effective coupling.

In case of mixing in the external fields, a more careful discussion is required. Even when the external fields are diagonal at the tree level,  loop contributions to the two-point vertices induce mixing between the mass eigenstates at higher orders. This transition has to be included as further external leg corrections,  in addition to the LSZ factors. In the MSSM and the THDM,  the mixing between the \CP-even Higgs bosons $h,H$ is the important issue. It is ascribed to a non-diagonal self-energy $\Sigma_{hH}$.

Conveniently, all external leg corrections can be written in form of a single matrix, the $Z$-matrix (see \cite{Frank:2006yh} for more details). It gives the relation between the external, asymptotical-free physical states and the tree-level mass eigenstates used for the calculation of the vertex correction. At the one-loop level, the MSSM relation reads
\begin{align}\label{LSZ_MSSM_Eq}
\begin{pmatrix} h^{\text{phys}} \\ H^{\text{phys}} \end{pmatrix} &=
\begin{pmatrix}
1+\frac{1}{2}\widehat\Sigma'_{hh}(m_h^2) & \frac{\widehat\Sigma_{hH}(m_h^2)}{m_h^2-m_H^2} \\
\frac{\widehat\Sigma_{hH}(m_H^2)}{m_H^2-m_h^2} & 1 +\frac{1}{2}\widehat\Sigma'_{HH}(m_H^2)
\end{pmatrix}
\begin{pmatrix} \widehat h \\ \widehat H\end{pmatrix} ,
\end{align}
where we used the symbol $\widehat{\hspace{.2cm}}$ to mark MSSM quantities.
$\Sigma_{hh}$ and $\Sigma_{HH}$ are the diagonal self-energies entering the LSZ factors.
The prime denotes the derivative with respect to the
external momentum. The corresponding relation in the THDM is written as follows,
\begin{align}\label{LSZ_THDM_Eq}
\begin{pmatrix} h^{\text{phys}} \\ H^{\text{phys}} \end{pmatrix} &=
\begin{pmatrix}
1+\frac{1}{2}\widetilde\Sigma'_{hh}(m_h^2) & \frac{\widetilde\Sigma_{hH}(m_h^2)}{m_h^2-m_H^2} \\
\frac{\widetilde\Sigma_{hH}(m_H^2)}{m_H^2-m_h^2} & 1 +\frac{1}{2}\widetilde\Sigma'_{HH}(m_H^2)
\end{pmatrix}
\begin{pmatrix} \widetilde h \\ \widetilde H\end{pmatrix},
\end{align}
where we used the symbol $\widetilde{\hspace{.2cm}}$ to mark THDM quantities.

\Eqs{LSZ_MSSM_Eq}{LSZ_THDM_Eq} yield the relation between the mass eigenstates of the MSSM and the THDM,
\begin{align}\label{Matching_hH_basis_Eq}
\begin{pmatrix} \widetilde h \\ \widetilde H\end{pmatrix} =
\begin{pmatrix}
1+\frac{1}{2}\Delta\Sigma'_{hh}(m_h^2) & \frac{\Delta\Sigma_{hH}(m_h^2)}{m_h^2-m_H^2} \\
\frac{\Delta\Sigma_{hH}(m_H^2)}{m_H^2-m_h^2} & 1 +\frac{1}{2}\Delta\Sigma'_{HH}(m_H^2)
\end{pmatrix}
\begin{pmatrix} \widehat h \\ \widehat H\end{pmatrix},
\end{align}
where the $\Delta\Sigma_{xy}$ summarize the
differences between the self-energies, for  $x,y\in \{h,H\}$,
\begin{align}
\Delta\Sigma_{xy}(p^2) \equiv \widehat\Sigma_{xy}(p^2) - \widetilde\Sigma_{xy}(p^2).
\end{align}
The mass eigenstates are related to the original field components $\phi_{1,2}$ via \Eq{Alpha_Rot_Eq},
\begin{align}
\begin{pmatrix} \widetilde h \\ \widetilde H\end{pmatrix} &=
U_{\widetilde\alpha}\begin{pmatrix} \widetilde\phi_1 \\ \widetilde\phi_2 \end{pmatrix}=
\begin{pmatrix}
-s_{\widetilde\alpha} & c_{\widetilde\alpha} \\
c_{\widetilde\alpha} & s_{\widetilde\alpha}
\end{pmatrix}
\begin{pmatrix} \widetilde\phi_1 \\ \widetilde\phi_2 \end{pmatrix}, \\
\begin{pmatrix} \widehat h \\ \widehat H\end{pmatrix} &=
U_{\widehat\alpha}\begin{pmatrix} \widehat\phi_1 \\ \widehat\phi_2 \end{pmatrix}=
\begin{pmatrix}
-s_{\widehat\alpha} & c_{\widehat\alpha} \\
c_{\widehat\alpha} & s_{\widehat\alpha}
\end{pmatrix}
\begin{pmatrix} \widehat\phi_1 \\ \widehat\phi_2 \end{pmatrix}.
\end{align}
With these relations, \Eq{Matching_hH_basis_Eq} can be transformed into a relation between the component fields $\phi_{1,2}$,
\begin{align}\label{PhiDiff_Eq}
\begin{pmatrix} \widetilde\phi_1 \\ \widetilde\phi_2\end{pmatrix} = U_{\widetilde\alpha}^T
\begin{pmatrix}
1+\frac{1}{2}\Delta\Sigma'_{hh}(m_h^2) & \frac{\Delta\Sigma_{hH}(m_h^2)}{m_h^2-m_H^2} \\
\frac{\Delta\Sigma_{hH}(m_H^2)}{m_H^2-m_h^2} & 1 +\frac{1}{2}\Delta\Sigma'_{HH}(m_H^2)
\end{pmatrix}
U_{\widehat\alpha}\begin{pmatrix} \widehat\phi_1 \\ \widehat\phi_2\end{pmatrix}.
\end{align}
In the THDM, the mixing angle $\widetilde\alpha$ is a free parameter. We fix it at lowest order by requiring that the THDM fields $\widetilde\phi_{1,2}$ are aligned with the MSSM fields $\widehat\phi_{1,2}$. Consequently, the two mixing angles $\widetilde\alpha$ and $\widehat\alpha$ are equal at the tree level. At the one-loop level we change the tree-level basis of the THDM slightly allowing for a small misalignment between the THDM and the MSSM fields,
\begin{align}
\Delta\alpha = \widehat\alpha - \widetilde\alpha.
\end{align}
Using this shift to replace $\widetilde\alpha$ by $\widehat\alpha$
in \Eq{PhiDiff_Eq} we obtain, expanded up to the one-loop level,
\begin{align}
\begin{pmatrix} \widetilde\phi_1 \\ \widetilde\phi_2\end{pmatrix} = U_{\widehat\alpha}^T
\begin{pmatrix}
1+\frac{1}{2}\Delta\Sigma'_{hh} & \frac{\Delta\Sigma_{hH}(m_h^2)}{m_h^2-m_H^2} - \Delta\alpha\\
\frac{\Delta\Sigma_{hH}(m_H^2)}{m_H^2-m_h^2} +\Delta\alpha & 1 +\frac{1}{2}\Delta\Sigma'_{HH}
\end{pmatrix}
U_{\widehat\alpha}\begin{pmatrix} \widehat\phi_1 \\ \widehat\phi_2\end{pmatrix}.
\end{align}
Next we expand $\Delta\Sigma_{hH}(m_H^2)$ around $p^2 = m_h^2$,
\begin{align}
\label{hHexpansionEq}
\Delta\Sigma_{hH}(m_H^2) = \Delta\Sigma_{hH}(m_h^2) + (m_H^2-m_h^2) \,\Delta\Sigma_{hH}'
 +\, \mathcal{O}(v/\msusy,M_A/\msusy).
\end{align}
All higher order derivatives of the $\Delta\Sigma_{xy}$ are suppressed by $\msusy$ and therefore negligible in the EFT calculation. For the same reason, we drop the specification of the external momentum in all derivatives of $\Delta\Sigma_{xy}$ in the following (which is always taken at $m_h^2$).

Using the expansion (\ref{hHexpansionEq}) and partly rewriting the self-energies yields
\begin{align}\label{DeltaAlphaEq}
\begin{pmatrix} \widetilde\phi_1 \\ \widetilde\phi_2\end{pmatrix} = \left[
\begin{pmatrix}
1+\frac{1}{2}\Delta\Sigma'_{11} & \frac{1}{2}\Delta\Sigma'_{12} \\
\frac{1}{2}\Delta\Sigma'_{12} & 1 +\frac{1}{2}\Delta\Sigma'_{22}
\end{pmatrix} + \left(\frac{\Delta\Sigma_{hH}(m_h^2)}{m_h^2-m_H^2} -\frac{1}{2}\Delta\Sigma_{hH}' - \Delta\alpha\right)
\begin{pmatrix}
0  & -1 \\
1 & 0
\end{pmatrix} \right]
\begin{pmatrix} \widehat\phi_1 \\ \widehat\phi_2\end{pmatrix}.
\end{align}
with the notation $\Delta\Sigma_{ij} \equiv \Delta\Sigma_{\phi_i\phi_j}$ for $i,j\in\{1,2$).

The second matrix corresponds to the one-loop part of a unitary matrix and thereby to a basis transformation by a rotation. It can be absorbed by adjusting $\Delta\alpha$ according to
\begin{align}
\Delta\alpha = \frac{\Delta\Sigma_{hH}(m_h^2)}{m_h^2-m_H^2} - \frac{1}{2}\Delta\Sigma_{hH}'.
\end{align}
The first matrix in \Eq{DeltaAlphaEq} is not unitary and hence cannot be removed by a basis transformation. Therefore, there is a remaining difference between the normalization of the $\phi_{1,2}$ fields in the MSSM and the THDM, given by the following relation,
\begin{align}\label{MatchingPhiBasis_Eq}
\begin{pmatrix} \widetilde\phi_1 \\ \widetilde\phi_2\end{pmatrix} =
\begin{pmatrix}
1+\frac{1}{2}\Delta\Sigma'_{11} & \frac{1}{2}\Delta\Sigma'_{12} \\
\frac{1}{2}\Delta\Sigma'_{12} & 1 +\frac{1}{2}\Delta\Sigma'_{22}
\end{pmatrix}
\begin{pmatrix} \widehat\phi_1 \\ \widehat\phi_2\end{pmatrix} ,
\end{align}
which corresponds to the one used in \cite{Haber:1993an}. As noted above, it is only valid at the one-loop level. We have to take care of this relation whenever we match a coupling involving an external Higgs field. This is achieved by rescaling the Higgs doublets of the THDM (or the MSSM) according to \Eq{MatchingPhiBasis_Eq}.

Since we rescale the whole doublets, a relation similar to \Eq{MatchingPhiBasis_Eq} also holds for the vevs,
\begin{align}
\begin{pmatrix} \widetilde v_1 \\ \widetilde v_2\end{pmatrix} =
\begin{pmatrix}
1+\frac{1}{2}\Delta\Sigma'_{11} & \frac{1}{2}\Delta\Sigma'_{12} \\
\frac{1}{2}\Delta\Sigma'_{12} & 1 +\frac{1}{2}\Delta\Sigma'_{22}
\end{pmatrix}
\begin{pmatrix} \widehat v_1 \\ \widehat v_2\end{pmatrix}.
\end{align}
This directly implies
\begin{align}\label{TB_Matching_Eq}
\widetilde\beta = \widehat\beta + \frac{1}{2}\left[\left(\Delta\Sigma'_{22}-\Delta\Sigma'_{11}\right)\sbe\cbe+\Delta\Sigma'_{12}c_{2\beta}\right] = \widehat\beta + \frac{1}{2}\Delta\Sigma'_{H_1 H_2},
\end{align}
or
\begin{align}
\label{tanbetaMatchingEq}
\tan\widetilde\beta = \tan\widehat\beta + \frac{1}{2\cbb} \Delta\Sigma'_{H_1H_2} ,
\end{align}
respectively, with $H_{1,2}$ being the fields of the Higgs basis defined in \Eq{HiggsBasisRotation_Eq}.

Following this procedure and including vertex corrections, we derived a full set of one-loop threshold corrections for all appearing effective couplings and hierarchies. Below, we list only the tree-level matching conditions and the dominant one-loop corrections, i.e. those proportional to the strong gauge coupling or the top Yukawa couplings. Full one-loop threshold corrections for all effective couplings including electroweak contributions are listed in \App{ThresholdApp}.

In addition to the calculation of matching conditions, we will also need \Eq{MatchingPhiBasis_Eq} for combining the diagrammatic fixed-order calculation and the EFT calculation.

%%%%%%%%%%%%%%%%%%%%%%%%%%%%%%%%%

\subsubsection*{Matching the THDM to the MSSM}

The Higgs self-couplings in the THDM scalar potential are fixed at the tree level by~\cite{Haber:1993an}
\begin{align}
\li(\msusy) = \lii(\msusy) =& \frac{1}{4}(g^2 + \gy^2), \\
\liii(\msusy) =& \frac{1}{4}(g^2 - \gy^2), \\
\liv(\msusy) =& -\frac{1}{2}g^2, \\
\lv(\msusy) = \lvi(\msusy) = \lvii(\msusy) =& 0,
\end{align}
where $g$ and $g'$ are the electroweak gauge couplings.

At one-loop order these relations receive additional contributions~\cite{Haber:1993an},
\begin{align}
\Delta\lambda_1 &= - \frac{1}{2}k h_t^4 \mf^4 + \mathcal{O}(g,g'),\\
\Delta\lambda_2 &= 6 k h_t^4 \at^2 \left(1-\frac{1}{12}\at^2\right) + \mathcal{O}(g,g'),\\
\Delta\lambda_3 &= \frac{1}{2}k\mf^2 h_t^4 (3-\at^2)  + \mathcal{O}(g,g'),\\
\Delta\lambda_4 &= \frac{1}{2}k\mf^2 h_t^4 (3-\at^2)  + \mathcal{O}(g,g'),\\
\Delta\lambda_5 &= -\frac{1}{2}k h_t^4\mf^2\at^2  + \mathcal{O}(g,g'),\\
\Delta\lambda_6 &= \frac{1}{2}k h_t^4 \mf^3\at  + \mathcal{O}(g,g'),\\
\Delta\lambda_7 &= \frac{1}{2}k h_t^4 \mf\at(\at^2-6)  + \mathcal{O}(g,g')
\end{align}
with $\mf = \mu/\msusy$ and $\at = A_t/\msusy$. $A_t$ is the stop trilinear coupling and $h_t$  the top Yukawa coupling of the MSSM.\footnote{For definiteness, we now assign  an explicit label for the Yukawa couplings $h_t, h'_t$ introduced in (\ref{YukawaTHDM}) for the THDM.}   The factor $k \equiv (4\pi)^{-2}$ is used to mark the loop-order. In addition to these one-loop corrections, we also include \order{\als\alt} threshold corrections, listed in \App{TLthresholds_App}.

For $\mf = 1$, the effective top Yukawa couplings are given by
\begin{align}
h_t^{\text{THDM}}(\msusy) =& h_t\Bigg\{1+k\bigg[\frac{4}{3} g_3^2(1-\at) - \frac{1}{4}h_t^2\at^2\bigg]\Bigg\}+ \mathcal{O}(g,g'),\label{htMSSMthreshold_Eq}\\
{h_t'}^{\text{\,THDM}}(\msusy) =& h_t k\Bigg\{\frac{4}{3} g_3^2  + \frac{1}{4} h_t^2\at\Bigg\}+ \mathcal{O}(g,g').\label{htpMSSMthreshold_Eq}
\end{align}
The full expressions for $\mf \neq 1$ are given in \App{ThresholdApp}.

The threshold correction for $\tan\beta$  is obtained
from~\Eq{tanbetaMatchingEq} yielding
\begin{align}\label{THDMtoMSSMtbEq}
\tbe^\THDM(\msusy) &= \tbe^\MSSM(\msusy)\left[1 + \frac{1}{4}k h_t^2 (\at-\mf/\tbe)(\at+\mf\tbe)+ \mathcal{O}(g,g')\right].
\end{align}

%%%%%%%%%%%%%%%%%%%%%%%%%%%%%%%%%

\subsubsection*{Matching the THDM+EWinos to the MSSM}

Neglecting the weak gauge couplings, the relations for matching the THDM to the MSSM are also valid when the THDM+EWinos is matched to the MSSM. The additional effective Higgs-Higgsino-Gaugino couplings of the THDM+EWinos fulfill the tree-level relations
\begin{align}
\hgiuu(\msusy) &= \hgidd(\msusy) = \gy, \\
\hgiiuu(\msusy) &= \hgiidd(\msusy) = g, \\
\hgiud(\msusy) &= \hgidu(\msusy) = \hgiiud(\msusy) = \hgiidu(\msusy) = 0.
\end{align}

%%%%%%%%%%%%%%%%%%%%%%%%%%%%%%%%%

\subsubsection*{Matching the THDM to the THDM+EWinos}

Matching the THDM to the THDM+EWinos, the Higgs self-couplings, the gauge couplings, the top Yukawa couplings and $\tbe$ are not modified at the tree level. If the weak gauge couplings are neglected, there are also no loop corrections. The full one-loop corrections including the weak gauge couplings are listed in \App{ThresholdApp}.

%%%%%%%%%%%%%%%%%%%%%%%%%%%%%%%%%

\subsubsection*{Matching the SM to the THDM}

In this specific case, the characteristic scale for all the couplings below is the mass $M_A$.
In the decoupling limit $M_A\gg M_Z$ ($\alpha\rightarrow\beta-\frac{\pi}{2}$), which is assumed when the heavy Higgs bosons are integrated out, the Higgs self-coupling $\lambda$ of the SM is obtained by
\begin{align}\label{Lambda_TreeLevel_Matching}
\lambda(M_A) =&  \;  c_{11}  +\, \Delta \lambda \, ,
\end{align}
with $c_{11}$ from \Eq{c11}, $\beta = \beta^\THDM$, and the one-loop correction
\begin{align}
\Delta\lambda =& - 3 k\left\{(\lvi+\lvii)c_{2\beta}+ (\lvi-\lvii)c_{4\beta}-\left(\li\cbb - \lii\sbb - (\liii+\liv+\lv)c_{2\beta}\right)s_{2\beta}\right\}^2.
\end{align}
The THDM top Yukawa couplings are related to the SM top Yukawa coupling $y_t$ via
\begin{align}\label{topYukawaSMtoTHDM_Eq}
y_t(M_A) =& (h_t^{\text{THDM}} \sbe + h_t'^{\text{\,THDM}} \cbe)\left[1- \frac{3}{8}k \left(h_t^{\text{THDM}}\cbe - h_t'^{\,\text{THDM}}\sbe\right)^2\right].
\end{align}
As stated above, the SM top Yukawa coupling is extracted from the top pole mass at the scale $M_t$. The top Yukawa couplings of the THDM are then determined by numerically solving the system of RGEs with the boundary conditions given in Eqs.~(\ref{topYukawaSMtoTHDM_Eq}),~(\ref{htMSSMthreshold_Eq}) and~(\ref{htpMSSMthreshold_Eq}) (see also \Eqs{htTHDMvsMSSM_Eq}{htpTHDMvsMSSM_Eq} for more general expressions).

%%%%%%%%%%%%%%%%%%%%%%%%%%%%%%%%%

\subsubsection*{Matching the SM+EWinos to the THDM+EWinos}

Neglecting the weak gauge couplings, the relations for matching the SM to the THDM are also valid when the SM+EWinos is matched to the THDM+EWinos. At the tree level, the effective Higgs-Higgsino-Gaugino couplings of the SM+EWinos and the THDM+EWinos are related by
\begin{align}
\tgiu &= \hgiuu\sbe + \hgidu\cbe, \hspace{1cm} \tgiiu = \hgiiuu\sbe + \hgiidu\cbe, \\
\tgid &= \hgidd\cbe + \hgiud\sbe, \hspace{1cm} \tgiid = \hgiidd\cbe + \hgiiud\sbe.
\end{align}
One-loop corrections proportional to the electroweak gauge couplings can be found in \App{ThresholdApp}.

%%%%%%%%%%%%%%%%%%%%%%%%%%%%%%%%%

\subsubsection*{Matching the SM to the SM+EWinos or the MSSM}

The matching conditions of the SM to the SM+EWinos or to the MSSM are well-known and can be found in \cite{Giudice:2011cg,Bagnaschi:2014rsa}.

%%%%%%%%%%%%%%%%%%%%%%%%%%%%%%%%%

\subsubsection*{Matching EFTs without and with gluino}

If the gluino is integrated out, no threshold corrections arise at the one-loop level. At the two-loop level however, the matching conditions of the scalar self-couplings between the THDM and the MSSM are modified if a gluino is added to the THDM. Corresponding expressions are listed in \App{TLthresholds_App}.

%%%%%%%%%%%%%%%%%%%%%%%%%%%%%%%%%

\subsection{Calculation of pole masses in the EFT approach}

The proper way to calculate the physical masses of the \CP-even Higgs bosons in the EFT framework depends on the mass hierarchy. For $M_A\gg M_t$, the low-energy theory is the SM (or the SM+EWinos). Therefore, the procedure described e.g.~in \cite{Bahl:2017aev} can be applied. For $M_A \sim M_t$, though, there is no need to integrate out the non-standard Higgs bosons and the low-energy theory is better described by a THDM (or a THDM+EWinos). In this case, the physical masses of the \CP-even Higgs bosons are obtained by finding the poles of the propagators, i.e.\ the zeroes of the determinant of the inverse propagator matrix, depicted here in the Higgs basis as a possible choice,
\begin{align}\label{InversePropMatrixEFT_Eq}
-i\Delta^{-1}_{\widetilde H\widetilde H}(p^2) =
\begin{pmatrix}
p^2 - \widetilde m_{H_1H_1}^2 + \widetilde\Sigma_{\widetilde H_1\widetilde H_1}(p^2) & - \widetilde m_{H_1H_2}^2 + \widetilde\Sigma_{\widetilde H_1\widetilde H_2}(p^2) \\[1ex]
- \widetilde m_{H_1H_2}^2 + \widetilde\Sigma_{\widetilde H_1\widetilde H_2}(p^2) & p^2 - \widetilde m_{H_2H_2}^2 + \widetilde\Sigma_{\widetilde H_2\widetilde H_2}(p^2)
\end{pmatrix}.
\end{align}
The widetilde $\widetilde{\hspace{.2cm}}$ indicates, as in~\Sec{MatchingProcedure_Sec}, that the corresponding quantities are those of the THDM, at the scale $M_A$. The quantitites $\widetilde m_{H_iH_j}^2$ are the entries of the matrix $\mathcal{M}_{HH}$ defined in \Eq{HiggsBasisMassMatrix_Eq}, and the various $\widetilde \Sigma$'s denote the corresponding self-energies of the THDM (or the THDM+EWinos) renormalized in the $\MS$ scheme.

In situations where $M_A$ is larger than $M_t$, but the separation is also not too large, e.g. $M_A - M_t \sim 100\gev$, it is difficult to decide if the SM should be used as low-energy theory or the THDM might the better choice. Therefore, a smooth transition between the two cases is beneficial. To implement such a transition, we follow a procedure similar to the one introduced in \cite{Lee:2015uza}: We include the contribution of the running between $M_A$ and $M_t$,
\begin{align}
\Delta(M_A\rightarrow M_t) = 2 v^2 \left(\lambda(M_t) - \lambda(M_A)\right),
\end{align}
into the $H_1H_1$ element of \Eq{InversePropMatrixEFT_Eq}. The same contribution is in addition added to the $H_1H_2$ and $H_2H_1$ entries with a prefactor $1/\tbe$ and to the $H_2H_2$ element with a prefactor $1/\tbb$,\footnote{Corresponding to the additional factor $1/\tbe$ in the top Yukawa coupling for $H_2$, which is responsible for the dominant contribution to $\Delta(M_A\rightarrow M_t)$ (see also \cite{Hahn:2013ria}).}
\begin{align}
-i\Delta^{-1}_{\widetilde H\widetilde H}(p^2) \rightarrow -i\Delta^{-1}_{\widetilde H\widetilde H}(p^2) - \Delta(M_A\rightarrow M_t)
\begin{pmatrix}
1 & \frac{1}{\tbe}\\[1ex]
\frac{1}{\tbe} & \frac{1}{\tbb}
\end{pmatrix}.
\end{align}
In this way both limits, $M_A\gg M_t$ and  $M_A\sim M_t$, are properly recovered.\footnote{Note that in addition it is necessary to ensure that logarithms of $M_A$ over $M_t$ contained in $\Delta(M_A\rightarrow M_t)$ as well as in the THDM self-energies $\widetilde\Sigma_{\widetilde H_i\widetilde H_j}$ are not double-counted.}

%%%%%%%%%%%%%%%%%%%%%%%%%%%%%%%%%%%%%%%%%%%%%%%%%%%%%%%%%%%%%%%%%%%%%%%%%%%%%%
%%%%%%%%%%%%%%%%%%%%%%%%%%%%%%%%%%%%%%%%%%%%%%%%%%%%%%%%%%%%%%%%%%%%%%%%%%%%%%

\section{Combination of fixed-order and EFT calculation}\label{Combination_Section}

The program \FH already contains a state-of-the-art fixed-order calculation, i.e., it comprises full one-loop and \order{\alt\als,\alb\als,\alt^2,\alt\alb,\alb^2}  higher-order corrections to the Higgs self-energies~\cite{Heinemeyer:1998yj,Heinemeyer:1998np,Degrassi:2001yf,Brignole:2001jy,Brignole:2002bz,Degrassi:2002fi,Dedes:2003km,Heinemeyer:2004xw,Frank:2006yh,Heinemeyer:2007aq,Hahn:2009zz,Hollik:2014bua}. For these corrections, a mixed OS/$\DR$ scheme is employed (see~\cite{Frank:2006yh} for more details), with the stop sector renormalized by default using the OS scheme. With version {\tt 2.14.0}, the possibility of using the $\DR$ scheme for the renormalization of the stop sector was introduced~\cite{Bahl:2017aev}. Field renormalization of the Higgs doublets and thereby the renormalization of $\tan\beta$ is always performed in the $\DR$ scheme, independent of the renormalization of the stop sector.

Our goal is to combine the result of this diagrammatic fixed-order calculation with the EFT calculation described in \Sec{EFT_Sec}. This combination is done in several steps. First, we have to relate the quantities computed in the EFT approach, namely the entries of the inverse propagator matrix, the two-point vertex function, to those in the fixed-order approach. Second, proper subtraction terms have to be identified and subtracted such that double-counting of terms appearing in the two results is avoided. Finally, differences in input parameters resulting from different renormalization schemes have to be considered by proper conversion of the parameters.

We choose to perform the combination in the gauge eigenstate basis. Therefore, we need to know the relation between the two-point vertex function matrix in the full MSSM, denoted by $\Delta^{-1}_{\widehat\phi\widehat\phi}$, and in the effective THDM, labeled as $\Delta^{-1}_{\widetilde\phi\widetilde\phi}$. Again, as in~\Sec{MatchingProcedure_Sec}, the symbol $\widehat{\hspace{.2cm}}$ is used to mark quantities in the full MSSM, and $\widetilde{\hspace{.2cm}}$ to mark quantities in the effective THDM. The two matrices have to be equal in case of Higgs fields with the same normalization in either of the models. In our case, however, the Higgs field normalization is different, as specified by~\Eq{MatchingPhiBasis_Eq}, which leads to the relation
\begin{align}\label{InvPropMatRel_Eq}
\Delta^{-1}_{\widehat\phi\widehat\phi}(p^2)&=
\begin{pmatrix}
1+\frac{1}{2}\Delta\Sigma'_{11} & \frac{1}{2}\Delta\Sigma'_{12} \\
\frac{1}{2}\Delta\Sigma'_{12} & 1 +\frac{1}{2}\Delta\Sigma'_{22}
\end{pmatrix}
\Delta^{-1}_{\widetilde\phi\widetilde\phi}(p^2)
\begin{pmatrix}
1+\frac{1}{2}\Delta\Sigma'_{11} & \frac{1}{2}\Delta\Sigma'_{12} \\
\frac{1}{2}\Delta\Sigma'_{12} & 1 +\frac{1}{2}\Delta\Sigma'_{22}
\end{pmatrix}.
\end{align}
As noted in \Sec{MatchingProcedure_Sec} this formula is valid only in the decoupling limit of $\msusy\gg M_t$ and at the one-loop level. Explicit formulae for the $\Delta\Sigma'_{ij}$ are listed in \App{AppDiffNormalization}.

In the combination of the EFT and  fixed-order results, it is convenient to take account of \Eq{InvPropMatRel_Eq} by introducing a finite shift in the field renormalization constants of the fixed-order result. Originally, the MSSM Higgs fields are renormalized by the scale transformation (up to two-loop order)
\begin{align}
\begin{pmatrix}
\widehat{\phi}_1 \\ \widehat{\phi}_2
\end{pmatrix}
\rightarrow
\begin{pmatrix}
1 + \frac{1}{2}\delta^{(1)}Z_{11} + \frac{1}{2}\Delta^{(2)}Z_{11} & \frac{1}{2}\delta^{(1)}Z_{12} + \frac{1}{2}\Delta^{(2)}Z_{12} \\
\frac{1}{2}\delta^{(1)}Z_{12} + \frac{1}{2}\Delta^{(2)}Z_{12} & 1 + \frac{1}{2}\delta^{(1)}Z_{22} + \frac{1}{2}\Delta^{(2)}Z_{22}
\end{pmatrix}
\begin{pmatrix}
\widehat{\phi}_1 \\ \widehat{\phi}_2
\end{pmatrix}
\end{align}
with
\begin{align}
\Delta^{(2)}Z_{ij} = \delta^{(2)}Z_{ij}-\frac{1}{4}\left(\delta^{(1)}Z_{ij}\right)^2.
\end{align}
The divergent pieces are fixed via the $\DR$ prescription in terms of the one- and two-loop self-energies,
\begin{align}
\delta^{(1)}Z_{11}\Big|_\text{div} &= - \mathfrak{Re}\left[\widehat\Sigma_{11}^{(1)\prime}\right]_\text{div},\hspace{.3cm} \delta^{(1)}Z_{22}\Big|_\text{div} = - \mathfrak{Re}\left[\widehat\Sigma_{22}^{(1)\prime}\right]_\text{div},\hspace{.3cm} \delta^{(1)}Z_{12}\Big|_\text{div} = 0, \\
\delta^{(2)}Z_{11}\Big|_\text{div} &= -\mathfrak{Re}\left[\widehat\Sigma_{11}^{(2)\prime}\right]_\text{div},\hspace{.3cm}
\delta^{(2)}Z_{22}\Big|_\text{div} = -\mathfrak{Re}\left[\widehat\Sigma_{22}^{(2)\prime}\right]_\text{div},\hspace{.3cm} \delta^{(2)}Z_{12}\Big|_\text{div} = 0 .
\end{align}
In \FH so far, the $\DR$ definition of the field renormalization constants is employed. We now add finite pieces to compensate for the different normalization of the MSSM and THDM Higgs doublets, redefining
\begin{align}
\delta^{(1)}Z_{ij} & = \delta^{(1)}Z_{ij}\Big|_\text{div} + \delta^{(1)}Z_{ij}\Big|_\text{fin}
\end{align}
with the proper choice, according to  \Eq{InvPropMatRel_Eq},
\begin{align}
\label{fieldredefinitionEq}
\delta^{(1)}Z_{11}\Big|_\text{fin} &= - \Delta\Sigma'_{11},\hspace{.3cm} \delta^{(1)}Z_{22}\Big|_\text{fin} = -\Delta\Sigma'_{22},\hspace{.3cm} \delta^{(1)}Z_{12}\Big|_\text{fin} = -\Delta\Sigma'_{12}.
\end{align}
Since \Eq{InvPropMatRel_Eq} is valid only at the one-loop level, it cannot be applied for the two-loop field counterterms $\delta^{(2)}Z_{ij}$. These two-loop terms, however, drop out  completely (see \App{FieldRenApp} for more details).

With the additional finite parts introduced in the field renormalization constants, the inverse propagator matrix of the MSSM becomes equal to that of effective THDM (with restriction to the same perturbative order). Hence, the combination of the fixed-order (MSSM) and the EFT (THDM) approach is straightforward, which means that the MSSM inverse propagator matrix is replaced by
\begin{align}
\Delta^{-1}_{\widehat\phi\widehat\phi} \rightarrow \Delta^{-1}_{\widehat\phi\widehat\phi} + \Delta^\text{EFT},
\end{align}
where $\Delta^\text{EFT}$ contains the resummed logarithms and corresponding subtraction terms,
\begin{align}
\Delta^\text{EFT} =  \Delta^{-1}_{\widetilde\phi\widetilde\phi}\Big|_{\text{logs}} - \Delta^{-1}_{\widehat\phi\widehat\phi}\Big|_{\text{logs}} .
\end{align}
We checked numerically that the logarithms of the EFT calculation properly recover the logarithmic behavior of the full fixed-order result when restricted to the same perturbative order. For more details on the calculation of the subtraction terms we refer to \cite{Bahl:2016brp,Bahl:2017aev}.

%%%%%%%%%%%%%%%%%%%%%%%%%%%%%%%%%%%%%%%%%%%%%%%%%%%%%%%%%%%%%%%%%%%%%%%%%%%%%%

\subsection{Redefinition of \texorpdfstring{$\tan\beta$}{tanb}}\label{InputtanbSec}

As mentioned above, in \FH by default the $\DR$-scheme is employed for field renormalization of the Higgs doublets and for the renormalization of $\tan\beta$. Thus, there is a renormalization scale entering the diagrammatic calculation. By default, it is chosen to be equal to the pole mass $M_t$ of the top quark. This in particular means that $\tan\beta$ is normally a MSSM  $\DR$ quantity defined at the scale $M_t$.

The redefinition of the field renormalization constants by a finite shift, as described above, has an impact on the renormalization and hence the conceptual definition of $\tan\beta$. In presence of an off-diagonal field renormalization constant, the counterterm of $\tan\beta$ is given by (assuming still $\delta^{(i)}v_1/v_1 = \delta^{(i)}v_2/v_2$)
\begin{align}
\delta^{(1)}\tbe ={}& \frac{1}{2}\tbe\left(\delta^{(1)}Z_{22}-\delta^{(1)}Z_{11}\right) +\frac{1}{2}\left(1-\tbb\right)\,\delta^{(1)}Z_{12}.
\end{align}
For the corresponding two-loop counterterm, see \App{FieldRenApp}. With the finite parts of the field renormalization constants in \Eq{fieldredefinitionEq} and switching to the Higgs basis, we find
\begin{align}
\delta^{(1)}\tbe\Big|_\text{fin} = - \frac{1}{2\cbb}\Delta\Sigma'_{H_1H_2}.
\end{align}
Comparing this result to \Eq{tanbetaMatchingEq}, we realize that $\tan\beta$ by now is not a MSSM quantity anymore, but instead a quantity of the THDM. Furthermore, the scale is changed to $M_A$, since the THDM part in $\Delta\Sigma'_{H_1H_2}$ is evaluated at the scale $M_A$. In conclusion, the finite shift in the field normalization constants of the MSSM leads to the conversion
\begin{align}
\tbe^\MSSM(M_t) \rightarrow \tbe^\THDM(M_A).
\end{align}
Hence, $\tbe^\THDM(M_A)$ is the proper input parameter of the fixed-order calculation.

%%%%%%%%%%%%%%%%%%%%%%%%%%%%%%%%%%%%%%%%%%%%%%%%%%%%%%%%%%%%%%%%%%%%%%%%%%%%%%

\subsection{Conversion of input parameters}\label{Conversion_Sec}

The diagrammatic calculation implemented in \FH employs either the OS or the $\DR$ scheme for the renormalization of the stop sector. In case of an OS renormalization, this means in particular that the stop masses and the stop mixing angle are renormalized on-shell. For the EFT calculation however, respective $\DR$ quantities are needed. Therefore, the parameters have to be converted. As argued in~\cite{Bahl:2016brp}, one-loop conversion including only logarithmic terms is sufficient  to reproduce the diagrammatic OS expressions  from the EFT $\DR$ result. Any further terms in the conversion induce higher order contributions which are presently not under control.

It was noticed in \cite{Hahn:2013ria,Bahl:2016brp} that the conversion of the stop mass scale does not involve large logarithms; only the stop mixing parameter $X_t$ was found to be affected by logarithmic terms. In that previous analysis, a common scale $M_A=\msusy$  was assumed. Here, we extend the conversion formulas to the case of $M_A\ll \msusy$. As in the case of $M_A = \msusy$, we find no large logarithms in the conversion of the stop mass scale $M_S =\msusy$. For the stop mixing parameter, however, additional large logarithms appear in the conversion formula,
\begin{align}\label{XtConv_Eq}
X_t^\DR(M_S) =&
X_t^{\text{OS}}\Bigg\{1+\bigg[\frac{\alpha_s}{\pi}-\frac{3\alpha_t}{16\pi}\big(1-\xf^2\big)\bigg]L-\frac{3}{16\pi}\frac{\alpha_t}{\tb^2}\big(1-\yf^2\big)L_A\Bigg\} ,
\end{align}
using the abbreviations
\begin{align}
\label{abbreviations}
L = \ln\left(\frac{M_S^2}{M_t^2}\right),\quad L_A =
\ln\left(\frac{M_S^2}{M_A^2}\right),   \quad
\xf = \frac{X_t}{M_S}  = \at - \frac{\mf}{t_\beta} \, , \quad
\yf = \at + \mf \tbe \, .
\end{align}
More details and full one-loop expressions for the parameter conversion are given in~\App{ConversionApp}.

%%%%%%%%%%%%%%%%%%%%%%%%%%%%%%%%%%%%%%%%%%%%%%%%%%%%%%%%%%%%%%%%%%%%%%%%%%%%%%
%%%%%%%%%%%%%%%%%%%%%%%%%%%%%%%%%%%%%%%%%%%%%%%%%%%%%%%%%%%%%%%%%%%%%%%%%%%%%%

\section{Comparison to other codes}\label{OtherCodes_Sec}

There are two other publicly available codes for calculating the Higgs pole masses via a THDM matched to the MSSM: the \MhEFT package~\cite{Gabe:2016}, based on \cite{Lee:2015uza}, and the program {\tt FlexibleSUSY}~\cite{Athron:2014yba} in the recent version~\cite{Athron:2017fvs}, based on~\cite{Bagnaschi:2015pwa}. As pointed out in~\cite{Athron:2017fvs}, agreement has been found with the \MhEFT results. We therefore restrict ourselves to a comparison of \FH to \MhEFT (version {\tt 1.1}).

\medskip

The basis of \MhEFT is a pure EFT calculation. Therefore, terms suppressed by heavy scales are absent. Apart from this obvious distinction, there are a few more differences to \FH:
\begin{itemize}
\item
\MhEFT does not employ the $\DR$ scheme for renormalization of the SUSY parameters. Instead, $\MS$ renormalization is used. Therefore, conversion of the input parameters is needed for the comparison with \FH. Corresponding conversion formulas can be found in \cite{Draper:2016pys}.

Although, as argued in \cite{Bahl:2017aev}, this conversion will induce unwanted higher order terms, it is currently the only way to compare the two results,  since neither \FH offers the possibility of a $\MS$ renormalization nor \MhEFT the possibility of a $\DR$ renormalization. In practice it is a viable method since the numerical impact of the conversion is almost negligible, owing to the small numerical difference between $\MS$ and $\DR$ parameters.

\item
The EFT calculations entering \FH and  \MhEFT differ in various aspects. \MhEFT assumes a type~II THDM as the effective THDM in the evolution equations. Furthermore, EWino contributions to the various threshold corrections are neglected. Also in the RGEs, EWino contributions are neglected at the two-loop level and only taken into account in approximate form at the one-loop level. In addition the one-loop threshold corrections between the SM and the THDM are neglected for the top Yukawa coupling and approximated for the SM Higgs self-coupling (i.e., the heavy Higgs contribution to the one-loop threshold correction between the SM and the MSSM is used). On the other hand, \MhEFT has implemented an approximation for the \order{\alt^2} threshold corrections for the quartic couplings by including the known \order{\alt^2} threshold correction from matching the SM to the MSSM in $\lambda_2$, whereas all other self-couplings receive no \order{\alt^2} threshold correction.

\item
In \MhEFT, the THDM self-energies $\widetilde\Sigma_{\widetilde H_1\widetilde H_2}$ and $\widetilde\Sigma_{\widetilde H_2\widetilde H_2}$ (see \Eq{InversePropMatrixEFT_Eq}) are neglected. Thereby, terms of \order{M_t/M_A} are missed.
\end{itemize}
These differences should be kept in mind, when interpreting the numerical results of the comparison presented in \Sec{NumResults_Sec}.

%%%%%%%%%%%%%%%%%%%%%%%%%%%%%%%%%%%%%%%%%%%%%%%%%%%%%%%%%%%%%%%%%%%%%%%%%%%%%%
%%%%%%%%%%%%%%%%%%%%%%%%%%%%%%%%%%%%%%%%%%%%%%%%%%%%%%%%%%%%%%%%%%%%%%%%%%%%%%

\section{Numerical results}\label{NumResults_Sec}

In this Section, we investigate the numerical impact of the implementation of an effective THDM into \FH. This means in practice that we compare the results from the latest version {\tt  FeynHiggs2.14.1} to those from the  calculation presented in this paper, which is implemented in a still private \FH version based on {\tt FeynHiggs2.14.1}. In addition, we show results from {\tt FeynHiggs2.14.0} to point out the impact of the non-degenerate \order{\alt^2} threshold corrections~\cite{Bagnaschi:2017xid} , which were implemented as a  new feature in {\tt FeynHiggs2.14.1}. The degenerate \order{\alt^2} threshold corrections~\cite{Vega:2015fna}, used in {\tt FeynHiggs2.14.0}, implicitly assume $M_A = \msusy$. We furthermore compare the results of the calculation presented in this paper to those of \MhEFT.

For illustration of the  numerical effects, we investigate simplified scenarios with  a common mass scale $M_S$ for all sfermions, and $M_\chi$ for the EWinos, setting (if not stated otherwise)
\begin{align}
M_S &\equiv \msusy, \hspace{.5cm}  M_\chi \equiv M_1 = M_2 = \mu , \hspace{.5cm} A_{e,\mu,\tau,u,d,c,s,b} = 0.
\end{align}
Also the gluino mass $M_{\tilde g}$ is set equal to $\msusy$\footnote{Note that our EFT calculation also allows to treat scenarios with $M_{\tilde g}$ as an independent parameter. The numerical effect of the additional threshold, however, is small since the dominant two-loop effect is already captured by the fixed-order calculation (see also \cite{Bahl:2016brp})}. As default values for the figures, we set $\msusy = 100\tev$ and $M_\chi = 500\gev$. In combination with low $M_A$ and $\tan\beta$ values, this choice maximizes the numerical impact of the effective THDM in the phenomenologically most interesting region of $M_h \sim 125\gev$.

The numerical impact of the effective THDM can also get large for $\msusy \sim 1\tev$ and moderate values of $\tan\beta$, if $\mu > \msusy$. This corresponds, however, to a hierarchy which we did not cover in this paper.

For the SUSY parameters, we use the $\DR$-scheme with the corresponding renormalization scale being $\msusy$. The $\DR$ scheme is also used for $X_t$ (except in \Fig{FH_XTOS_Fig}, where the OS scheme is used). $\tan\beta$ is defined as $\tan\beta^\THDM(M_A)$, unless stated otherwise.

Aside from the simplified scenarios, we also study a more complicated situation, the ``low-$\tan\beta$-high'' scenario proposed by the LHC Higgs Cross Section Working Group in~\cite{Bagnaschi:2015hka}.

%%%%%%%%%%%%%%%%%%%%%%%%%%%%%%%%%%%%%%%%%%%%%%%%%%%%%%%%%%%%%%%%%%%%%%%%%%%%%%

\subsection{Shifts from \texorpdfstring{$\tan\beta$}{tanb} definition}

As explained in \Sec{Combination_Section}, we account for the different normalization of the Higgs doublets in the full MSSM and the effective THDM by introducing a finite shift in the field renormalization constants of the fixed-order calculation. This changes the definition of $\tan\beta$: from a MSSM quantity to one of the THDM, along with a  change of the renormalization scale from $M_t$ (the default of \FH) to $M_A$.

%%%%%%%%%%%%%%%%%%%%%%%%% F I G U R E %%%%%%%%%%%%%%%%%%%%%%%%%%%%%%%%%%%%%%%%
\begin{figure}\centering
\begin{minipage}{.48\textwidth}\centering
\includegraphics[width=\textwidth]{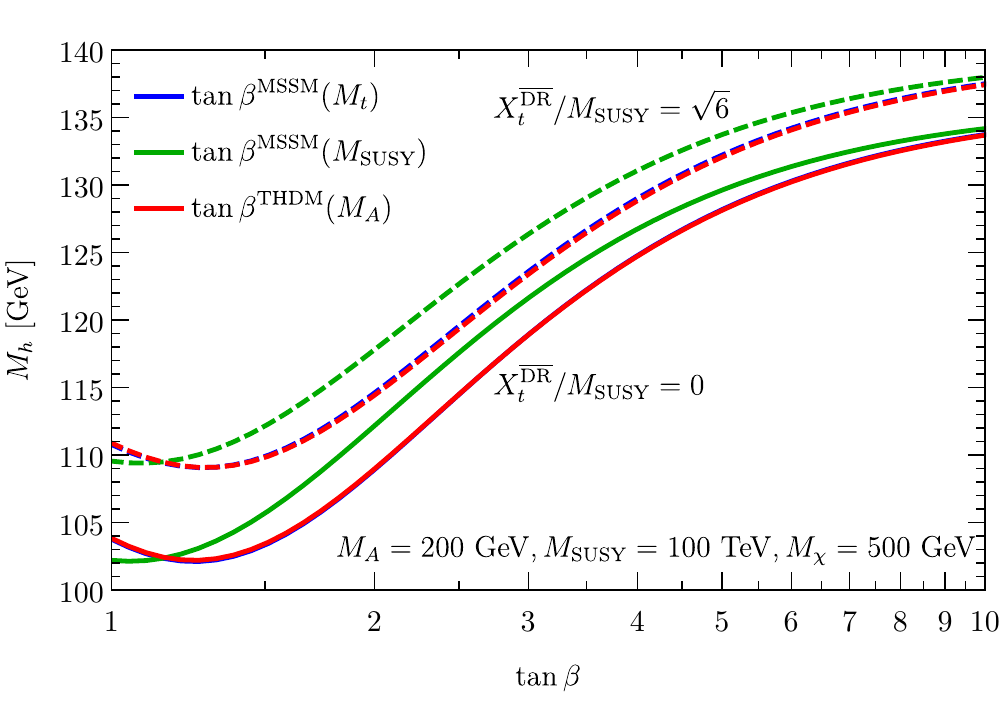}
\end{minipage}
\begin{minipage}{.48\textwidth}\centering
\includegraphics[width=\textwidth]{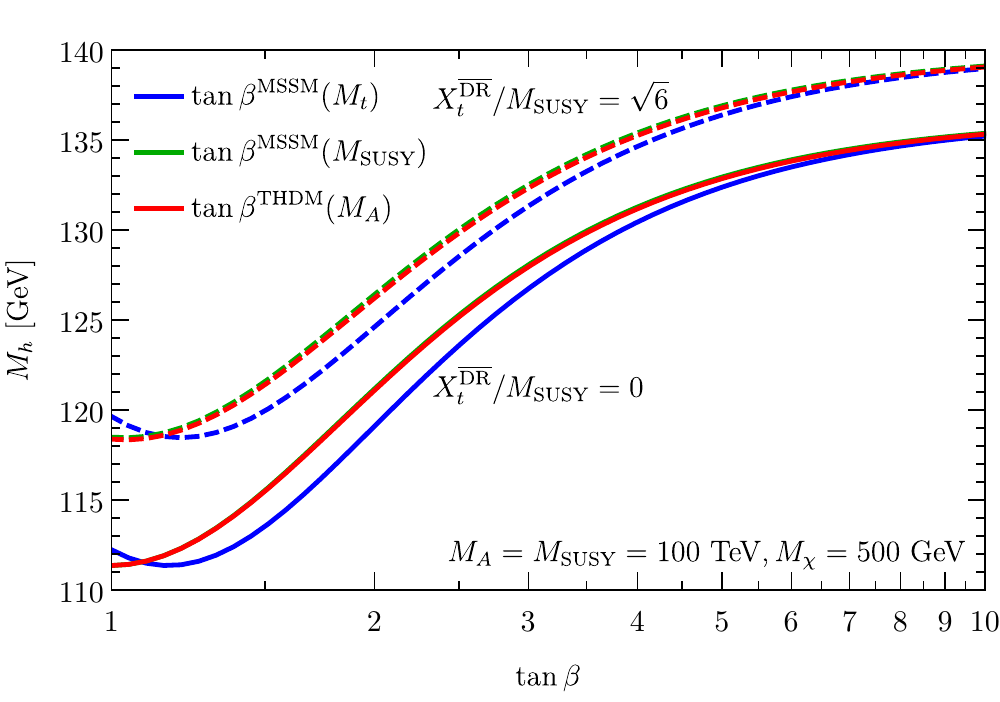}
\end{minipage}
\caption{Left: $M_h$ as function of $\tan\beta$ for $X_t^\DR/\msusy = 0$ (solid) and $X_t^\DR/\msusy = \sqrt{6}$ (dashed) in a scenario with a low $M_A$  and with different definitions of $\tan\beta$: in the MSSM at the scale $M_t$ (blue) and at the scale $\msusy$ (red, overlapping with blue),  and in the THDM at the scale $M_A$ (green). Right: Same signature, but for $M_A = \msusy$ (overlapping red and green curves).}
\label{tanb_redef_Fig}
\end{figure}
%%%%%%%%%%%%%%%%%%%%%%%%% F I G U R E %%%%%%%%%%%%%%%%%%%%%%%%%%%%%%%%%%%%%%%%

We analyze the numerical effect of this redefinition in \Fig{tanb_redef_Fig}. It shows results of \FH for $M_h$ using different definitions of $\tan\beta$: $\tan\beta^\MSSM(M_t)$ (default definition in \FH), $\tan\beta^\THDM(M_A)$ (default definition in this Section) and, for comparison, $\tan\beta^\MSSM(\msusy)$ (by shifting the renormalization scale to $\msusy$). Accordingly, the meaning of the horizontal axis is not the same for the different curves.

The left panel displays a low-$M_A$ scenario. The curves for $\tan\beta^\MSSM(M_t)$ and  $\tan\beta^\THDM(M_A)$ are very close to each other. This is essentially due to $M_A \sim M_t$, the additional non-logarithmic threshold correction of $\tan\beta$ between the THDM and the MSSM in  \Eq{THDMtoMSSMtbEq} has only a small numerical impact. In contrast, there is a large hierarchy between $M_t$ (or $M_A$) and $\msusy$. Therefore, the third curve for $\tan\beta^\MSSM(\msusy)$ is shifted upwards for low $\tan\beta$, by up to $\sim 2 \gev$ for $\tan\beta\gtrsim 1.2$. This shift  shrinks for rising $\tan\beta$, as a consequence of the decreasing dependence of $M_h$ on $\tan\beta$. For $\tan\beta \lesssim 1.2$ a small downwards shift of up to 2 GeV is visible.

In the right panel, the same set of curves is displayed, but now for $M_A$ equal to $\msusy$. Therefore, the curves using $\tan\beta^\THDM(M_A)$ and $\tan\beta^\MSSM(\msusy)$ are very close; again, the additional non-logarithmic threshold correction of $\tan\beta$ between the THDM and the MSSM turns out to be negligible. Due to the large scale separation between $M_t$ and $\msusy$ the curve using $\tan\beta^\MSSM(M_t)$ is shifted downwards by up to 2~GeV between $\tan\beta\sim 1.2$ and $\tan\beta\sim 6$. For $\tan\beta\lesssim 1.2$, a small upwards shift  up to 1 GeV is visible.

Note that for the rest of this section, $\tan\beta$ is defined as $\tan\beta^\THDM(M_A)$ for all displayed results.

%%%%%%%%%%%%%%%%%%%%%%%%%%%%%%%%%%%%%%%%%%%%%%%%%%%%%%%%%%%%%%%%%%%%%%%%%%%%%%

\subsection{Impact of the effective THDM}\label{NumTHDM_Sec}

Having investigated the numerical effect of different definitions of $\tan\beta$, we now scrutinize the impact of the main result of this paper -- the implementation of an effective THDM into the hybrid framework of \FH.

%%%%%%%%%%%%%%%%%%%%%%%%% F I G U R E %%%%%%%%%%%%%%%%%%%%%%%%%%%%%%%%%%%%%%%%
\begin{figure}\centering
\begin{minipage}{.48\textwidth}\centering
\includegraphics[width=\textwidth]{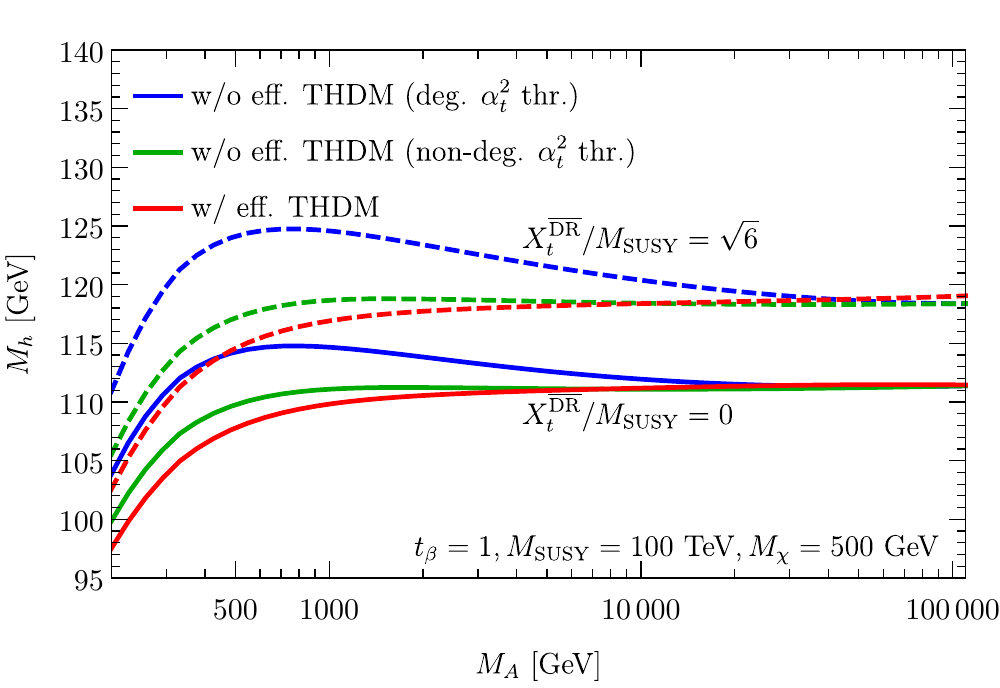}
\end{minipage}
\begin{minipage}{.48\textwidth}\centering
\includegraphics[width=\textwidth]{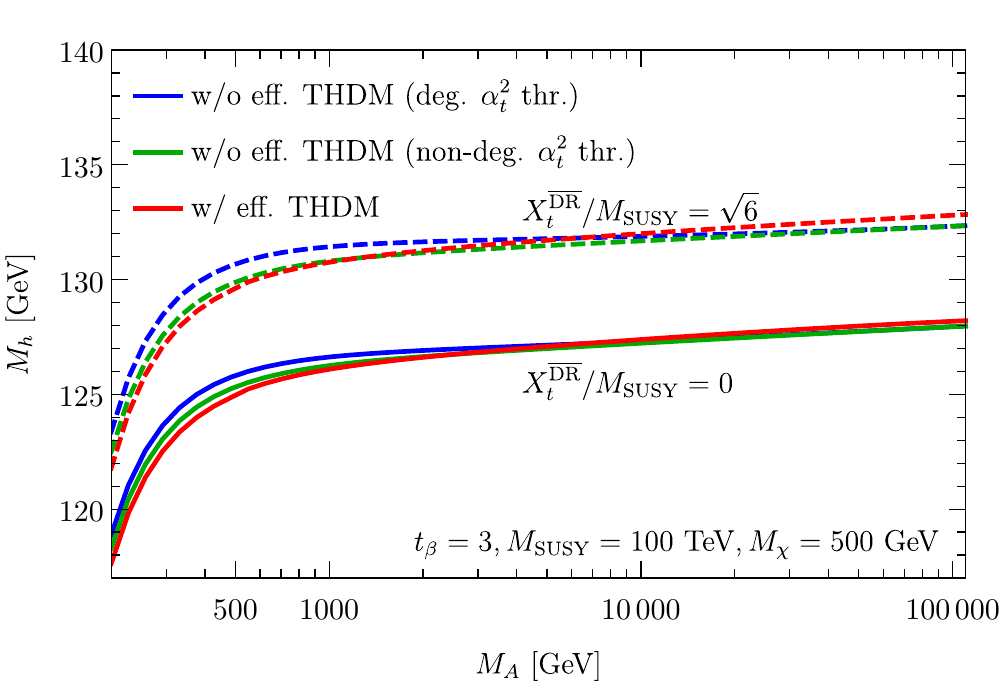}
\end{minipage}
\caption{$M_h$ as a function of $M_A$ for $X_t^\DR/\msusy = 0$ (solid) and $X_t^\DR/\msusy = \sqrt{6}$ (dashed). Left: $\tan\beta = 1$.  Right: $\tan\beta = 3$. The results of \FH without effective THDM -- using the degenerate \order{\alt^2} threshold correction (blue) and using the non-degenerate \order{\alt^2} threshold correction (green) -- are compared with the results of \FH with effective THDM (red).}
\label{FH_MAvar_Fig}
\end{figure}
%%%%%%%%%%%%%%%%%%%%%%%%% F I G U R E %%%%%%%%%%%%%%%%%%%%%%%%%%%%%%%%%%%%%%%%

In \Fig{FH_MAvar_Fig}, we compare the results of various stages of \FH by showing $M_h$ in dependence of $M_A$: the previous version without an intermediate effective THDM using degenerate \order{\alt^2} threshold corrections (corresponding to version {\tt 2.14.0}) as well as using non-degenerate \order{\alt^2} threshold corrections (corresponding to version {\tt 2.14.1}), and the new version with the effective THDM implemented. One observes that the curves of \FH with and without effective THDM converge to each other for rising $M_A$. This is expected since for $M_A = \msusy$, the SM+EWinos can be matched directly to the MSSM and no effective THDM is needed. The small remaining deviation of the THDM curve for $M_A=\msusy$ and $X_t^\DR/\msusy =\sqrt{6}$ is caused by the \order{\alt^2} threshold correction, which is part of the current \FH (without effective THDM) but not available for the THDM-modified version. For $M_A \ll \msusy$ we observe sizeable shifts, in particular in the left panel where $\tan\beta$ is set to 1. The step from degenerate to non-degenerate \order{\alt^2} threshold corrections already induces a downwards shift of up to 5 GeV  for vanishing stop mixing and of up to 7 GeV for $X_t^\DR/\msusy =\sqrt{6}$. Implementing now the effective THDM leads to a further shift downwards by up to 2 GeV for vanishing stop mixing and up to 3 GeV for $X_t^\DR/\msusy =\sqrt{6}$.

In the right panel with $\tan\beta = 3$, the curves show the same qualitative behavior, i.e.\ for low $M_A$ the implementation of an effective THDM shifts $M_h$ downwards, but in comparison to the results with $\tan\beta=1$, the effects are less pronounced ($\lesssim 1.5 \gev$).

%%%%%%%%%%%%%%%%%%%%%%%%% F I G U R E %%%%%%%%%%%%%%%%%%%%%%%%%%%%%%%%%%%%%%%%
\begin{figure}\centering
\begin{minipage}{.48\textwidth}\centering
\includegraphics[width=\textwidth]{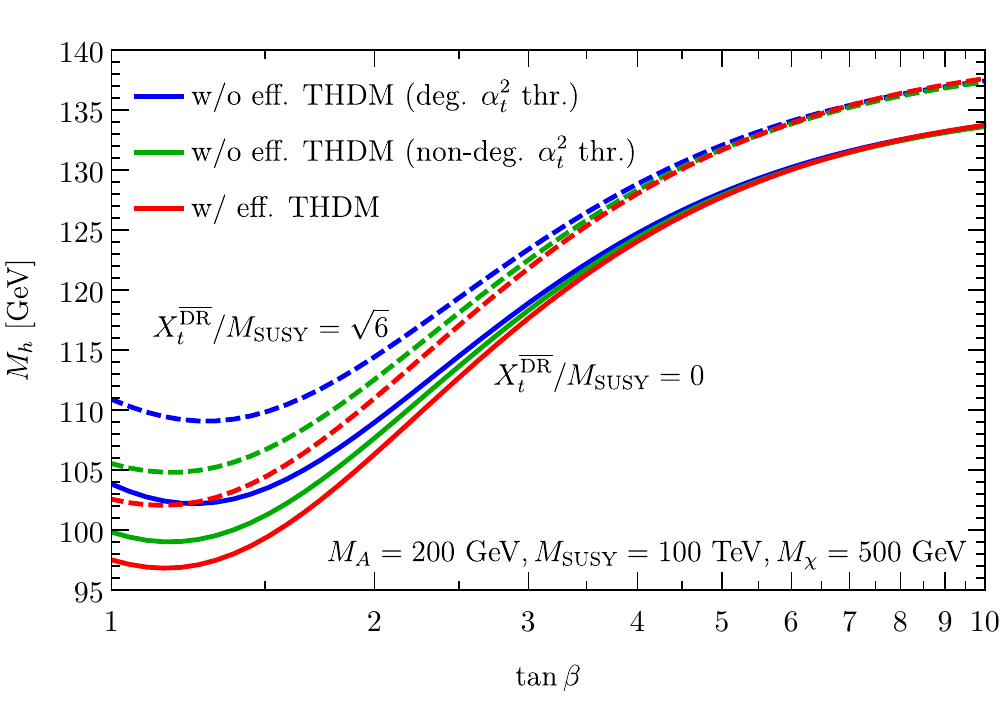}
\end{minipage}
\begin{minipage}{.48\textwidth}\centering
\includegraphics[width=\textwidth]{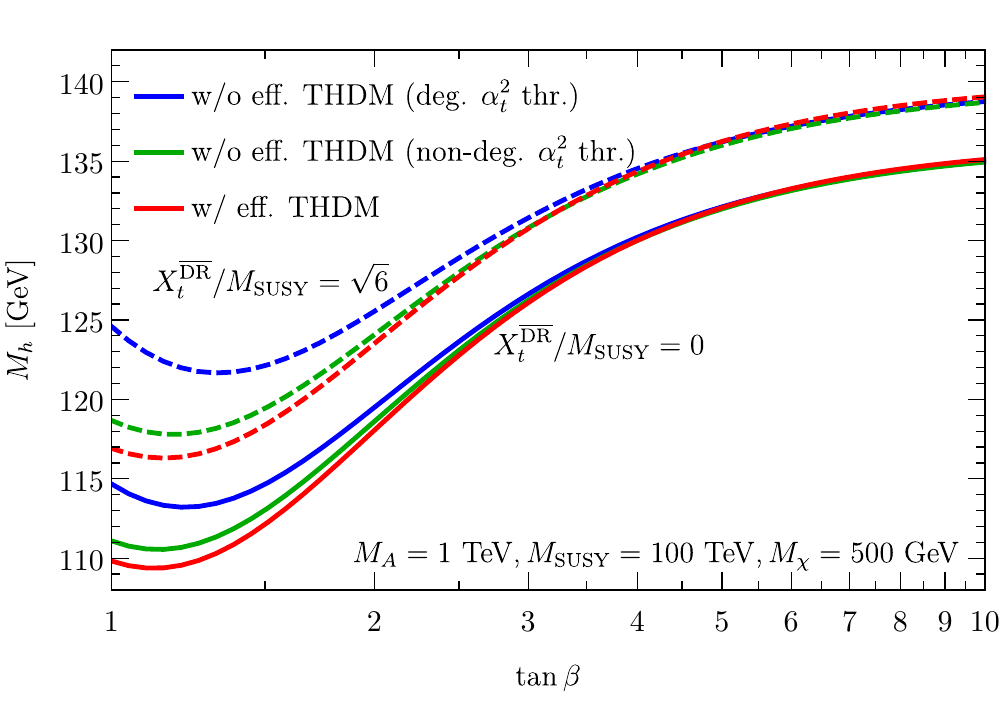}
\end{minipage}
\caption{$M_h$ as a function of $\tan\beta$ for $X_t^\DR/\msusy = 0$ (solid) and $X_t^\DR/\msusy = \sqrt{6}$ (dashed). Left: $M_A = 200\gev$. Right: $M_A = 1\tev$. The results of \FH without effective THDM -- using the degenerate \order{\alt^2} threshold correction (blue) and using the non-degenerate \order{\alt^2} threshold correction (green) -- are compared with the results of \FH with effective THDM (red).}
\label{FH_TBvar_Fig}
\end{figure}
%%%%%%%%%%%%%%%%%%%%%%%%% F I G U R E %%%%%%%%%%%%%%%%%%%%%%%%%%%%%%%%%%%%%%%%

\medskip
This strong dependence on $\tan\beta$ is visualized more specifically in \Fig{FH_TBvar_Fig},  where  $M_h$ is shown versus $\tan\beta$ for the same cases as in \Fig{FH_MAvar_Fig}. In the left panel, the difference  between \FH with and without effective THDM is displayed for $M_A = 200\gev$ and in the right panel for a larger value $M_A=1$~TeV. The effects of the various steps of improvement are most pronounced for low  $\tan\beta$ and shrink quickly  for increasing values; for $\tan\beta \gtrsim 5$, the shifts are negligible. Again, the use of the non-degenerate \order{\alt^2} threshold correction brings the result without effective THDM closer to that with effective THDM. The curves in the left and right panel behave very similar; the overall $M_h$ values are higher for larger $M_A$, but the shifts remain of the same size despite the slightly reduced hierarchy between $M_A$ and $\msusy$.

%%%%%%%%%%%%%%%%%%%%%%%%% F I G U R E %%%%%%%%%%%%%%%%%%%%%%%%%%%%%%%%%%%%%%%%
\begin{figure}\centering
\begin{minipage}{.48\textwidth}\centering
\includegraphics[width=\textwidth]{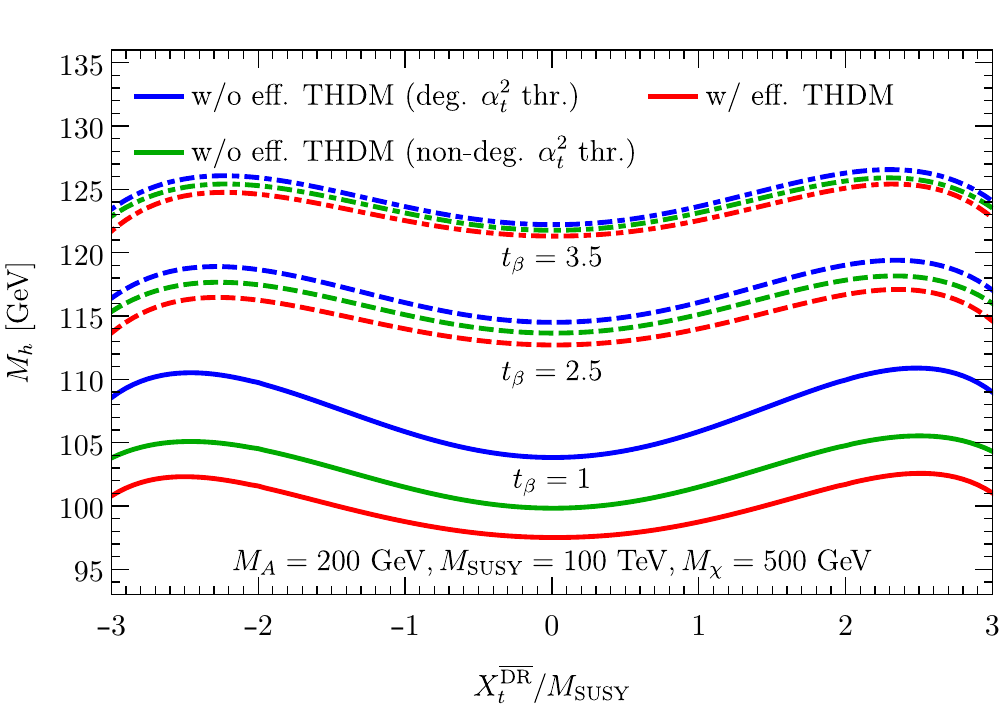}
\end{minipage}
\begin{minipage}{.48\textwidth}\centering
\includegraphics[width=\textwidth]{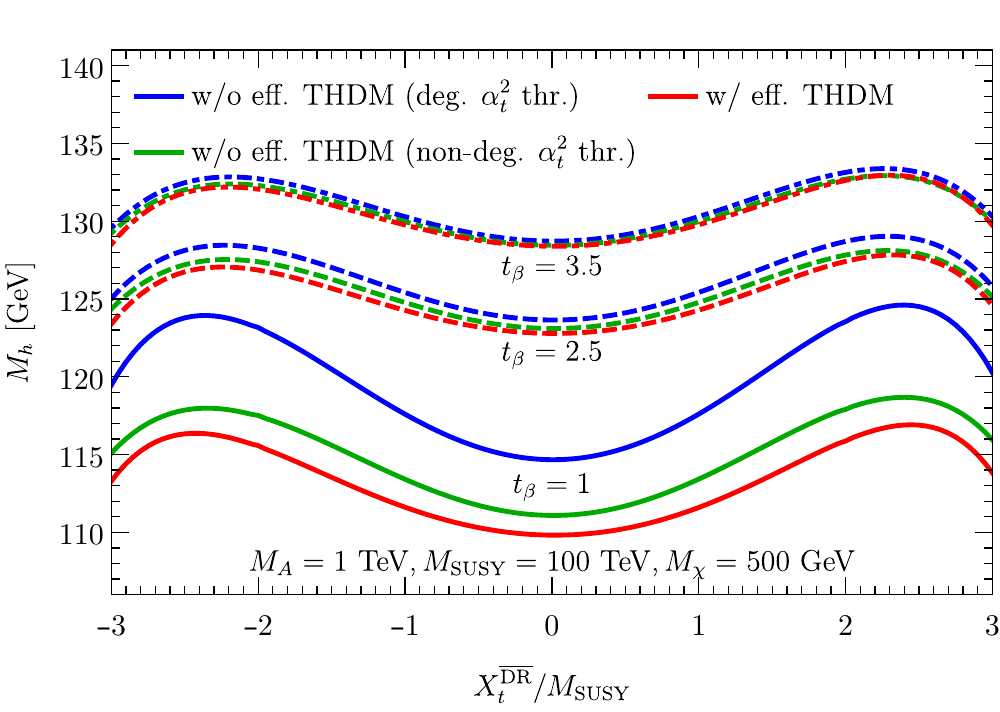}
\end{minipage}
\caption{$M_h$ as a function of $X_t^\DR/\msusy$ for $\tan\beta = 1$, (solid) $\tan\beta = 2.5$ (dashed), and $\tan\beta = 3.5$ (dotdashed). Left: $M_A = 200\gev$. Right:  $M_A = 1\tev$. The results of \FH without effective THDM -- using the degenerate \order{\alt^2} threshold correction (blue) and using the non-degenerate \order{\alt^2} threshold correction (green) -- are compared with the results of \FH with effective THDM (red).}
\label{FH_XTvar_Fig}
\end{figure}
%%%%%%%%%%%%%%%%%%%%%%%%% F I G U R E %%%%%%%%%%%%%%%%%%%%%%%%%%%%%%%%%%%%%%%%

\medskip

Next, the dependence on  the stop-mixing parameter $X_t^\DR$ is analyzed in \Fig{FH_XTvar_Fig}, presenting $M_h$  versus $X_t^\DR/\msusy$ for two different mass scales $M_A = 200\gev$ (left) and $M_A = 1$~TeV (right). As one can see, the difference between $M_h$ predicted by \FH with and without effective THDM is only mildly dependent on $X_t^\DR/\msusy$. For all values, the effect of including the THDM is a downwards shift of $M_h$, becoming smaller for increasing $\tan\beta$.

From a phenomenological point of view, shifting the curves according to the various levels of improvement is relevant for the proper determination of the parameter range that predicts $M_h$ compatible with the measurement.  We have kept in all the figures the case with degenerate \order{\alt^2} threshold correction in the version without THDM in order to point out the significance of going to the non-degenerate \order{\alt^2} threshold correction (realized in {\tt FeynHiggs2.14.1}) which already accounts for a substantial part of the shift when turning to the new version with the effective THDM.

%%%%%%%%%%%%%%%%%%%%%%%%% F I G U R E %%%%%%%%%%%%%%%%%%%%%%%%%%%%%%%%%%%%%%%%
\begin{figure}\centering
\begin{minipage}{.48\textwidth}\centering
\includegraphics[width=\textwidth]{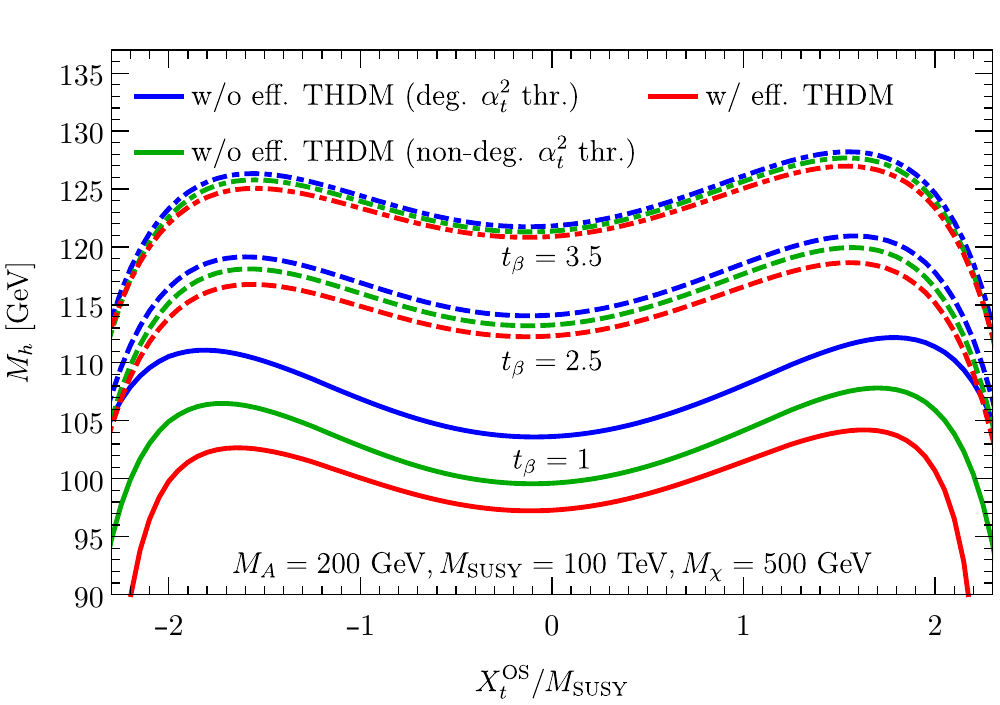}
\end{minipage}
\begin{minipage}{.48\textwidth}\centering
\includegraphics[width=\textwidth]{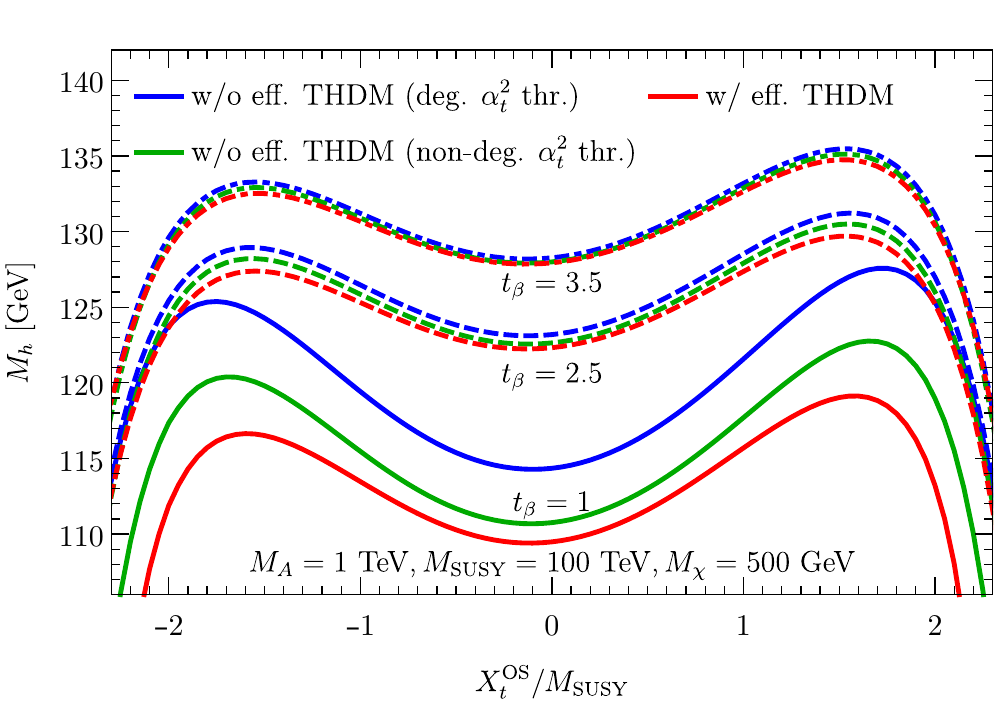}
\end{minipage}
\caption{$M_h$ as a function of $X_t^\OS/\msusy$ for $\tan\beta = 1$ (solid), $\tan\beta = 2.5$ (dashed), and $\tan\beta = 3.5$ (dotdashed). Left: $M_A = 200 \gev$. Right: $M_A = 1\tev$. The results of \FH without effective THDM -- using the degenerate \order{\alt^2} threshold correction (blue) and using the non-degenerate \order{\alt^2} threshold correction (green) -- are compared with the results of \FH with effective THDM (red).}
\label{FH_XTOS_Fig}
\end{figure}
%%%%%%%%%%%%%%%%%%%%%%%%% F I G U R E %%%%%%%%%%%%%%%%%%%%%%%%%%%%%%%%%%%%%%%%

\medskip

So far, all the numerical results refer to the $\DR$ scheme for the stop-sector renormalization. As a distinct feature of \FH, also the OS scheme can be used for renormalizing the stop input parameters. In order to illustrate the use of  OS renormalization, we include \Fig{FH_XTOS_Fig} as the equivalent  of \Fig{FH_XTvar_Fig}, now in the OS scheme, displaying the $M_h$ dependence on $X_t^\OS/\msusy$ for $M_A = 200\gev$ (left) and for $M_A = 1\tev$ (right). The overall behavior of the results is similar to the results obtained in the $\DR$ scheme; also the shifts when turning to the THDM case are similar in size, although slighty more pronounced in the OS scheme.

Here, it is however important to note that the shift between \FH with and without effective THDM depends sensitively on the Higgsino mass parameter $\mu$ when the OS scheme is used\footnote{$\mu$ is set to $M_\chi = 500\gev$ in \Fig{FH_XTOS_Fig}}. This is due to the needed conversion of $X_t$ between the $\DR$ and the OS scheme, according to~\Eq{XtConv_Eq}, which involves an extra term  that can become large for $M_A \ll \msusy$, low $\tan\beta$, $\mu \sim \msusy$ and $X_t^\text{OS}/\msusy \sim 2$, inducing large differences between $X_t^\OS$ and $X_t^\DR$. This signals that in those regions the one-loop conversion is insufficient yielding unreliable results for $M_h$, and recommends the use of the $\DR$ scheme.

%%%%%%%%%%%%%%%%%%%%%%%%% F I G U R E %%%%%%%%%%%%%%%%%%%%%%%%%%%%%%%%%%%%%%%%
\begin{figure}\centering
\begin{minipage}{.48\textwidth}\centering
\includegraphics[width=\textwidth]{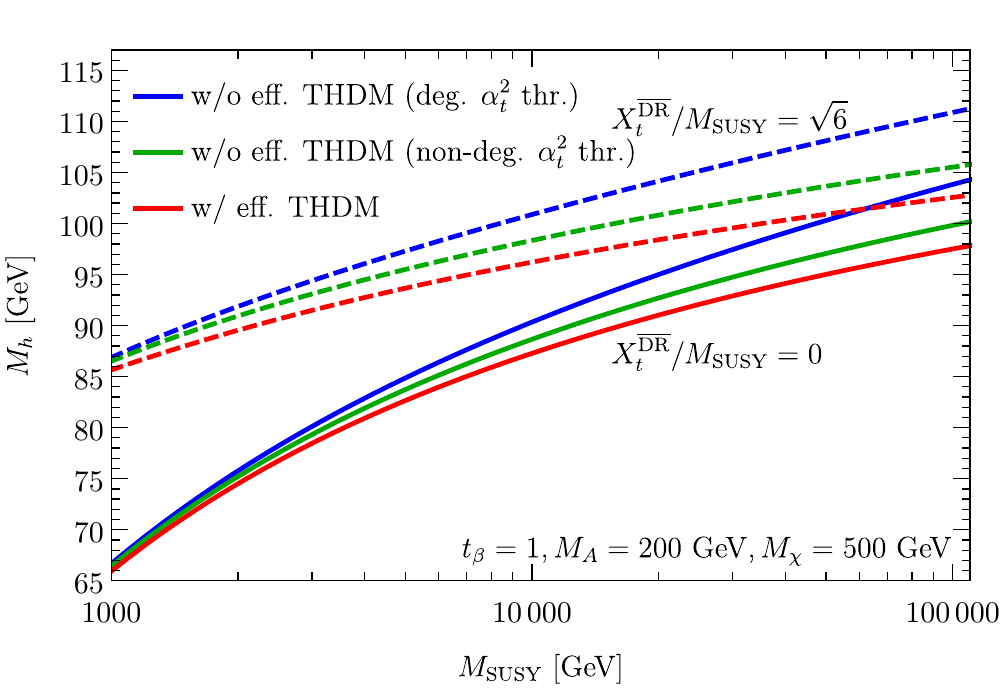}
\end{minipage}
\begin{minipage}{.48\textwidth}\centering
\includegraphics[width=\textwidth]{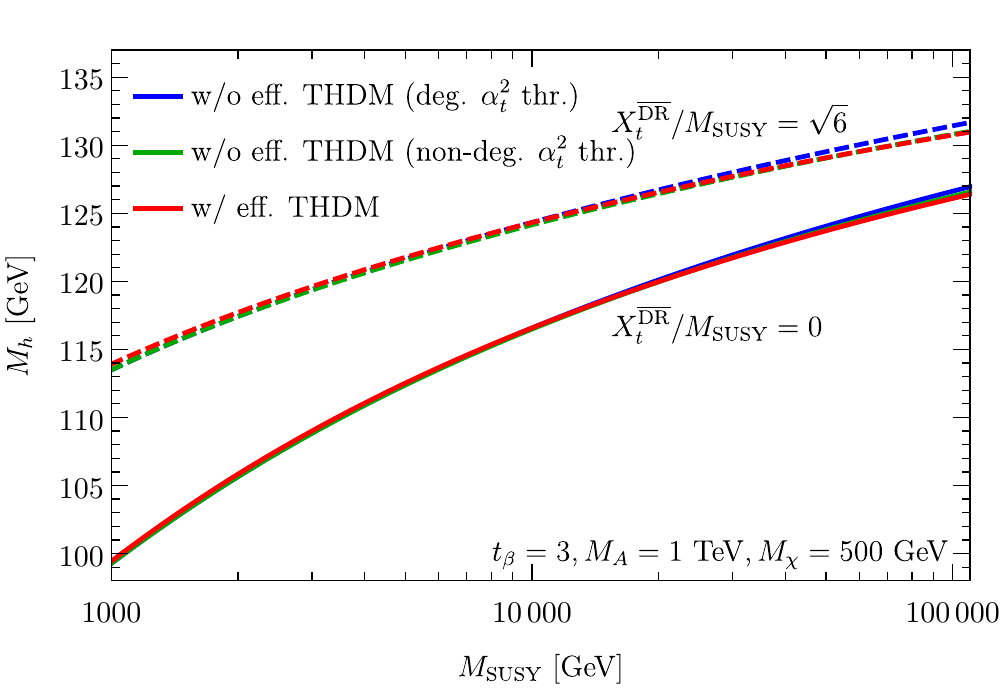}
\end{minipage}
\caption{$M_h$ as a function of $\msusy$ for $X_t^\DR/\msusy = 0$ (solid) and $X_t^\DR/\msusy = \sqrt{6}$ (dashed). Left: $\tan\beta =1$ and $M_A = 200 \gev$. Right: $\tan\beta = 3$ and $M_A = 1\tev$. The results of \FH without effective THDM -- using the degenerate \order{\alt^2} threshold correction (blue) and using the non-degenerate \order{\alt^2} threshold correction (green) -- are compared with the results of \FH with effective THDM (red).}
\label{FH_MSUSYvar_Fig}
\end{figure}
%%%%%%%%%%%%%%%%%%%%%%%%% F I G U R E %%%%%%%%%%%%%%%%%%%%%%%%%%%%%%%%%%%%%%%%

\medskip

The $\msusy$ scale dependence of the effect from implementing the THDM is explicitly shown in \Fig{FH_MSUSYvar_Fig}. In the left panel, we set $\tan\beta = 1$ and $M_A = 200\gev$ to maximize the shift for illustrational purposes. Even for $\msusy \sim$ few TeV, a sizeable shift occurs between the results with and without effective THDM, despite the small hierarchy between $M_A$ and $\msusy$. Phenomenologically this observation is, however, of less interest since the Higgs mass values reached are below 115 GeV over the whole considered range of $\msusy$.

The configuration in the right panel of \Fig{FH_MSUSYvar_Fig}, with $\tan\beta = 3$ and $M_A = 1\tev$, is more relevant for phenomenology since $M_h\sim 125\gev$ can be reached for $\msusy\sim 10\tev$ (and $X_t^\DR/\msusy = \sqrt{6}$). The difference between the results from \FH with and without effective THDM, however, is negligible for $\msusy \lesssim 20\tev$. We conclude that in the commonly considered scenarios with stop masses around the TeV scale, $M_\chi \le \msusy$ and the $h$ boson playing the role of the SM Higgs boson the additional corrections from an intermediate THDM are negligible.

%%%%%%%%%%%%%%%%%%%%%%%%%%%%%%%%%%%%%%%%%%%%%%%%%%%%%%%%%%%%%%%%%%%%%%%%%%%%%%

\subsection{Results for the heavier Higgs bosons}

The role of the SM-like Higgs boson can not only be played by the $h$ boson, also the $H$ boson is a potential candidate (see~\cite{Bechtle:2016kui,Haber:2017erd} for recent studies) and deserves a closer inspection. In the following, we investigate the prediction for the mass of $H$ boson within our hybrid approach.

In this class of scenarios $M_A$ is smaller than $M_t$. In consequence, the proper EFT at the electroweak scale is the THDM and not the SM. In the present study, we approximate the values of the SM $\MS$ couplings ($y_t, g_1,g_2,g_3$)  at the scale $M_t$ computed in \cite{Buttazzo:2013uya} as boundary values for the EFT calculation. Thus, the EFT at the scale $M_t$ is replaced by the SM, which is then matched to the THDM. This procedure avoids the detailed calculation of the THDM $\MS$ couplings at the electroweak scale, but neglects THDM-specific terms (i.e., terms of order \order{M_t/M_A}).

%%%%%%%%%%%%%%%%%%%%%%%%% F I G U R E %%%%%%%%%%%%%%%%%%%%%%%%%%%%%%%%%%%%%%%%
\begin{figure}\centering
\begin{minipage}{.48\textwidth}\centering
\includegraphics[width=\textwidth]{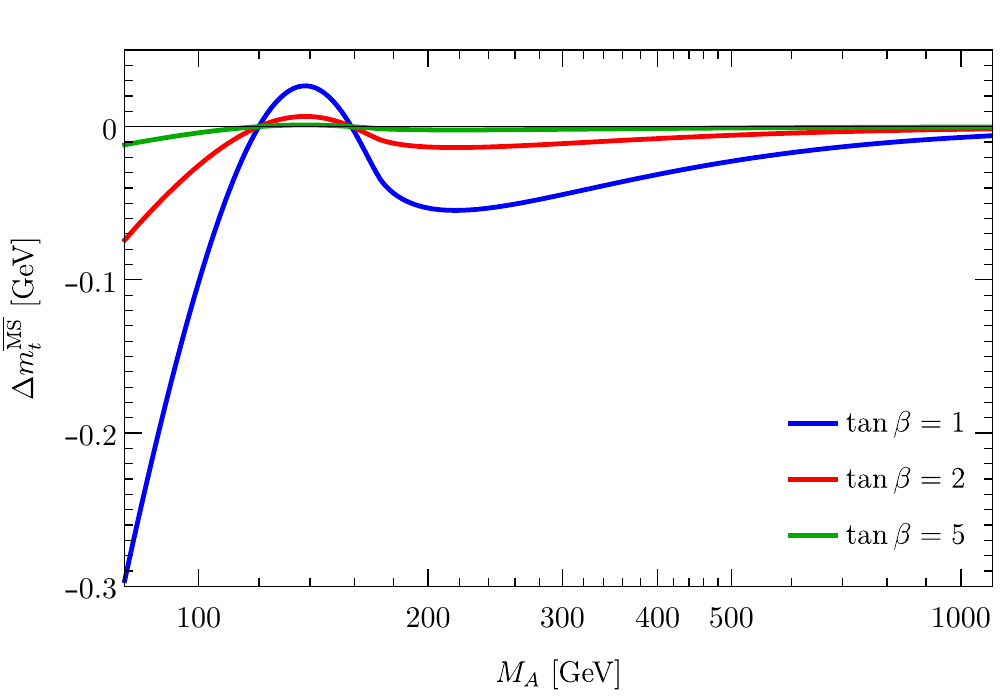}
\end{minipage}
\caption{Shifts to the SM $\MS$ top mass induced by non-SM Higgs bosons as a function of $M_A$ for $\tan\beta = 1$ (blue), $\tan\beta = 2$ (red) and $\tan\beta = 5$ (green).}
\label{THDM_dMT_Fig}
\end{figure}
%%%%%%%%%%%%%%%%%%%%%%%%% F I G U R E %%%%%%%%%%%%%%%%%%%%%%%%%%%%%%%%%%%%%%%%

In order to estimate the uncertainty arising from this approximate determination of the boundary values, we investigate the numerical effect of the presence of extra Higgs bosons for the determination of the $\MS$ top mass, as the parameter with the strongest impact in the Higgs-boson mass calculation. As a rule of thumb, a shift of 1 GeV in the  top mass implies a shift of the same size in the Higgs masses. As displayed in \Fig{THDM_dMT_Fig}, the shift induced by the presence of extra non-SM Higgs bosons is at most 300 MeV. This value is reached if $M_A = 80\gev$ and $\tan\beta =1$. For larger $M_A$ and/or larger $\tan\beta$, the shift is quickly diminished below 100 MeV. Accordingly, we estimate the uncertainty induced by neglecting the non-SM Higgs bosons when extracting the $\MS$ couplings to be below $\order{0.5\gev}$.

%%%%%%%%%%%%%%%%%%%%%%%%% F I G U R E %%%%%%%%%%%%%%%%%%%%%%%%%%%%%%%%%%%%%%%%
\begin{figure}\centering
\begin{minipage}{.48\textwidth}\centering
\includegraphics[width=\textwidth]{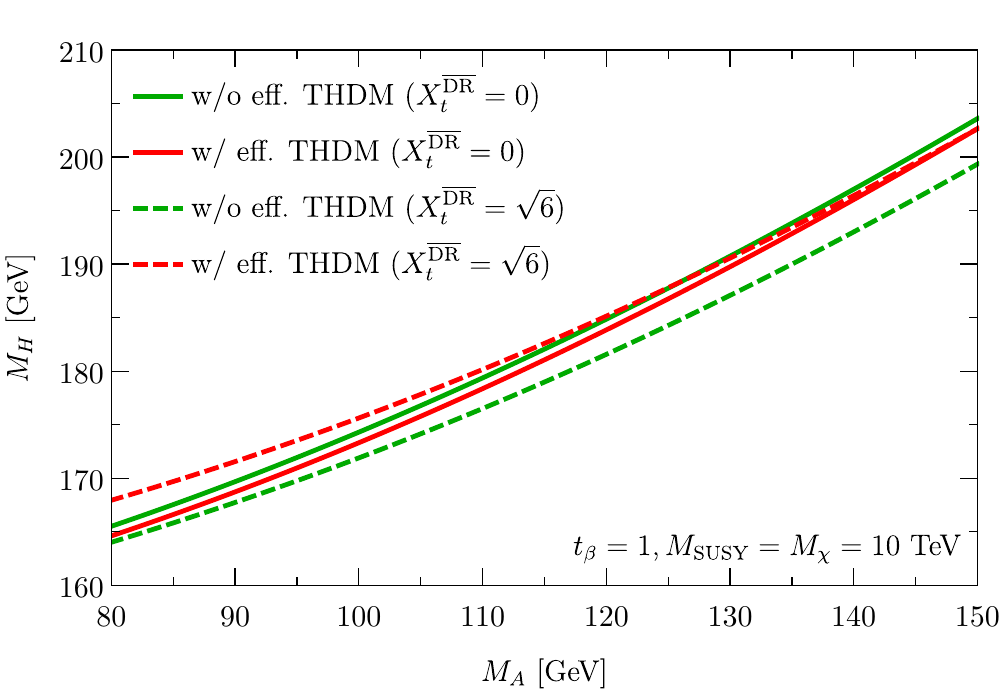}
\end{minipage}
\begin{minipage}{.48\textwidth}\centering
\includegraphics[width=\textwidth]{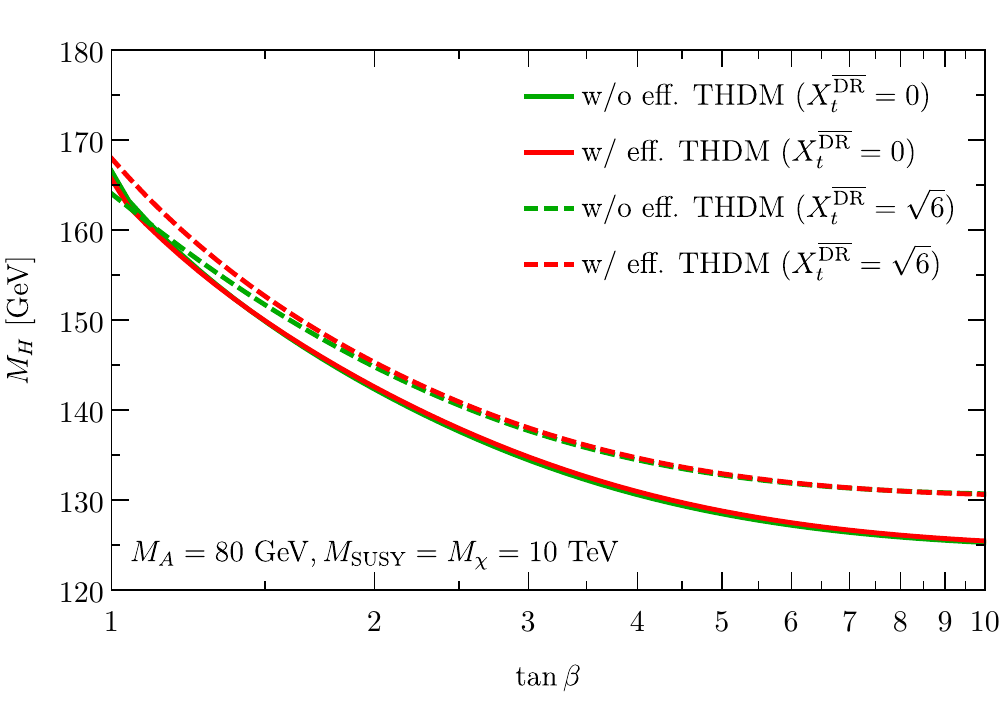}
\end{minipage}
\caption{Left: $M_H$ as a function of $M_A$ for $\tan\beta = 1$. Right: $M_H$ as a function of $\tan\beta$ for $M_A = 80 \gev$. The results of \FH without effective THDM using the non-degenerate \order{\alt^2} threshold correction (green) and with effective THDM (red) are compared. $X_t^\DR/\msusy = 0$ (solid) and $X_t^\DR/\msusy = \sqrt{6}$ (dashed).}
\label{FH_MH_Fig}
\end{figure}
%%%%%%%%%%%%%%%%%%%%%%%%% F I G U R E %%%%%%%%%%%%%%%%%%%%%%%%%%%%%%%%%%%%%%%%

In \Fig{FH_MH_Fig}, the dependence of $M_H$ on $M_A$ (left) and on $\tan\beta$ (right) is presented. In contrast to the parameters in the previous figures, we set $M_\chi=\msusy = 10\tev$ to reduce the overall size of $M_H$. The left panel illustrates the situation for  $\tan\beta =1$, when the differences between the various versions are sizeable. We find an approximately constant shift between the results with and without effective THDM (employing the non-degenerate \order{\alt^2} threshold correction), of about $1\gev$ for unmixed top squarks and $4\gev$ for $X_t^\DR/\msusy = \sqrt{6}$. For the range of input quantities, however, $M_H$ is too large for $H$ playing the role of the SM Higgs boson.

$M_H$ can only be significantly decreased by raising $\tan\beta$. This possibility is analyzed in the right plot of \Fig{FH_MH_Fig}, where $M_A$ is set to 80 GeV. The shift between the results with and without effective THDM shrinks for rising $\tan\beta$, as was the case for $M_h$. To reach the desired value of $ 125\gev$ for $M_H$, $\tan\beta$ has to be at least $>7$. In this region, however, the difference between the results with and without the effective THDM is completely negligible. Also the uncertainty induced by not including contributions from non-SM Higgs bosons in the extraction of the low-energy couplings, estimated above, is totally negligible.

In addition, we also investigated the impact of the effective THDM on the prediction of the charged Higgs mass $M_{H^\pm}$. For the calculation of $M_{H^\pm}$ no resummation of large logarithms was available before. Nevertheless, we only find negligible shifts below 1 GeV in the scenarios considered above.

As noted above, the numerical impact of the effective THDM on the heavier Higgs boson masses might be enhanced in case of $\mu > \msusy$, which is not covered in this work.

%%%%%%%%%%%%%%%%%%%%%%%%%%%%%%%%%%%%%%%%%%%%%%%%%%%%%%%%%%%%%%%%%%%%%%%%%%%%%%

\subsection{The ``low-\texorpdfstring{$\tan\beta$}{tanb}-high'' scenario}
\label{sec:lowtbhigh}

In the ``low-$\tan\beta$-high'' scenario, defined in \cite{Bagnaschi:2015hka}, all soft SUSY-breaking sfermion masses, as well as the gluino mass, are set equal to $\msusy$. The value of $\msusy$ is chosen such that the result for $M_h$ is close to the experimentally determined mass and varies between a few TeV (in case of large $M_A$ or $\tan\beta$) and 100 TeV (in case of small $M_A$ or $\tan\beta$). In its original definition, the OS scheme was employed for renormalization, with the OS stop mixing parameter varying with $\tan\beta$ as follows,
\begin{align}\label{lowtanbhigh_OS_Eq}
X_t^\OS/\msusy =
\begin{cases}
      \hfil 2                             & \text{for}\ \tan\beta \le 2 \\
      0.0375\tan^2\beta-0.7\tan\beta+3.25 & \text{for}\ 2<\tan\beta\le 8.6\\
      \hfil 0                             & \text{for}\ 8.6<\tan\beta
\end{cases}
\end{align}
Owing to the problems with OS parameters in scenarios with low $M_A$ mentioned in \Sec{NumTHDM_Sec},  we define all parameters as $\DR$ quantitites\footnote{The use of the $\DR$ scheme will be also be beneficial when comparing with \MhEFT in the next subsection.}. Accordingly, we modify the values for $X_t$,
\begin{align}
X_t^\DR/\msusy =
\begin{cases}
      0.0375\tan^2\beta-0.7\tan\beta+3.25 & \text{for}\ \tan\beta\le 8.6\\
      \hfil 0                             & \text{for}\ 8.6<\tan\beta
\end{cases}.
\end{align}
In this way, $X_t^\DR/\msusy$ will be close to the value which maximizes $M_h$ when $\tan\beta=1$ is approached.

The remaining parameters are given by
\begin{align}
\mu = 1.5 \tev,\hspace{.5cm} M_2 = 2 \tev, \hspace{.5cm} A_{b,c,s,u,d} &= 2 \tev.
\end{align}
$M_1$ is fixed via the GUT relation $M_1 = \frac{5}{3}\tan^2\theta_W M_2\approx 0.5 M_2$.

%%%%%%%%%%%%%%%%%%%%%%%%% F I G U R E %%%%%%%%%%%%%%%%%%%%%%%%%%%%%%%%%%%%%%%%
\begin{figure}\centering
\begin{minipage}{.48\textwidth}\centering
\includegraphics[width=\textwidth]{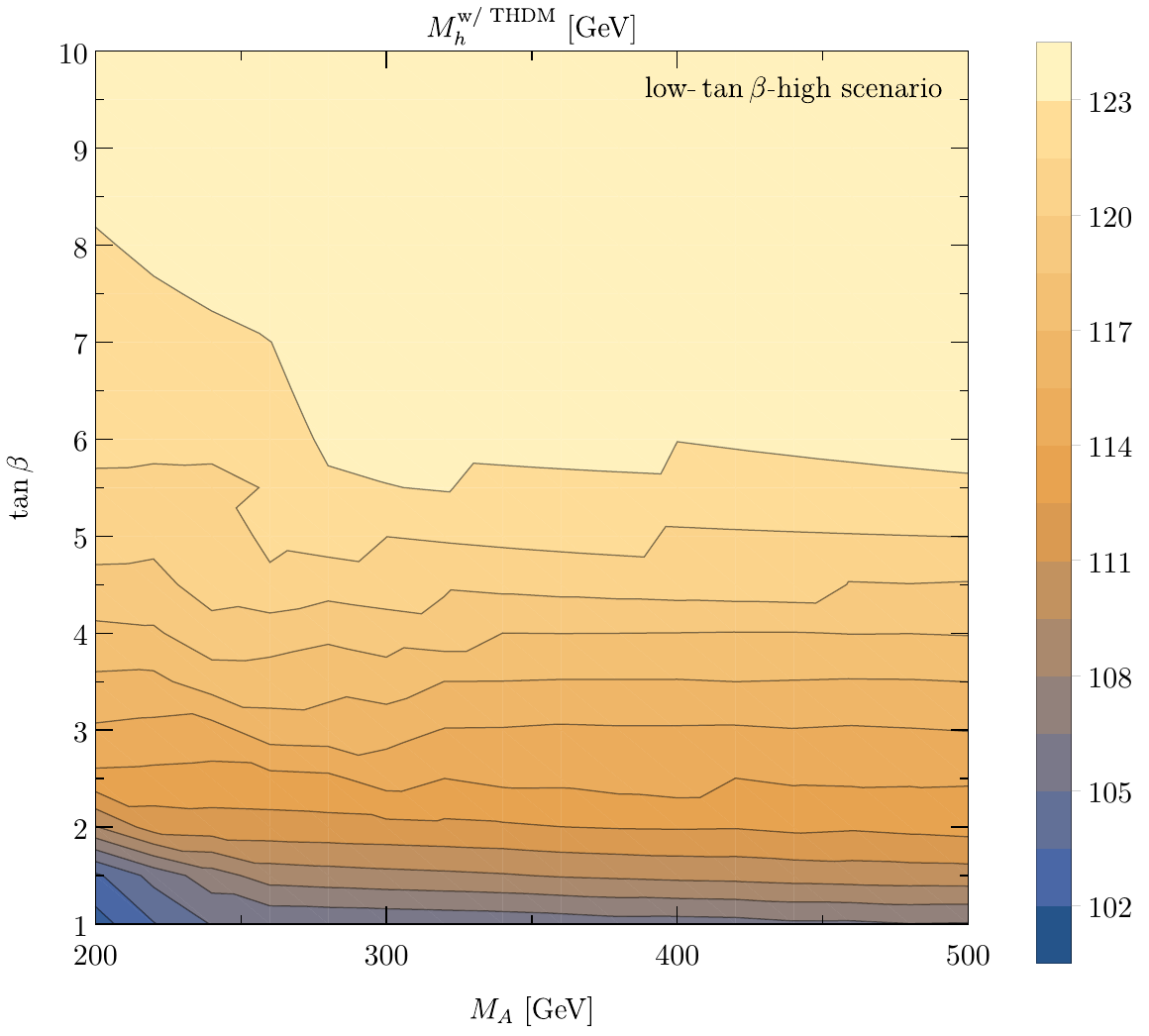}
\end{minipage}
\begin{minipage}{.48\textwidth}\centering
\includegraphics[width=\textwidth]{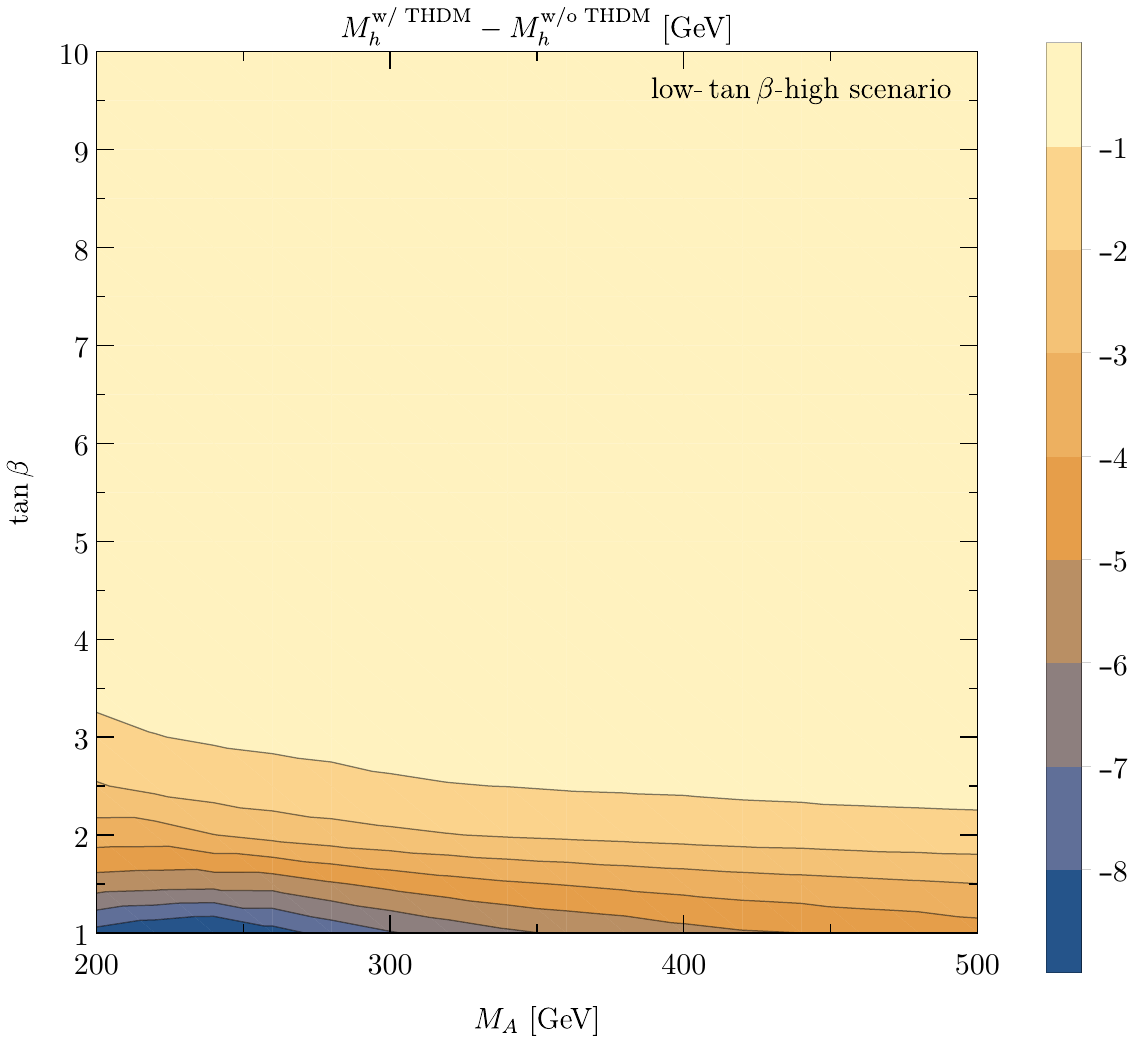}
\end{minipage}
\caption{Left: $M_h$ computed with \FH including the effective THDM as a function of $M_A$ and $\tan\beta$ in the low-$\tan\beta$-high scenario. Right: Difference between the results with and without effective THDM ({\tt FeynHiggs2.14.0}).}
\label{FH_lowtbhigh_Fig}
\end{figure}
%%%%%%%%%%%%%%%%%%%%%%%%% F I G U R E %%%%%%%%%%%%%%%%%%%%%%%%%%%%%%%%%%%%%%%%

The left panel of \Fig{FH_lowtbhigh_Fig} contains $M_h$ obtained from the \FH version including the THDM, in dependence of $\tan\beta$ and $M_A$. One finds that $M_h$ comes close to the experimental value of $125\gev$ only in the upper part of the plot where $\tan\beta \gtrsim 6$. For lower values of $\tan\beta$, $M_h$ drops down to the region around $105\gev$. If additionally $M_A$ is small ($\sim 200\gev$), $M_h$ is even below 102  GeV. In comparison with the results shown in Fig.~3 of~\cite{Bagnaschi:2015hka}, $M_h$ is reduced by several GeV.

The results in~\cite{Bagnaschi:2015hka} were produced using {\tt FeynHiggs2.10.4}. Since then, many additional improvements were implemented in {\tt FeynHiggs} (see also the discussions in~\cite{Bahl:2016brp,Bahl:2017aev} of important changes that have entered the versions {\tt 2.13.0} and {\tt 2.14.0}). To point out the effect of the most recent developments since {\tt FeynHiggs2.14.0},   we show the difference between the most topical version of \FH with effective THDM and the non-THDM version {\tt 2.14.0} in the right panel of \Fig{FH_lowtbhigh_Fig}. The diagram shows that for the considered scenario the $M_h$ values obtained with an effective THDM are below the values obtained without effective THDM. For $\tan\beta \gtrsim 3$, the downwards shift is small (below $1\gev$). For smaller $\tan\beta$, the shift increases to about $4\gev$ for $M_A = 500\gev$. If in addition also $M_A$ is small ($\sim 200\gev$), the difference amounts to even more than 8~GeV.

%%%%%%%%%%%%%%%%%%%%%%%%%%%%%%%%%%%%%%%%%%%%%%%%%%%%%%%%%%%%%%%%%%%%%%%%%%%%%%
\newpage
\subsection{Comparison to \MhEFT}

After investigating the numerical impact of an effective THDM on the hybrid calculation of \texttt{FeynHiggs}, we compare our results to \MhEFT (version {\tt 1.1}).

%%%%%%%%%%%%%%%%%%%%%%%%% F I G U R E %%%%%%%%%%%%%%%%%%%%%%%%%%%%%%%%%%%%%%%%
\begin{figure}\centering
\begin{minipage}{.48\textwidth}\centering
\includegraphics[width=\textwidth]{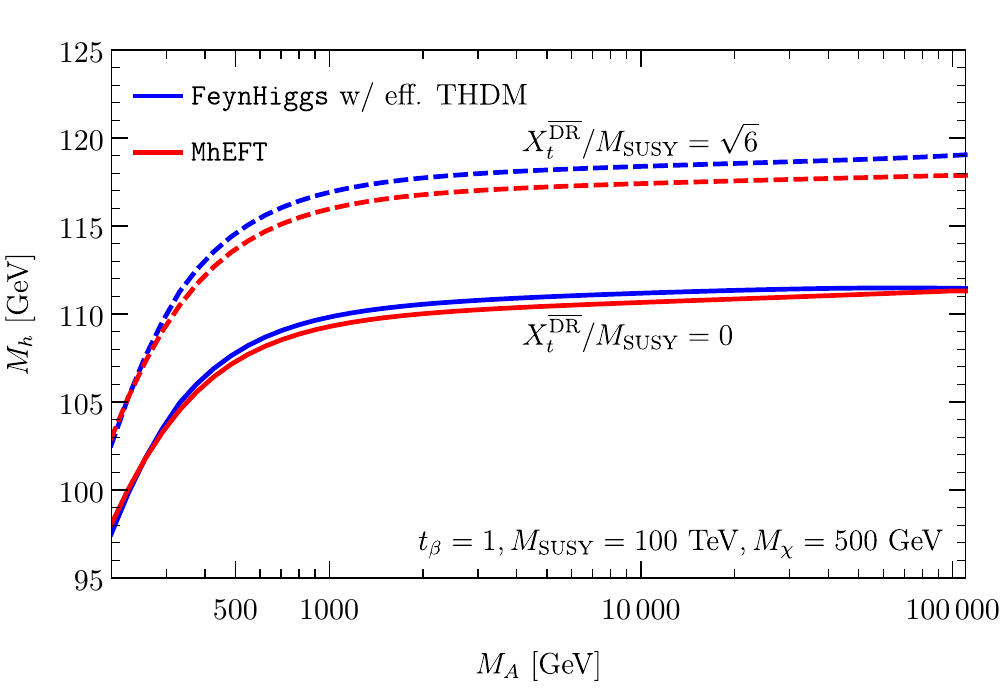}
\end{minipage}
\begin{minipage}{.48\textwidth}\centering
\includegraphics[width=\textwidth]{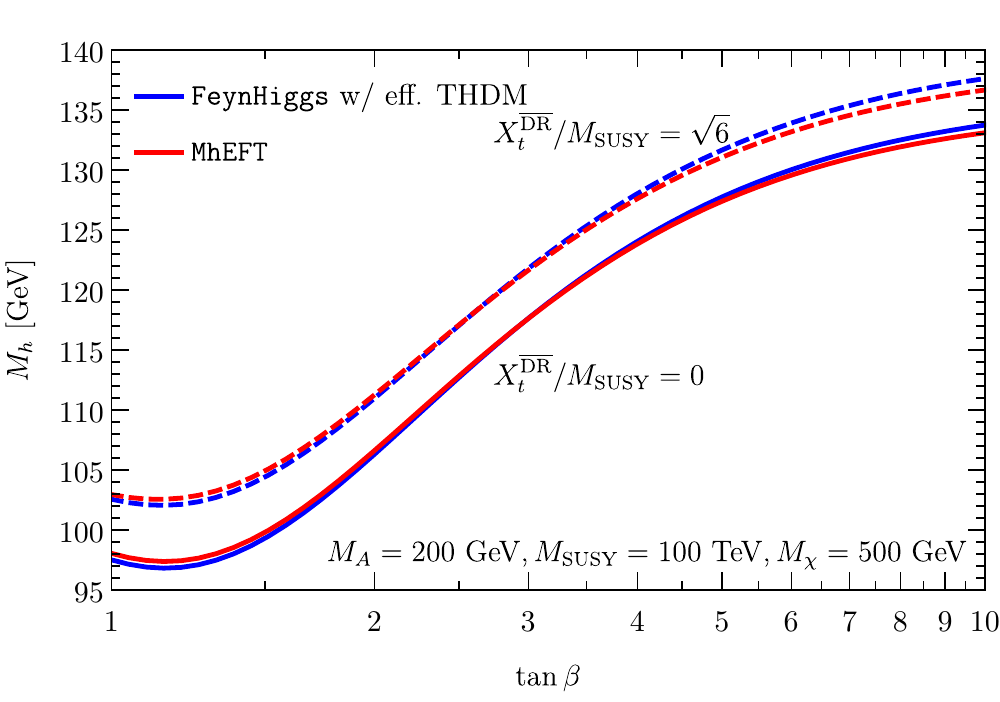}
\end{minipage}
\caption{Left: $M_h$ as a function of $M_A$ for $\tan\beta = 1$. Right: $M_h$ as a function of $\tan\beta$ for $M_A = 200\gev$. The results of \FH with effective THDM (blue) and \MhEFT (red) are compared for $X_t^\DR/\msusy = 0$ (solid) and $X_t^\DR/\msusy = \sqrt{6}$ (dashed).}
\label{FHvsMhEFT_MAvar_TBvar_Fig}
\end{figure}
%%%%%%%%%%%%%%%%%%%%%%%%% F I G U R E %%%%%%%%%%%%%%%%%%%%%%%%%%%%%%%%%%%%%%%%

\medskip
First, we compare the results for $M_h$ in dependence of $M_A$ (see left panel of \Fig{FHvsMhEFT_MAvar_TBvar_Fig}). We choose $\tan\beta = 1$ to maximize the impact of the effective THDM. For vanishing stop mixing, \FH and \MhEFT are in close agreement. Also for $X_t^\DR/\msusy = \sqrt{6}$, the two codes agree within $\sim 1\gev$. The remaining deviation is caused by the different parameterization of non-logarithmic terms (see \cite{Bahl:2017aev} for an extensive discussion). For low $M_A$ this constant shift is compensated by terms of $\mathcal{O}(M_t/M_A)$ originating from the THDM self-energies (see \Eq{InversePropMatrixEFT_Eq}) which are included in \FH but not in \MhEFT.

\medskip
In the right panel of \Fig{FHvsMhEFT_MAvar_TBvar_Fig}, the results are compared as a function of $\tan\beta$, setting $M_A = 200\gev$. The overall good agreement is confirmed. Especially around $\tan\beta \sim 3$ the two results are very close to each other, whereas the agreement is slightly worse for smaller or higher values of $\tan\beta$ (but still within 1 GeV). Reasons for the disagreement are again the different parameterization of non-logarithmic terms as well as terms of $\mathcal{O}(M_t/M_A)$.

%%%%%%%%%%%%%%%%%%%%%%%%% F I G U R E %%%%%%%%%%%%%%%%%%%%%%%%%%%%%%%%%%%%%%%%
\begin{figure}\centering
\begin{minipage}{.48\textwidth}\centering
\includegraphics[width=\textwidth]{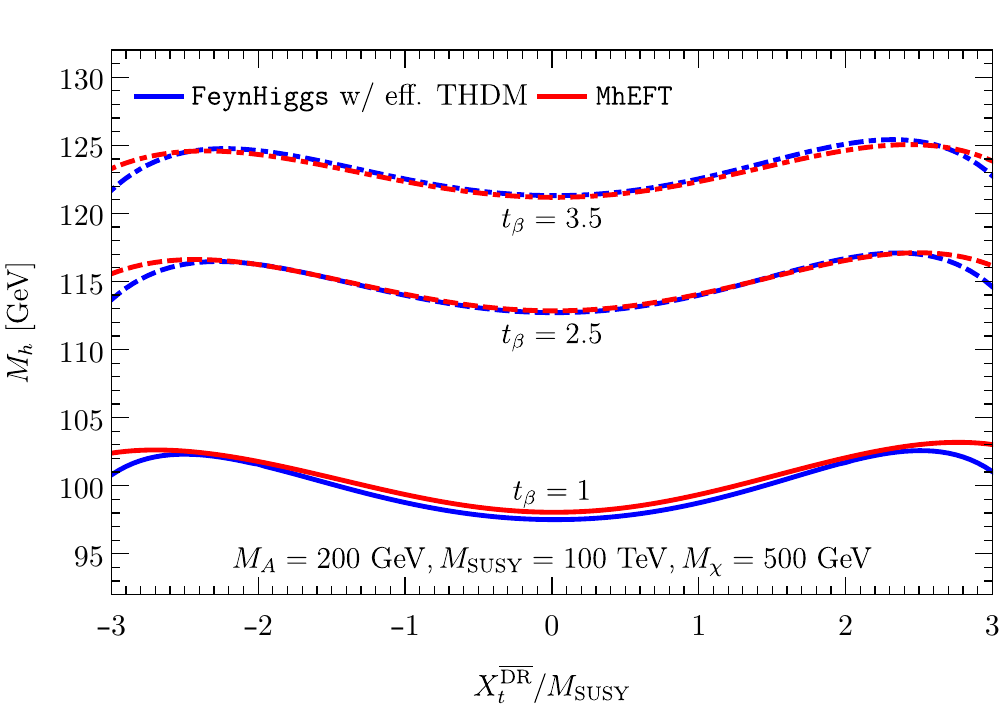}
\end{minipage}
\begin{minipage}{.48\textwidth}\centering
\includegraphics[width=\textwidth]{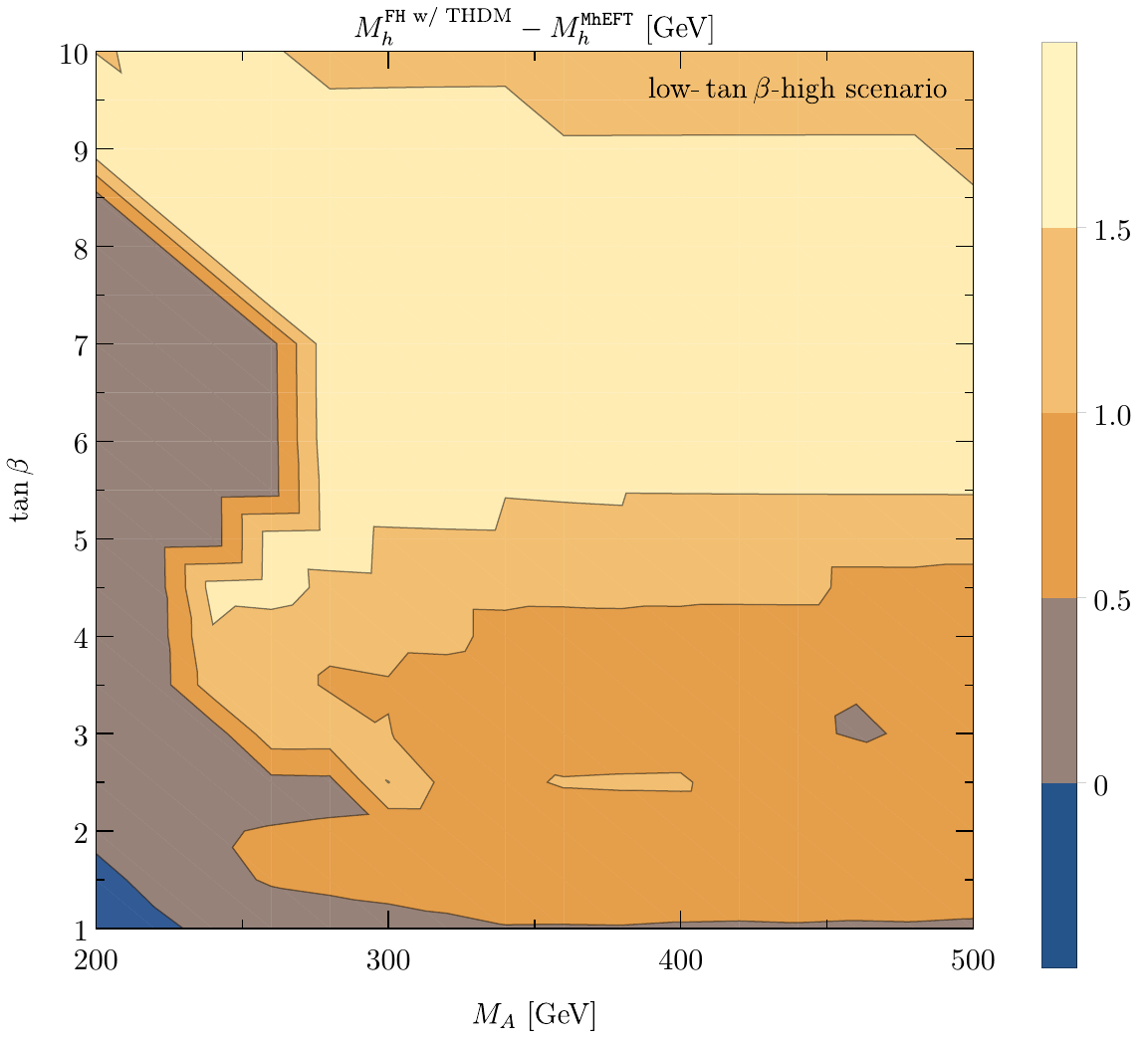}
\end{minipage}
\caption{Left: $M_h$ as a function of $X_t$ for $\tan\beta = 1$
  (solid), $\tan\beta = 2.5$ (dashed), and $\tan\beta = 3.5$
  (dotdashed). $M_A = 200 \gev$ is chosen.  The results of \FH with
  effective THDM (blue) and \MhEFT (red) are compared. Right: $M_h$
  in the ``low-$\tan\beta$-high'' scenario.
 The difference between \FH with effective THDM and \MhEFT is displayed.}
\label{FHvsMhEFT_XTvar_Fig}
\end{figure}
%%%%%%%%%%%%%%%%%%%%%%%%% F I G U R E %%%%%%%%%%%%%%%%%%%%%%%%%%%%%%%%%%%%%%%%

\medskip
This behavior is also reflected in the left panel of \Fig{FHvsMhEFT_XTvar_Fig} showing $M_h$ as a function of $X_t^\DR$. For $\tan\beta = 2.5$ and $\tan\beta = 3.5$, \FH and \MhEFT nearly superpose each other. Only for $|X_t^\DR/\msusy|>2.5$, small deviations are visible which originate from the different parameterizations of non-logarithmic terms. These terms become large for large $|X_t^\DR/\msusy|$. For $\tan\beta = 1$, a deviation of $\lesssim 1\gev$ is visible for $|X_t^\DR/\msusy|<2.5$, which is mainly caused by $\mathcal{O}(M_t/M_A)$ terms.

\medskip
In the right panel of \Fig{FHvsMhEFT_XTvar_Fig}, we have another look at the ``low-$\tan\beta$-high'' scenario using the $\DR$ scheme, as defined in Section~\ref{sec:lowtbhigh}. In the whole $M_A$--$\tan\beta$ plane the difference between the two codes is smaller than 2 GeV. Especially for low $M_A$ or low $\tan\beta$ the two codes agree very well, whereas \FH yields slightly larger results than \MhEFT in the rest of the parameter plane.

\medskip
Finally, we comment on the comparison between \FH and \MhEFT shown in~\cite{Lee:2015uza} (see  Fig.~10 and~11 therein). The authors of \cite{Lee:2015uza} compared the two codes in the low-$\tan\beta$-high scenario and found deviations of up to 15 GeV. According to their claim, this discrepancy was mainly caused by the missing implementation of an effective THDM in \FH. In our \Fig{FH_lowtbhigh_Fig}, right panel, we found, however, the effective THDM to induce shifts of not more than 8 GeV. This raises the question for the origin of the remaining difference of $\sim 7\gev$. One reason is certainly the fact that \FH has evolved a lot since version~{\tt2.10.2}, which was taken for the comparison in \cite{Lee:2015uza}. A second more important reason is the parameter conversion used for the comparison, which was done for the ``low-$\tan\beta$-high'' scenario defined with OS parameters, \Eq{lowtanbhigh_OS_Eq}. Therefore, the OS stop mixing parameter had to be converted to the $\MS$ scheme which is employed in \MhEFT. In this conversion,  $M_A=\msusy$ was assumed. Thereby, an important logarithmic contribution was missed (last term in \Eq{XtConv_Eq}), which is especially large for low $\tan\beta$ and low $M_A$, thus exactly in the parameter region where the largest deviation between \FH and \MhEFT was observed.

%%%%%%%%%%%%%%%%%%%%%%%%%%%%%%%%%%%%%%%%%%%%%%%%%%%%%%%%%%%%%%%%%%%%%%%%%%%%%%
%%%%%%%%%%%%%%%%%%%%%%%%%%%%%%%%%%%%%%%%%%%%%%%%%%%%%%%%%%%%%%%%%%%%%%%%%%%%%%

\section{Conclusions}\label{Conclusions_Sec}

In this paper, we discussed the implementation of an effective THDM into the hybrid framework of \FH for the calculation of the MSSM Higgs boson mass spectrum. Our new EFT calculation allows to treat the case of light non-SM Higgs bosons as well as of light EWinos and a light gluino. Furthermore, it includes complete one-loop and dominant two-loop threshold corrections and takes all appearing effective couplings fully into account. In this context, we also discussed how the matching between the various EFT versions is performed paying special attention to the different normalization of the Higgs doublets in the MSSM and the THDM.

This difference in field normalization plays a crucial role in the combination of the existing fixed-order calculation in \FH with the new EFT calculation for low $M_A$. Our accounting of the different normali\-za\-tions is done by introducing finite shifts in the field renormalization constants of the fixed-order calculation, which affects also the conceptual definition of $\tan\beta$ as an input parameter. Moreover, we investigated the effect of a low $M_A$ in the scheme conversion of the parameters for the stop sector, which is necessary if OS input parameters are used.

In our numerical study, we compared {\tt FeynHiggs2.14.0} and {\tt FeynHiggs2.14.1}, both with the SM as the EFT, to our new computation with an effective THDM, which is implemented in a still private \FH version based on {\tt 2.14.1}. We found the switch to an effective THDM to cause a negative shift in $M_h$ of up to 3 GeV with respect to {\tt FeynHiggs2.14.1}. This maximal value is reached when $\tan\beta \sim 1$ and the hierarchy between the SUSY scale and $M_A$ is large ($\msusy/M_A\sim 10^3$). The shift shrinks quickly when $\tan\beta$ is increased. For $\tan\beta\gtrsim 7$, the effects resulting fom the THDM are almost completely negligible. Similarly, the shift decreases when $M_A$ is increased or $\msusy$ is lowered. Larger shifts, up to 10 GeV, are found when comparing to {\tt FeynHiggs2.14.0}. In that version, the implemented \order{\alt^2} threshold correction implicitly assumed $M_A$ to be equal to $\msusy$, leading to an overestimate of $M_h$ in scenarios with $M_A \ll \msusy$.

We also investigated predictions for the mass of the second \CP-even Higgs boson $H$. In the phenomenologically most interesting parameter region, where the $H$ boson can play the role of the SM Higgs boson, we found the shift induced by an effective THDM to be negligible. Also the prediction of the charged Higgs boson mass is only marginally affected. In addition, we looked at the ``low-tanb-high'' benchmark scenario developed by the LHCHXSWG. For this scenario, we found corrections of up to -8 GeV for $\tan\beta\lesssim 3$ with the consequence that the updated $M_h$ prediction is too low for meeting the experimental Higgs boson mass. Finally, we compared our results with those of the code \MhEFT finding good agreement within 1 GeV throughout the considered parameter space.

Our calculation will become publicly available as part of the code \FH in a future version. We leave possible improvements of the present work, like the implementation of threshold corrections valid for arbitrary masses of the decoupled particles, or \order{\alt^2} threshold corrections, for future work.

%%%%%%%%%%%%%%%%%%%%%%%%%%%%%%%%%%%%%%%%%%%%%%%%%%%%%%%%%%%%%%%%%%%%%%%%%%%%%%
%%%%%%%%%%%%%%%%%%%%%%%%%%%%%%%%%%%%%%%%%%%%%%%%%%%%%%%%%%%%%%%%%%%%%%%%%%%%%%

\section*{Acknowledgments}
\sloppy{
We thank Sven Heinemeyer, Gabriel Lee, Pietro Slavich, Carlos Wagner and Georg Weiglein for useful discussions. H.B.\ is thankful to Thomas Hahn for his invaluable help
concerning all issues related to \FH and to Alexander Voigt for sharing {\tt Mathematica} expressions. H.B.\ and W.H.\ gratefully acknowledge support by the Deutsche Forschungsgemeinschaft (DFG) under Grant No.\ EXC-153 (Excellence Cluster ``Structure and Origin of the Universe'').
}

%%%%%%%%%%%%%%%%%%%%%%%%%%%%%%%%%%%%%%%%%%%%%%%%%%%%%%%%%%%%%%%%%%%%%%%%%%%%%%
%%%%%%%%%%%%%%%%%%%%%%%%%%%%%%%%%%%%%%%%%%%%%%%%%%%%%%%%%%%%%%%%%%%%%%%%%%%%%%
%%%%%%%%%%%%%%%%%%%%%%%%%%%%%%%%%%%%%%%%%%%%%%%%%%%%%%%%%%%%%%%%%%%%%%%%%%%%%%

\appendix

\section{Threshold corrections}\label{ThresholdApp}

In this Appendix one-loop formulas for matching the various EFTs to each other are provided. All expressions are derived under the assumption that all particles that are integrated out have masses equal to the matching scale. The couplings on the right hand side of all following expressions have to be evaluated at the scale given on the left hand side of the corresponding expressions. Couplings not listed do not receive any one-loop contributions to the matching conditions.

In addition, two-loop \order{\als\alt} corrections for the matching of the THDM quartic couplings to the full MSSM are given.

Expression for matching the SM to the MSSM and the SM to the SM+EWinos are listed e.g. in \cite{Bagnaschi:2014rsa}.

%%%%%%%%%%%%%%%%%%%%%%%%%%%%%%%%%%%%%%%%%%%%%%%%%%%%%%%%%%%%%%%%%%%%%%%%%%%%%%

\subsection{Matching the SM+EWinos to the MSSM}

The threshold corrections for matching the SM+EWinos to the MSSM are also known (see e.g. \cite{Bagnaschi:2014rsa}). We extend the known expressions for the effective Higgs-Higgsino-Gaugino couplings $\tilde g_{1u,1d,2u,2d}$ by including also terms owing to the external Higgs wave-function renormalization, which are proportional to $\xf^2$. They have been neglected in \cite{Bagnaschi:2014rsa}, because of $|X_t|\ll \msusy$ in the split-SUSY scenarios considered there.
We split up the matching expressions into four pieces,
\begin{subequations}
\begin{align}
\tgiu(\msusy) &= \gy\sbe + \Delta_{\tilde f}\tgiu + \Delta_{H}\tgiu + \Delta_{\DR\rightarrow\MS}\tgiu, \\
\tgiiu(\msusy) &= g\sbe + \Delta_{\tilde f}\tgiiu + \Delta_{H}\tgiiu + \Delta_{\DR\rightarrow\MS}\tgiiu, \\
\tgid(\msusy) &= \gy\cbe + \Delta_{\tilde f}\tgid + \Delta_{H}\tgid + \Delta_{\DR\rightarrow\MS}\tgid, \\
\tgiid(\msusy) &= g\cbe + \Delta_{\tilde f}\tgiid + \Delta_{H}\tgiid + \Delta_{\DR\rightarrow\MS}\tgiid.
\end{align}
\end{subequations}
The sfermion contributions are given by
\begin{subequations}
\begin{align}
\Delta_{\tilde f}\tgiu &= \gy\sbe k\left(-\frac{5}{2}\gy^2 + \frac{1}{4}h_t^2 (9 - \sbb\xf^2)\right), \label{SM+EWinos_MSSM_g12ud_Sf_Eqa}\\
\Delta_{\tilde f}\tgiiu &= g\sbe k\left(-\frac{3}{2}g^2 + \frac{1}{4}h_t^2 (9 - \sbb\xf^2)\right),\\
\Delta_{\tilde f}\tgid &= - \gy\cbe k\left(\frac{5}{2} \gy^2 + \frac{1}{4}h_t^2 \sbb\xf^2\right),\\
\Delta_{\tilde f}\tgiid &= - g\cbe k\left(\frac{3}{2} g^2 + \frac{1}{4}h_t^2 \sbb\xf^2\right).\label{SM+EWinos_MSSM_g12ud_Sf_Eqb}
\end{align}
\end{subequations}
Note that the new wave-function renormalization contributions proportional to $\hat X_t^2$ have been already implemented in \FH from version {\tt 2.13.0} on.

Integrating out the heavy Higgs yields
\begin{subequations}
\begin{align}
\Delta_{H}\tgiu &= \frac{1}{16}\gy\sbe k\left( 21 g^2 \cbb + \gy^2(-2+7\cbb)\right),\\
\Delta_{H}\tgiiu &= \frac{1}{16}g\sbe k\left( -g^2(2+11\cbb) + 7 \gy^2\cbb \right),\\
\Delta_{H}\tgid &= \frac{1}{16}\gy\cbe k\left( 21 g^2 \sbb + \gy^2(-2+7\sbb)\right),\\
\Delta_{H}\tgiid &= \frac{1}{16}g\cbe k\left( -g^2(2+11\sbb) + 7 \gy^2\sbb\right),
\end{align}
\end{subequations}
where we neglected terms of \order{M_\chi/M_A}.

Changing the regularization scheme from DRED for $Q>\msusy$ to DREG for $Q<\msusy$ gives rise to
\begin{subequations}
\begin{align}
\Delta_{\DR\rightarrow\MS}\tgiu &= - \frac{1}{8}\gy\sbe k(3 g^2+\gy^2), \\
\Delta_{\DR\rightarrow\MS}\tgiiu &= \frac{1}{24}g\sbe k(23 g^2 - 3 \gy^2),\\
\Delta_{\DR\rightarrow\MS}\tgid &= - \frac{1}{8}\gy\cbe k(3 g^2+\gy^2),\\
\Delta_{\DR\rightarrow\MS}\tgiid &= \frac{1}{24}g\cbe k(23 g^2 - 3 \gy^2).
\end{align}
\end{subequations}
See e.g. \cite{Martin:1993yx} for more details on the origin of these contributions.

%%%%%%%%%%%%%%%%%%%%%%%%%%%%%%%%%%%%%%%%%%%%%%%%%%%%%%%%%%%%%%%%%%%%%%%%%%%%%%

\subsection{Matching the SM to the THDM}\label{SMtoTHDM}

The SM Higgs self-coupling is obtained in terms of the $\lambda_i$ of the THDM by
\begin{align}\label{lambdaSMtoTHDM}
\lambda(M_A) =& \lambda_{\text{tree}} + \Delta\lambda
\end{align}
with
\begin{align}
\lambda_{\text{tree}} =& \li \cbe^4 + \lii \sbe^4 + 2 (\liii+\liv+\lv)\cbb\sbb + 4\lvi \cbe^3\sbe+4\lvii \cbe\sbe^3, \\
\Delta\lambda =& - 3k\left\{(\lvi+\lvii)c_{2\beta}+ (\lvi-\lvii)c_{4\beta}-\left(\li\cbb - \lii\sbb - (\liii+\liv+\lv)c_{2\beta}\right)s_{2\beta}\right\}^2.
\end{align}
Plugging in the tree-level expressions for the $\lambda_i$ from the matching of the THDM to the MSSM, we recover the heavy Higgs contribution to the matching condition of the SM Higgs self-coupling to the full MSSM given in Eq.~(10) of \cite{Bagnaschi:2014rsa}. %comparison to \cite{Cheung:2014hya}: Completely different $\Delta\lambda$.

The top Yukawa coupling of the SM $y_t$ is related to the top Yukawa couplings of the THDM via
\begin{align}
y_t(M_A) =& (h_t \sbe + h_t' \cbe)\left[1- \frac{3}{8}k \left(h_t\cbe - h_t'\sbe\right)^2\right].
%\Delta y_t =& - \frac{1}{8}k \left(3\frac{\cbb}{\sbb}y_t^2 + \left(\frac{\sbb}{\cbb}+8\right)y_b^2\right).
\end{align}
This correction corresponds to the heavy Higgs contribution to the threshold of the top Yukawa coupling when matching the SM to the MSSM given in Eq.~(24) of \cite{Bagnaschi:2014rsa}.

%%%%%%%%%%%%%%%%%%%%%%%%%%%%%%%%%%%%%%%%%%%%%%%%%%%%%%%%%%%%%%%%%%%%%%%%%%%%%%

\subsection{Matching the THDM to the MSSM}

At tree level the Higgs self-couplings of the THDM are given by
\begin{subequations}
\begin{align}
\li_{,\text{tree}}(\msusy) = \lii_{,\text{tree}}(\msusy) =& \frac{1}{4}(g^2 + \gy^2), \\
\liii_{,\text{tree}}(\msusy) =& \frac{1}{4}(g^2 - \gy^2), \\
\liv_{,\text{tree}}(\msusy) =& -\frac{1}{2}g^2, \\
\lv_{,\text{tree}}(\msusy) = \lvi_{,\text{tree}}(\msusy) = \lvii_{,\text{tree}}(\msusy) =& 0.
\end{align}
\end{subequations}
At one-loop level corrections arise from integrating out the stops, EWinos, as well as from the transition from $\DR$ to $\MS$. We split up the stop contribution into one part originating from vertex corrections and another part originating from the wave function renormalization (WFR) of the Higgs fields,
\begin{align}
\lambda_i(\msusy) = \lambda_{i,\text{tree}}+\Delta_{\text{Ver.Cor.}}\lambda_i+\Delta_{\text{WFR}}\lambda_i+\Delta_{\text{EWinos}}\lambda_i+\Delta_{\DR\rightarrow\MS}\lambda_i.
\end{align}
The stop contributions have originally been calculated in \cite{Haber:1993an}; they are listed here for completeness. The vertex corrections from box and triangle diagrams are given by
\begin{subequations}
\begin{align}\label{THDMtoMSSMlambdasVerCorr}
\Delta_{\text{Ver.Cor.}}\lambda_1 &= - \frac{1}{2}k h_t^4 \mf^4 + \frac{3}{4}k(g^2+\gy^2)h_t^2\mf^2,\\
\Delta_{\text{Ver.Cor.}}\lambda_2 &= 6 k h_t^4 \at^2 \left(1-\frac{1}{12}\at^2\right)-\frac{3}{4}(g^2+\gy^2)h_t^2\at^2,\\
\Delta_{\text{Ver.Cor.}}\lambda_3 &= \frac{1}{2}k\mf^2 h_t^4 (3-\at^2) - \frac{3}{8}k(g^2-\gy^2)h_t^2(\at^2-\mf^2),\\
\Delta_{\text{Ver.Cor.}}\lambda_4 &= \frac{1}{2}k\mf^2 h_t^4 (3-\at^2) + \frac{3}{4}k g^2 h_t^2(\at^2-\mf^2),\\
\Delta_{\text{Ver.Cor.}}\lambda_5 &= -\frac{1}{2}h_t^4\mf^2\at^2,\\
\Delta_{\text{Ver.Cor.}}\lambda_6 &= \frac{1}{2}k h_t^4 \mf^3\at - \frac{3}{8}k(g^2+\gy^2)h_t^2\mf\at,\\
\Delta_{\text{Ver.Cor.}}\lambda_7 &= \frac{1}{2}k h_t^4 \mf\at(\at^2-6) + \frac{3}{8}k(g^2+\gy^2)h_t^2\mf\at,
\end{align}
\end{subequations}
whereas the WFR corrections read
\begin{subequations}
\begin{align}
\Delta_{\text{WFR}}\lambda_1 &= -2(\hat\Sigma'_{11}\li + \hat\Sigma'_{12}\lvi),\label{liWFR} \\
\Delta_{\text{WFR}}\lambda_2 &= -2(\hat\Sigma'_{22}\lii + \hat\Sigma'_{12}\lvii), \\
\Delta_{\text{WFR}}\lambda_3 &= -(\hat\Sigma'_{11}+\hat\Sigma'_{22})\liii - \hat\Sigma'_{12}(\lvi+\lvii),\\
\Delta_{\text{WFR}}\lambda_4 &= -(\hat\Sigma'_{11}+\hat\Sigma'_{22})\liv - \hat\Sigma'_{12}(\lvi+\lvii),\\
\Delta_{\text{WFR}}\lambda_5 &= -(\hat\Sigma'_{11}+\hat\Sigma'_{22})\lv - \hat\Sigma'_{12}(\lvi+\lvii),\\
\Delta_{\text{WFR}}\lambda_6 &= -\frac{1}{2}(3 \hat\Sigma'_{11}+\hat\Sigma'_{22})\lvi - \frac{1}{2}\hat\Sigma'_{12}(\li+\liii+\liv+\lv),\\
\Delta_{\text{WFR}}\lambda_7 &= -\frac{1}{2}(\hat\Sigma'_{11}+ 3 \hat\Sigma'_{22})\lvii - \frac{1}{2}\hat\Sigma'_{12}(\lii+\liii+\liv+\lv),\label{lviiWFR}
\end{align}
\end{subequations}
where the $\hat\Sigma'_{ij}=\left(\frac{\partial}{\partial p^2}\hat\Sigma_{\phi_i\phi_j}\right)|_{p^2=0}$ are given by
\begin{subequations}
\begin{align}
\hat\Sigma'_{11} &= \frac{1}{2}k h_t^2\mf^2,\\
\hat\Sigma'_{22} &= \frac{1}{2}k h_t^2\at^2,\\
\hat\Sigma'_{12} &= -\frac{1}{2}k h_t^2\at\mf.
\end{align}
\end{subequations}
The scheme change from $\DR$ to $\MS$ yields the additional contributions
\begin{subequations}
\begin{align}
\Delta_{\DR\rightarrow\MS}\lambda_{1,2} &= - \frac{1}{12}k (7 g^4 + 6 g^2 \gy^2 + 3 \gy^4), \\
\Delta_{\DR\rightarrow\MS}\liii &= - \frac{1}{12}k (7 g^4 - 6 g^2 \gy^2 + 3 \gy^4), \\
\Delta_{\DR\rightarrow\MS}\liv &= - \frac{1}{3}k g^2(g^2+3\gy^2), \\
\Delta_{\DR\rightarrow\MS}\lambda_{5,6,7} &= 0,
\end{align}
\end{subequations}
which have already been calculated in \cite{Gorbahn:2009pp}.

The EWino corrections can be obtained by replacing the effective Higgs-Higgsino-Gaugino couplings $\hat g_{1uu,1ud,..}$ in the expression for matching the THDM to the THDM+EWinos given below by their tree-level values.

Due to the wave-function renormalization, also $\beta$ receives a threshold correction,
\begin{align}
\beta_{\text{THDM}} &= \beta_{\text{MSSM}} + \frac{1}{2}\Delta\Sigma_{H_1 H_2}'.
\end{align}
$\Delta\Sigma_{H_1 H_2}'$ receives corrections from sfermions and EWinos,
\begin{align}
\Delta\Sigma_{H_1 H_2}'\Big|_{\tilde f} &= \frac{1}{4}k h_t^2 s_{2\beta}(\at-\mf/\tbe)(\at+\mf\tbe),\\
\Delta\Sigma_{H_1 H_2}'\Big|_{\text{EWino}} &= -\frac{1}{6}k (3 g^2+\gy^2) c_{2\beta}.
\end{align}
Only when taking into account this threshold correction, can the well known one-loop matching condition of $\lambda$ (when matching the SM to the MSSM) be recovered from \Eq{lambdaSMtoTHDM} considering the limit $M_A\rightarrow \msusy$.

The top Yukawa couplings are obtained at the one-loop level via
\begin{align}
h_t^{\text{THDM}}(\msusy) =& h_t\Bigg\{1+k\bigg[\frac{4}{3} g_3^2(1-\at)+ h_t^2\Big(\mathcal{F}_5(\mf)-\frac{1}{4}\at^2\Big) \nonumber\\
&\hspace{2.3cm}+ g^2 \left(\mathcal{F}_1(\mf)-\frac{3}{8}\right)+ \gy^2 \left(\mathcal{F}_3(\mf)-\frac{1}{72}\right)\bigg]\Bigg\}, \label{htTHDMvsMSSM_Eq}\\
(h'_t)^{\text{THDM}}(\msusy) =& h_t k\Bigg\{\frac{4}{3} g_3^2 \mf + \frac{1}{4} h_t^2\at\mf+ g^2 \mathcal{F}_2(\mf)+ \gy^2 \mathcal{F}_4(\mf)\Bigg\}. \label{htpTHDMvsMSSM_Eq}
\end{align}
Here, we implicitly assume that $M_1 = M_2 = \mu$.

The appearing functions are given by
\begin{subequations}
\begin{align}
\mathcal{F}_1(\mf)=&\frac{3}{16(1-\mf^2)^2}\Big[7-4\mf^2-3\mf^4+2\mf^2(8-3\mf^2)\ln\mf^2\Big],\\
\mathcal{F}_2(\mf)=&\frac{3\mf^2}{2(1-\mf^2)^2}\Big[1-\mf^2+\ln\mf^2\Big],\\
\mathcal{F}_3(\mf)=&\frac{1}{144(1-\mf^2)^2}\Big[(55-32\at\mf+51\mf^2)(1-\mf^2)+2\mf^2(72-16\at\mf-19\mf^2)\ln\mf^2\Big],\\
\mathcal{F}_4(\mf)=&\frac{\mf^2}{18(1-\mf^2)^2}\Big[13(1-\mf^2)+(9+4\mf^2)\ln\mf^2\Big],\\
\mathcal{F}_5(\mf)=&\frac{3}{8(1-\mf^2)^2}\Big[-1+4\mf^2-3\mf^4+2\mf^4\ln\mf^2\Big],
\end{align}
\end{subequations}
with
\begin{subequations}
\begin{alignat}{3}
\mathcal{F}_1(0)&= \frac{21}{16}, \hspace{3cm}&&\mathcal{F}_1(1)&&= -\frac{3}{4},\\
\mathcal{F}_2(0)&= 0, &&\mathcal{F}_2(1)&&= -\frac{3}{4},\\
\mathcal{F}_3(0)&= \frac{55}{144}, &&\mathcal{F}_3(1)&&= -\frac{1}{36}(9+4\at),\\
\mathcal{F}_4(0)&= 0, &&\mathcal{F}_4(1)&&= -\frac{5}{36},\\
\mathcal{F}_5(0)&= -\frac{3}{8}, &&\mathcal{F}_5(1)&&= 0
\end{alignat}
\end{subequations}
as limiting values.

%%%%%%%%%%%%%%%%%%%%%%%%%%%%%%%%%%%%%%%%%%%%%%%%%%%%%%%%%%%%%%%%%%%%%%%%%%%%%%

\subsection{Matching the THDM to the THDM+EWinos}

We again split up the matching conditions for the Higgs self-couplings into a piece due to vertex corrections and a piece due to wave-function renormalization,
\begin{align}
\lambda_i^{\text{THDM}}(M_\chi) = \lambda_i^{\text{THDM+EWinos}} + \Delta_{\text{Ver.Cor.}}\lambda_i + \Delta_{\text{WFR}}\lambda_i .
\end{align}
The vertex corrections read
\begin{subequations}
\begin{align}
\Delta_{\text{Ver.Cor.}}\lambda_1 =& -\frac{1}{12} k \bigg[7 \hgidd^4+16 \hgidd^3 \hgidu+2 \hgidd^2 (9 \hgidu^2+7 \hgiidd^2+8 \hgiidd \hgiidu+\hgiidu^2)\nonumber\\
&\hspace{1.3cm}+16 \hgidd \hgidu\Big(\hgidu^2+(\hgiidd+\hgiidu)^2\Big)+7 \hgidu^4\nonumber\\
&\hspace{1.3cm}+2 \hgidu^2 (\hgiidd^2+8 \hgiidd \hgiidu+7 \hgiidu^2)\nonumber\\
&\hspace{1.3cm}+3 (\hgiidd+\hgiidu)^2 (9 \hgiidd^2-2\hgiidd \hgiidu+9 \hgiidu^2)\bigg],\\
\Delta_{\text{Ver.Cor.}}\lambda_2 =& -\frac{1}{12} k \bigg[7 \hgiud^4+16 \hgiud^3 \hgiuu+2 \hgiud^2 (9 \hgiuu^2+7 \hgiiud^2+8 \hgiiud \hgiiuu+\hgiiuu^2)\nonumber\\
&\hspace{1.3cm}+16 \hgiud \hgiuu\Big(\hgiuu^2+(\hgiiud+\hgiiuu)^2\Big)+7 \hgiuu^4\nonumber\\
&\hspace{1.3cm}+2 \hgiuu^2 (\hgiiud^2+8 \hgiiud \hgiiuu+7 \hgiiuu^2)\nonumber\\
&\hspace{1.3cm}+3 (\hgiiud+\hgiiuu)^2 (9 \hgiiud^2-2\hgiiud \hgiiuu+9 \hgiiuu^2)\bigg],\\
\Delta_{\text{Ver.Cor.}}\lambda_3 =& -\frac{1}{12} k \bigg[\hgidd^2 (7 \hgiud^2+8 \hgiud \hgiuu+7 \hgiuu^2+10 \hgiiud^2+8 \hgiiud \hgiiuu+4 \hgiiuu^2)\nonumber\\
&\hspace{1.3cm}+2 \hgidd \Big(2 \hgidu (2\hgiud^2+\hgiud \hgiuu+2 \hgiuu^2+2 \hgiiud^2+\hgiiud \hgiiuu+2 \hgiiuu^2)\nonumber\\
&\hspace{1.3cm}-3 (\hgiud \hgiidd \hgiiud+\hgiud \hgiidu \hgiiuu+\hgiuu\hgiidd \hgiiuu-3 \hgiuu \hgiidu \hgiiud)\Big)\nonumber\\
&\hspace{1.3cm}+\hgidu^2 (7 \hgiud^2+8 \hgiud \hgiuu+7 \hgiuu^2+4 \hgiiud^2+8 \hgiiud \hgiiuu+10\hgiiuu^2)\nonumber\\
&\hspace{1.3cm}-6 \hgidu (-3 \hgiud \hgiidd \hgiiuu+\hgiud \hgiidu \hgiiud+\hgiuu \hgiidd \hgiiud+\hgiuu \hgiidu \hgiiuu)\nonumber\\
&\hspace{1.3cm}+10 \hgiud^2
   \hgiidd^2+8 \hgiud^2 \hgiidd \hgiidu+4 \hgiud^2 \hgiidu^2+8 \hgiud \hgiuu \hgiidd^2\nonumber\\
&\hspace{1.3cm}+4 \hgiud \hgiuu \hgiidd \hgiidu+8 \hgiud \hgiuu
   \hgiidu^2+4 \hgiuu^2 \hgiidd^2+8 \hgiuu^2 \hgiidd \hgiidu\nonumber\\
&\hspace{1.3cm}+10 \hgiuu^2 \hgiidu^2+27 \hgiidd^2 \hgiiud^2+24 \hgiidd^2 \hgiiud \hgiiuu+27 \hgiidd^2
   \hgiiuu^2\nonumber\\
&\hspace{1.3cm}+24 \hgiidd \hgiidu \hgiiud^2-12 \hgiidd \hgiidu \hgiiud \hgiiuu+24 \hgiidd \hgiidu \hgiiuu^2\nonumber\\
&\hspace{1.3cm}+27 \hgiidu^2 \hgiiud^2+24 \hgiidu^2 \hgiiud
   \hgiiuu+27 \hgiidu^2 \hgiiuu^2\bigg],\\
\Delta_{\text{Ver.Cor.}}\lambda_4 =& -\frac{1}{12} k \bigg[\hgidd^2 (7 \hgiud^2+8 \hgiud \hgiuu+4 \hgiuu^2-5 \hgiiud^2-4 \hgiiud \hgiiuu-2 \hgiiuu^2)\nonumber\\
&\hspace{1.3cm}+2 \hgidd \Big(\hgidu (4\hgiud^2+5 \hgiud \hgiuu+4 \hgiuu^2-2 \hgiiud^2-\hgiiud \hgiiuu-2 \hgiiuu^2)\nonumber\\
&\hspace{1.3cm}+3\big(\hgiud (2 \hgiidd+\hgiidu) (2 \hgiiud+\hgiiuu)\nonumber\\
&\hspace{1.3cm}+\hgiuu (2\hgiidd \hgiiud+4 \hgiidd \hgiiuu-\hgiidu \hgiiud+2 \hgiidu \hgiiuu)\big)\Big)\nonumber\\
&\hspace{1.3cm}+\hgidu^2 (4 \hgiud^2+8 \hgiud \hgiuu+7 \hgiuu^2-2 \hgiiud^2-4\hgiiud \hgiiuu-5 \hgiiuu^2)\nonumber\\
&\hspace{1.3cm}+6 \hgidu \Big(\hgiud (2 \hgiidd \hgiiud-\hgiidd \hgiiuu+4 \hgiidu \hgiiud+2 \hgiidu \hgiiuu)\nonumber\\
&\hspace{1.3cm}+\hgiuu (\hgiidd+2\hgiidu) (\hgiiud+2 \hgiiuu)\Big)-5 \hgiud^2 \hgiidd^2-4 \hgiud^2 \hgiidd \hgiidu\nonumber\\
&\hspace{1.3cm}-2 \hgiud^2 \hgiidu^2-4 \hgiud \hgiuu \hgiidd^2-2 \hgiud \hgiuu\hgiidd \hgiidu-4 \hgiud \hgiuu \hgiidu^2\nonumber\\
&\hspace{1.3cm}-2 \hgiuu^2 \hgiidd^2-4 \hgiuu^2 \hgiidd \hgiidu-5 \hgiuu^2 \hgiidu^2+27 \hgiidd^2 \hgiiud^2\nonumber\\
&\hspace{1.3cm}+24 \hgiidd^2\hgiiud \hgiiuu+24 \hgiidd \hgiidu \hgiiud^2+42 \hgiidd \hgiidu \hgiiud \hgiiuu\nonumber\\
&\hspace{1.3cm}+24 \hgiidd \hgiidu \hgiiuu^2+24 \hgiidu^2 \hgiiud \hgiiuu+27
   \hgiidu^2 \hgiiuu^2\bigg],\\
\Delta_{\text{Ver.Cor.}}\lambda_5 =& -\frac{1}{12} k \bigg[\hgidd^2 (7 \hgiud^2+8 \hgiud \hgiuu-2 \hgiuu^2+2 \hgiiud^2+4 \hgiiud \hgiiuu-\hgiiuu^2)\nonumber\\
&\hspace{1.3cm}+2 \hgidd \Big(\hgidu (4\hgiud^2+11 \hgiud \hgiuu+4 \hgiuu^2+2 \hgiiud^2+7 \hgiiud \hgiiuu+2 \hgiiuu^2)\nonumber\\
&\hspace{1.3cm}+\hgiud \big(\hgiidd (5 \hgiiud+2 \hgiiuu)+2 \hgiidu(\hgiiud+\hgiiuu)\big)\nonumber\\
&\hspace{1.3cm}+\hgiuu \big(\hgiidd (2 \hgiiud-\hgiiuu)+2 \hgiidu (\hgiiud+\hgiiuu)\big)\Big)\nonumber\\
&\hspace{1.3cm}+\hgidu^2 (-2 \hgiud^2+8 \hgiud \hgiuu+7\hgiuu^2-\hgiiud^2+4 \hgiiud \hgiiuu+2 \hgiiuu^2)\nonumber\\
&\hspace{1.3cm}+2 \hgidu \Big(\hgiud \big(2 \hgiidd (\hgiiud+\hgiiuu)-\hgiidu (\hgiiud-2 \hgiiuu)\big)\nonumber\\
&\hspace{1.3cm}+\hgiuu \big(2 \hgiidd (\hgiiud+\hgiiuu)+\hgiidu (2 \hgiiud+5 \hgiiuu)\big)\Big)+2 \hgiud^2 \hgiidd^2\nonumber\\
&\hspace{1.3cm}+4 \hgiud^2 \hgiidd \hgiidu-\hgiud^2 \hgiidu^2+4 \hgiud \hgiuu \hgiidd^2+14\hgiud \hgiuu \hgiidd \hgiidu\nonumber\\
&\hspace{1.3cm}+4 \hgiud \hgiuu \hgiidu^2-\hgiuu^2 \hgiidd^2+4 \hgiuu^2 \hgiidd \hgiidu+2 \hgiuu^2 \hgiidu^2\nonumber\\
&\hspace{1.3cm}+27 \hgiidd^2\hgiiud^2+24 \hgiidd^2 \hgiiud \hgiiuu-6 \hgiidd^2 \hgiiuu^2+24 \hgiidd \hgiidu \hgiiud^2\nonumber\\
&\hspace{1.3cm}+54 \hgiidd \hgiidu \hgiiud \hgiiuu+24 \hgiidd \hgiidu   \hgiiuu^2-6 \hgiidu^2 \hgiiud^2\nonumber\\
&\hspace{1.3cm}+24 \hgiidu^2 \hgiiud \hgiiuu+27 \hgiidu^2 \hgiiuu^2\bigg],\\
\Delta_{\text{Ver.Cor.}}\lambda_6 =& -\frac{1}{12} k \bigg[\hgidd^3 (7 \hgiud+4 \hgiuu)+\hgidd^2 (12 \hgidu \hgiud+9 \hgidu \hgiuu\nonumber\\
&\hspace{1.3cm}+7 \hgiidd \hgiiud+4 \hgiidd \hgiiuu+4 \hgiidu\hgiiud+\hgiidu \hgiiuu)\nonumber\\
&\hspace{1.3cm}+\hgidd \Big(3 \hgidu^2 (3 \hgiud+4 \hgiuu)+8 \hgidu (\hgiidd+\hgiidu) (\hgiiud+\hgiiuu)\nonumber\\
&\hspace{1.3cm}+(\hgiidd+\hgiidu)\big(\hgiud (7\hgiidd+\hgiidu)+4 \hgiuu (\hgiidd+\hgiidu)\big)\Big)\nonumber\\
&\hspace{1.3cm}+\hgidu^3 (4 \hgiud+7 \hgiuu)+\hgidu^2 (\hgiidd \hgiiud+4 \hgiidd \hgiiuu+4 \hgiidu \hgiiud\nonumber\\
&\hspace{1.3cm}+7 \hgiidu \hgiiuu)+\hgidu (\hgiidd+\hgiidu)\Big(4 \hgiud (\hgiidd+\hgiidu)\nonumber\\
&\hspace{1.3cm}+\hgiuu (\hgiidd+7 \hgiidu)\Big)+3 (\hgiidd+\hgiidu) \Big(\hgiidd^2 (9 \hgiiud+4\hgiiuu)\nonumber\\
&\hspace{1.3cm}+3 \hgiidd \hgiidu (\hgiiud+\hgiiuu)+\hgiidu^2 (4 \hgiiud+9 \hgiiuu)\Big)\bigg],\\
\Delta_{\text{Ver.Cor.}}\lambda_7 =& -\frac{1}{12} k \bigg[\hgidd \Big(7 \hgiud^3+12 \hgiud^2 \hgiuu+\hgiud (9 \hgiuu^2+7 \hgiiud^2+8 \hgiiud \hgiiuu+\hgiiuu^2)\nonumber\\
&\hspace{1.3cm}+4 \hgiuu\big(\hgiuu^2+(\hgiiud+\hgiiuu)^2\big)\Big)\nonumber\\
&\hspace{1.3cm}+\hgidu \Big(4 \hgiud^3+9 \hgiud^2 \hgiuu+4 \hgiud \big(3 \hgiuu^2+(\hgiiud+\hgiiuu)^2\big)\nonumber\\
&\hspace{1.3cm}+\hgiuu (7 \hgiuu^2+\hgiiud^2+8 \hgiiud \hgiiuu+7 \hgiiuu^2)\Big)+7 \hgiud^2 \hgiidd \hgiiud\nonumber\\
&\hspace{1.3cm}+4 \hgiud^2 \hgiidd \hgiiuu+4 \hgiud^2 \hgiidu\hgiiud+\hgiud^2 \hgiidu \hgiiuu+8 \hgiud \hgiuu \hgiidd \hgiiud\nonumber\\
&\hspace{1.3cm}+8 \hgiud \hgiuu \hgiidd \hgiiuu+8 \hgiud \hgiuu \hgiidu \hgiiud+8 \hgiud\hgiuu \hgiidu \hgiiuu\nonumber\\
&\hspace{1.3cm}+\hgiuu^2 \hgiidd \hgiiud+4 \hgiuu^2 \hgiidd \hgiiuu+4 \hgiuu^2 \hgiidu \hgiiud+7 \hgiuu^2 \hgiidu \hgiiuu\nonumber\\
&\hspace{1.3cm}+27 \hgiidd\hgiiud^3+36 \hgiidd \hgiiud^2 \hgiiuu+21 \hgiidd \hgiiud \hgiiuu^2+12 \hgiidd \hgiiuu^3\nonumber\\
&\hspace{1.3cm}+12 \hgiidu \hgiiud^3+21 \hgiidu \hgiiud^2 \hgiiuu+36 \hgiidu\hgiiud \hgiiuu^2+27 \hgiidu \hgiiuu^3\bigg].
\end{align}
\end{subequations}
The WFR corrections are identical to those listed in \Eqss{liWFR}{lviiWFR}, but with
\begin{subequations}
\begin{align}
\hat\Sigma'_{11} &= -\frac{1}{6}k\Big[(\hgidd+\hgidu)^2+3(\hgiidd+\hgiidu)^2\Big]\\
\hat\Sigma'_{22} &= -\frac{1}{6}k\Big[(\hgiuu+\hgiud)^2+3(\hgiiuu+\hgiiud)^2\Big]\\
\hat\Sigma'_{12} &= -\frac{1}{6}k\Big[(\hgiuu+\hgiud)(\hgidd+\hgidu)+3(\hgiiuu+\hgiiud)(\hgiidd+\hgiidu)\Big]
\end{align}
\end{subequations}
The matching conditions of the top Yukawa coupling are purely due to wave-function renormalization,
\begin{subequations}
\begin{align}
h_t^{\text{THDM}}(M_\chi)&= h_t^{\text{THDM+EWinos}}-\frac{1}{2}h_t\hat\Sigma'_{22}-\frac{1}{2}h_t'\hat\Sigma'_{12} ,\\
(h_t')^{\text{THDM}}(M_\chi)&= (h_t')^{\text{THDM+EWinos}}-\frac{1}{2}h_t'\hat\Sigma'_{11}-\frac{1}{2}h_t\hat\Sigma'_{12}.
\end{align}
\end{subequations}
The threshold correction of $\beta$ reads
\begin{align}
\beta_{\text{THDM}}(M_A) &= \beta_{\text{THDM+EWinos}} + \frac{1}{2}\Delta\Sigma_{H_1 H_2}'
\end{align}
with
\begin{align}
\Delta\Sigma_{H_1 H_2}' &= \sbe\cbe\big(\hat\Sigma'_{22}-\hat\Sigma'_{11}\big)+c_{2\beta}\hat\Sigma'_{12}
\end{align}
In the limit of $M_\chi\rightarrow \msusy$, we cross-checked the threshold corrections of $\lambda_{1..7}$
against the expressions given in \cite{Gorbahn:2009pp} and found agreement.

%%%%%%%%%%%%%%%%%%%%%%%%%%%%%%%%%%%%%%%%%%%%%%%%%%%%%%%%%%%%%%%%%%%%%%%%%%%%%%

\subsection{Matching the SM+EWinos to the THDM+EWinos}

Matching the SM+EWinos to the THDM+EWinos, the threshold corrections for the SM Higgs self-coupling as well as the top Yukawa couplings are the same as in the case of matching the SM to the THDM (see \Sec{SMtoTHDM}), since no corresponding unsuppressed diagrams containing heavy Higgs as well as EWinos exist.

We split up the matching condition of the effective Higgs--Higgsino--Gaugino couplings into a part due to vertex corrections and another one due to wave-function renormalization,
\begin{align}
\tilde g_i (M_A) = \tilde g_{i,\text{tree}}+\Delta_{\text{Ver.Cor.}} \tilde g_i + \Delta_{\text{WFR}} \tilde g_i.
\end{align}
The vertex corrections are given by
\begin{subequations}
\begin{align}
\Delta_{\text{Ver.Cor.}}\tgiiu =& \frac{1}{2}(\hgiiud\cbe-\hgiidd\sbe)\bigg[(\hgiidd\hgiiuu-\hgidd\hgiuu)\cbb\nonumber\\
&\hspace{3.3cm}+(\hgidd\hgidu-\hgiud\hgiuu-\hgiidd\hgiidu+\hgiiud\hgiiuu)\sbe\cbe\nonumber\\
&\hspace{3.3cm}+(\hgidu\hgiud-\hgiidu\hgiiud)\sbb\bigg],\\
\Delta_{\text{Ver.Cor.}}\tgiid =&\frac{1}{2}(\hgiiuu\cbe-\hgiidu\sbe)\bigg[(\hgiidu\hgiiud-\hgidu\hgiud)\cbb\nonumber\\
&\hspace{3.3cm}+(\hgidd\hgidu-\hgiud\hgiuu-\hgiidd\hgiidu+\hgiiud\hgiiuu)\sbe\cbe\nonumber\\
&\hspace{3.3cm}+(\hgiuu\hgidd-\hgiidd\hgiiuu)\sbb\bigg],\\
\Delta_{\text{Ver.Cor.}}\tgiu =&\frac{1}{2}(\hgiud\cbe-\hgidd\sbe)\bigg[-(\hgiuu\hgidd+3\hgiiuu\hgiidd)\cbb\nonumber\\
&\hspace{3.3cm}+(\hgidd\hgidu-\hgiuu\hgiud-3\hgiiuu\hgiiud+3\hgiidd\hgiidu)\sbe\cbe\nonumber\\
&\hspace{3.3cm}+(\hgidu\hgiud+3\hgiidu\hgiiud)\sbb\bigg],\\
\Delta_{\text{Ver.Cor.}}\tgid =& \frac{1}{2}(\hgiuu\cbe-\hgidu\sbe)\bigg[-(\hgidu\hgiud+3\hgiidu\hgiiud)\cbb\nonumber\\
&\hspace{3.3cm}+(\hgidd\hgidu-\hgiuu\hgiud-3\hgiiuu\hgiiud+3\hgiidd\hgiidu)\sbe\cbe\nonumber\\
&\hspace{3.3cm}+(\hgiuu\hgidd+3\hgiiuu\hgiidd)\sbb\bigg].
\end{align}
\end{subequations}
The wave-function renormalization contributions read
\begin{subequations}
\begin{align}
\Delta_{\text{WFR}}\tgiiu =&-\frac{1}{16}(\hgiiuu\sbe+\hgiidu\cbe)\bigg[(\hgiuu^2+2\hgiiud^2+5\hgiiuu^2)\cbb\nonumber\\
&\hspace{3.3cm}-2(\hgiuu\hgidu+2\hgiidd\hgiiud+5\hgiiuu\hgiidu)\sbe\cbe\nonumber\\
&\hspace{3.3cm}+(\hgidu^2+2\hgiidd^2+5\hgiidu^2)\sbb\bigg],\\
\Delta_{\text{WFR}}\tgiid =&-\frac{1}{16}(\hgiidd\cbe+\hgiiud\sbe)\bigg[(\hgiud^2+5\hgiiud^2+2\hgiiuu^2)\cbb\nonumber\\
&\hspace{3.3cm}-2(\hgidd\hgiud+5\hgiidd\hgiiud+2\hgiiuu\hgiidu)\sbe\cbe\nonumber\\
&\hspace{3.3cm}+(\hgidd^2+5\hgiidd^2+2\hgiidu^2)\sbb\bigg],\\
\Delta_{\text{WFR}}\tgiu =&-\frac{1}{16}(\hgiuu\sbe+\hgidu\cbe)\bigg[(3\hgiuu^2+2\hgiud^2+3\hgiiuu^2)\cbb\nonumber\\
&\hspace{3.3cm}-2(2\hgidd\hgiud+3\hgiuu\hgidu+3\hgiiuu\hgiidu)\sbe\cbe\nonumber\\
&\hspace{3.3cm}+(2\hgidd^2+3\hgidu^2+3\hgiidu^2)\sbb\bigg],\\
\Delta_{\text{WFR}}\tgid =&-\frac{1}{16}(\hgidd\cbe+\hgiud\sbe)\bigg[(2\hgiuu^2+3\hgiud^2+3\hgiiud^2)\cbb\nonumber\\
&\hspace{3.3cm}-2(3\hgidd\hgiud+2\hgiuu\hgidu+3\hgiidd\hgiiud)\sbe\cbe\nonumber\\
&\hspace{3.3cm}+(3\hgidd^2+2\hgidu^2+3\hgiidd^2)\sbb\bigg].
\end{align}
\end{subequations}

%%%%%%%%%%%%%%%%%%%%%%%%%%%%%%%%%%%%%%%%%%%%%%%%%%%%%%%%%%%%%%%%%%%%%%%%%%%%%%

\subsection{Matching the THDM+EWinos to the MSSM}

The threshold corrections for $\beta$ and $\lambda_i$ are obtained by taking the respective ones from the matching of the THDM to the MSSM but removing the EWino contributions.

The matching conditions of the effective Higgs--Higgsino--Gaugino couplings,
 only receive corrections due to sfermions,  given by the expressions  (at the scale $\msusy$)
\begin{subequations}
\begin{align}
\Delta_{\tilde f}\hgiuu &= \gy k\left(-\frac{5}{2}\gy^2 + \frac{1}{4}h_t^2 (9 - \at^2)\right),\\
\Delta_{\tilde f}\hgiiuu &= g k\left(-\frac{3}{2}g^2 + \frac{1}{4}h_t^2 (9 - \at^2)\right),\\
\Delta_{\tilde f}\hgidd &= - \gy k\left(\frac{5}{2} \gy^2 + \frac{1}{4}h_t^2 \mf^2\right),\\
\Delta_{\tilde f}\hgiidd &= - g k\left(\frac{3}{2} g^2 + \frac{1}{4}h_t^2 \mf^2\right),
\end{align}
\end{subequations}
and
\begin{subequations}
\begin{align}
\Delta_{\tilde f}\hgiud &= \gy \cdot \frac{1}{4}k h_t^2 \at\mf,\\
\Delta_{\tilde f}\hgiiud &= g \cdot \frac{1}{4}k h_t^2 \at\mf,\\
\Delta_{\tilde f}\hgidu &= \gy \cdot \frac{1}{4}k h_t^2 \at\mf,\\
\Delta_{\tilde f}\hgiidu &= g \cdot \frac{1}{4}k h_t^2 \at\mf.
\end{align}
\end{subequations}
In the limit $M_A\rightarrow \msusy$, we recover the corresponding matching conditions of the SM+EWinos to the MSSM, given in \Eqs{SM+EWinos_MSSM_g12ud_Sf_Eqa}{SM+EWinos_MSSM_g12ud_Sf_Eqb} only if correctly taking into account the threshold corrections of $\tan\beta$.

The corrections due to the change of the regularization scheme read
\begin{subequations}
\begin{align}
\Delta_{\DR\rightarrow\MS}\hgiuu &= - \frac{1}{8}\gy k(3 g^2+\gy^2), \\
\Delta_{\DR\rightarrow\MS}\hgiiuu &= \frac{1}{24}g k(23 g^2 - 3 \gy^2),\\
\Delta_{\DR\rightarrow\MS}\hgidd &= - \frac{1}{8}\gy k(3 g^2+\gy^2),\\
\Delta_{\DR\rightarrow\MS}\hgiidd &= \frac{1}{24}g k(23 g^2 - 3 \gy^2), \\
\Delta_{\DR\rightarrow\MS}\hgidu &= \Delta_{\DR\rightarrow\MS}\hgiud =\Delta_{\DR\rightarrow\MS}\hgiidu = \Delta_{\DR\rightarrow\MS}\hgiiud=0.
\end{align}
\end{subequations}

%%%%%%%%%%%%%%%%%%%%%%%%%%%%%%%%%%%%%%%%%%%%%%%%%%%%%%%%%%%%%%%%%%%%%%%%%%%%%%

\subsection{Two-loop \texorpdfstring{$\mathcal{O}(\alpha_s\alpha_t)$}{O(asat)} threshold corrections}
\label{TLthresholds_App}

For deriving the \order{\als\alt} threshold corrections for the quartic couplings $\lambda_i$, we follow the strategy outlined in~\cite{Lee:2015uza}. As the authors of \cite{Lee:2015uza} pointed out, the \order{\als\alt} threshold corrections do not depend on $\tan\beta$. Therefore, they  can be extracted from the threshold correction for the SM quartic coupling $\lambda$ in the case $M_A\sim \msusy$ from matching the SM to the MSSM by selecting the coefficients of the various  $\beta$-dependent terms according to \Eq{Lambda_TreeLevel_Matching} and \Eq{c11}.

In contrast to the $\MS$ scheme employed in \cite{Lee:2015uza}, we use the $\DR$ scheme. Expressing the one-loop threshold corrections in terms of $X_t^\DR$ and the MSSM $\DR$-renormalized top Yukawa coupling $h_t^{\text{MSSM}}$, the two-loop $\mathcal{O}(\alpha_s\alpha_t)$ threshold correction for $\lambda$ at $\msusy$ reads as follows~\cite{Bagnaschi:2014rsa},
\begin{align}
\Delta_{\alpha_s\alpha_t}\lambda =-\frac{4}{3}k^2 g_3^2 h_t^4 \sbe^4 \xf\big(24-12\xf-4\xf^2+\xf^3\big).
\end{align}
Inserting $\hat{X}_t$ from \Eq{abbreviations} and selecting
the terms proportional to $(\cbe^4,\sbe^4,\cbb\sbb,\cbe^3\sbe,\cbe\sbe^3)$ yields
\begin{subequations}
\begin{alignat}{2}
&\Delta_{\alpha_s\alpha_t}\li &&= -\frac{4}{3}k^2 g_3^2 h_t^4\mf^4, \\
&\Delta_{\alpha_s\alpha_t}\lii &&= 16 k^2 g_3^2 h_t^4\bigg(-2\at+\at^2+\frac{1}{3}\at^3-\frac{1}{12}\at^4\bigg), \\
&\Delta_{\alpha_s\alpha_t}\lambda_{345} &&= 8 k^2 g_3^2 h_t^4\mf^2\bigg(1+\at-\frac{1}{2}\at^2\bigg), \\
&\Delta_{\alpha_s\alpha_t}\lvi &&= \frac{4}{3}k^2 g_3^2 h_t^4\mf^3\bigg(-1+\at\bigg), \\
&\Delta_{\alpha_s\alpha_t}\lvii &&= 4 k^2 g_3^2 h_t^4\mf\bigg(2-2\at-\at^2+\frac{1}{3}\at^3\bigg),
\end{alignat}
\end{subequations}
where $\lambda_{345}=\liii+\liv+\lv$. These expressions are valid under assumption of $M_{\tilde g} = \msusy$.

In the case $M_{\tilde g} \ll \msusy$, the SM--MSSM $\mathcal{O}(\alpha_s\alpha_t)$ threshold correction reads~\cite{Bagnaschi:2014rsa}
\begin{align}
\Delta_{\alpha_s\alpha_t}^\text{low $M_{\tilde g}$}\lambda =-\frac{8}{3}k^2 g_3^2 h_t^4 \sbe^4 \big(9-12\xf+\xf^4\big).
\end{align}
Selecting again the terms proportional to $(\cbe^4,\sbe^4,\cbb\sbb,\cbe^3\sbe,\cbe\sbe^3)$  yields
\begin{subequations}
\begin{alignat}{2}
&\Delta_{\alpha_s\alpha_t}^{\text{low $M_{\tilde g}$}}\li &&= -\frac{8}{3}k^2 g_3^2 h_t^4\mf^4, \\
&\Delta_{\alpha_s\alpha_t}^{\text{low $M_{\tilde g}$}}\lii &&= -\frac{8}{3} k^2 g_3^2 h_t^4\bigg(9-12\at^2+\at^4\bigg), \\
&\Delta_{\alpha_s\alpha_t}^{\text{low $M_{\tilde g}$}}\lambda_{345} &&= 8 k^2 g_3^2 h_t^4\mf^2\bigg(2-\at^2\bigg), \\
&\Delta_{\alpha_s\alpha_t}^{\text{low $M_{\tilde g}$}}\lvi &&= \frac{8}{3}k^2 g_3^2 h_t^4\at\mf^3, \\
&\Delta_{\alpha_s\alpha_t}^{\text{low $M_{\tilde g}$}}\lvii &&= -\frac{8}{3} k^2 g_3^2 h_t^4\mf\bigg(6-\at^2\bigg).
\end{alignat}
\end{subequations}
Using this method, we get only an information about the sum $\lambda_{345}$, leaving thus some arbitrariness. We follow the arrangement in~\cite{Lee:2015uza}, assigning
\begin{subequations}
\begin{alignat}{2}
&\Delta_{\alpha_s\alpha_t}\liii                             &&= \frac{1}{2}\Delta_{\alpha_s\alpha_t}\lambda_{345}, \\
&\Delta_{\alpha_s\alpha_t}\liv                              &&= \frac{1}{2}\Delta_{\alpha_s\alpha_t}\lambda_{345}, \\
&\Delta_{\alpha_s\alpha_t}\lv                               &&= 0, \\
&\Delta_{\alpha_s\alpha_t}^{\text{low $M_{\tilde g}$}}\liii &&= \frac{1}{2}\Delta_{\alpha_s\alpha_t}^{\text{low $M_{\tilde g}$}}\lambda_{345}, \\
&\Delta_{\alpha_s\alpha_t}^{\text{low $M_{\tilde g}$}}\liv  &&= \frac{1}{2}\Delta_{\alpha_s\alpha_t}^{\text{low $M_{\tilde g}$}}\lambda_{345}, \\
&\Delta_{\alpha_s\alpha_t}^{\text{low $M_{\tilde g}$}}\lv   &&= 0.
\end{alignat}
\end{subequations}
Other possible distributions yield numerically very similar results.

%%%%%%%%%%%%%%%%%%%%%%%%%%%%%%%%%%%%%%%%%%%%%%%%%%%%%%%%%%%%%%%%%%%%%%%%%%%%%%
%%%%%%%%%%%%%%%%%%%%%%%%%%%%%%%%%%%%%%%%%%%%%%%%%%%%%%%%%%%%%%%%%%%%%%%%%%%%%%

\section{Difference in field normalization}\label{AppDiffNormalization}

In this Appendix, we give explicit formulas for the difference of the field normalization between MSSM and THDM fields. The expressions are valid up to terms of $\mathcal{O}(M_t/\msusy,M_A/\msusy,M_t/M_A)$.

The contribution from sfermions is given by
\begin{align}
\Delta_{\tilde f}\Sigma_{11}' &= \frac{1}{2}k h_t^2\mf^2,\\
\Delta_{\tilde f}\Sigma_{12}' & = -\frac{1}{2}k h_t^2\at\mf,\\
\Delta_{\tilde f}\Sigma_{22}' & = \frac{1}{2}k h_t^2\at^2.
\end{align}
The contribution from electroweakinos reads
\begin{align}
\Delta_\chi\Sigma_{11}' &= -\frac{1}{6}k\left(3 g^2 + \gy^2\right)\left(1+3 \ln\frac{M_\chi^2}{\hat Q^2}\right),\\
\Delta_\chi\Sigma_{12}' & = -\frac{1}{6}k\left(3 g^2 + \gy^2\right),\\
\Delta_\chi\Sigma_{22}' & = -\frac{1}{6}k\left(3 g^2 + \gy^2\right)\left(1+3 \ln\frac{M_\chi^2}{\hat Q^2}\right).
\end{align}
In addition, also all non SUSY particles, i.e.~the particles of the THDM, yield a contribution if the renormalization scales of the THDM and the MSSM are not equal,
\begin{align}
\Delta_{\THDM}\Sigma_{11}' &= -\frac{1}{2}k\left(3 g^2 + \gy^2\right)\ln\frac{\widehat Q^2}{\widetilde Q^2},\\
\Delta_{\THDM}\Sigma_{12}' & = 0,\\
\Delta_{\THDM}\Sigma_{22}' & = -\frac{1}{2}k\left(3 g^2 + \gy^2\right)\ln\frac{\widehat Q^2}{\widetilde Q^2} + 3 k h_t^2\ln\frac{\widehat Q^2}{\widetilde Q^2},
\end{align}
with $\widehat Q$ being the renormalization scale of the MSSM and $\widetilde Q$ the scale of the THDM.

%%%%%%%%%%%%%%%%%%%%%%%%%%%%%%%%%%%%%%%%%%%%%%%%%%%%%%%%%%%%%%%%%%%%%%%%%%%%%%
%%%%%%%%%%%%%%%%%%%%%%%%%%%%%%%%%%%%%%%%%%%%%%%%%%%%%%%%%%%%%%%%%%%%%%%%%%%%%%

\section{Dependence on field renormalization constants}\label{FieldRenApp}

Here, we specify in more detail how the renormalized two-loop self-energies are influenced by field renormalization. The discussion is valid  in the limit of vanishing electroweak gauge couplings (gaugeless limit), which is the current approximation applied for the two-loop fixed-order corrections implemented in \FH. The notation follows closely that of~\cite{Hollik:2014bua}, where also more details about the renormalization as well as the applied approximations can be found.

Field renormalization is performed by rescaling the original MSSM Higgs fields, introducing  loop-expanded renormalization constants  up to the two-loop level,
\begin{align}
\begin{pmatrix}
\widehat{\phi}_1 \\ \widehat{\phi}_2
\end{pmatrix}
\rightarrow
\begin{pmatrix}
1 + \frac{1}{2}\delta^{(1)}Z_{11} + \frac{1}{2}\Delta^{(2)}Z_{11} & \frac{1}{2}\delta^{(1)}Z_{12} + \frac{1}{2}\Delta^{(2)}Z_{12} \\
\frac{1}{2}\delta^{(1)}Z_{12} + \frac{1}{2}\Delta^{(2)}Z_{12} & 1 + \frac{1}{2}\delta^{(1)}Z_{22} + \frac{1}{2}\Delta^{(2)}Z_{22}
\end{pmatrix}
\begin{pmatrix}
\widehat{\phi}_1 \\ \widehat{\phi}_2
\end{pmatrix}
\end{align}
with
\begin{align}
\Delta^{(2)}Z_{ij} := \delta^{(2)}Z_{ij}-\frac{1}{4}\left(\delta^{(1)}Z_{ij}\right)^2.
\end{align}
In extension of~\cite{Hollik:2014bua}, we also allow for the possibility of non-diagonal field renormalization terms.

Similarly, we introduce field renormalization constants in the mass eigenstate basis,
\begin{align}
\begin{pmatrix}
h\\ H
\end{pmatrix}
\rightarrow
\begin{pmatrix}
1 + \frac{1}{2}\delta^{(1)}Z_{hh} + \frac{1}{2}\delta^{(2)}Z_{hh} & \frac{1}{2}\delta^{(1)}Z_{hH} + \frac{1}{2}\delta^{(2)}Z_{hH} \\
\frac{1}{2}\delta^{(1)}Z_{hH} + \frac{1}{2}\delta^{(2)}Z_{hH} & 1 + \frac{1}{2}\delta^{(1)}Z_{HH} + \frac{1}{2}\delta^{(2)}Z_{HH}
\end{pmatrix}
\begin{pmatrix}
h \\ H
\end{pmatrix} .
\end{align}
These field renormalization constants are related to the ones in the gauge basis via
\begin{subequations}
\begin{align}
\delta^{(1)}Z_{hh} &= \saa\delta^{(1)}Z_{11} - s_{2\alpha}\delta^{(1)}Z_{12} + \caa\delta^{(1)}Z_{22}, \\
\delta^{(1)}Z_{hH} &= -\sae\cae\left(\delta^{(1)}Z_{11} - \delta^{(1)}Z_{22}\right) + c_{2\alpha}\delta^{(1)}Z_{12}, \\
\delta^{(1)}Z_{HH} &= \caa\delta^{(1)}Z_{11} + s_{2\alpha}\delta^{(1)}Z_{12}+ \saa\delta^{(1)}Z_{22}, \\
\delta^{(1)}Z_{AA} &= \sbb\delta^{(1)}Z_{11} - s_{2\beta}\delta^{(1)}Z_{12} + \cbb\delta^{(1)}Z_{22}, \\
\delta^{(1)}Z_{AG} &= -\sbe\cbe\left(\delta^{(1)}Z_{11} -  \delta^{(1)}Z_{22}\right) + c_{2\beta}\delta^{(1)}Z_{12} \, ,
\end{align}
\end{subequations}
and at the two-loop level,
\begin{subequations}
\begin{align}
\delta^{(2)}Z_{hh} &= \saa\Delta^{(2)}Z_{11} - s_{2\alpha}\Delta^{(2)}Z_{12} + \caa\Delta^{(2)}Z_{22}, \\
\delta^{(2)}Z_{hH} &= -\sae\cae\left(\Delta^{(1)}Z_{11} - \Delta^{(2)}Z_{22}\right) + c_{2\alpha}\Delta^{(2)}Z_{12}, \\
\delta^{(2)}Z_{HH} &= \caa\Delta^{(2)}Z_{11} + s_{2\alpha}\Delta^{(2)}Z_{12}+ \saa\Delta^{(2)}Z_{22}, \\
\delta^{(2)}Z_{AA} &= \sbb\Delta^{(2)}Z_{11} - s_{2\beta}\Delta^{(2)}Z_{12} + \cbb\Delta^{(2)}Z_{22}, \\
\delta^{(2)}Z_{AG} &= -\sbe\cbe\left(\Delta^{(2)}Z_{11} - \Delta^{(2)}Z_{22}\right) + c_{2\beta}\Delta^{(2)}Z_{12}.
\end{align}
\end{subequations}
In the following we set $\alpha = \beta - \frac{\pi}{2}$, according to the gaugeless limit.

Moreover, we set the external momentum $p^2$ to zero in the two-loop self-energies, as it is the default setting for the two-loop corrections in \FH\footnote{For the inclusion of non-zero external momentum at the two-loop level see \cite{Borowka:2014wla,Borowka:2015ura,Borowka:2018anu,Degrassi:2014pfa}}. The renormalized two-loop self-energies are composed of the unrenormalized self-energies and the corresponding two-loop counterterms,
\begin{subequations}
\label{renselftwoloop}
\begin{align}
\hat\Sigma_{hh}^{(2)}(0) &= \Sigma_{hh}^{(2)}(0)  - \delta^{(2)} m_h^{\mathbf{Z}}, \\
\hat\Sigma_{hH}^{(2)}(0) &= \Sigma_{hH}^{(2)}(0)  - \delta^{(2)} m_{hH}^{\mathbf{Z}}, \\
\hat\Sigma_{HH}^{(2)}(0) &= \Sigma_{HH}^{(2)}(0)  - \delta^{(2)} m_H^{\mathbf{Z}}.
\end{align}
\end{subequations}
The counterterms can be written in the following way,
\begin{subequations}
\label{mzCT}
\begin{align}
\delta^{(2)}m_h^{\textbf{Z}} =& \frac{1}{4}M_A^2\left(\delta^{(1)}Z_{hH}\right)^2+\delta^{(1)}\!Z_{hh}\,\delta^{(1)}m_h^2+\delta^{(1)}\!Z_{hH}\,\delta^{(1)}m_{hH}^2 + \delta^{(2)}m_h^2 ,\\
\delta^{(2)}m_{hH}^{\textbf{Z}} =& \frac{1}{2}\left[\left(\delta^{(1)}\!Z_{hh}+\delta^{(1)}\!Z_{HH}\right)\delta^{(1)}m_{hH}^2+\delta^{(1)}\!Z_{hH}\left(\delta^{(1)}m_h^2+\delta^{(1)}m_H^2\right)\right] \nonumber\\
&+\frac{1}{4}M_A^2\,\delta^{(1)}\!Z_{HH}\,\delta^{(1)}\!Z_{hH}+\frac{1}{2}M_A^2\,\delta^{(2)}\!Z_{hH}+\delta^{(2)}m_{hH}^2,\\
\delta^{(2)}m_H^{\textbf{Z}} =& M_A^2\left[\delta^{(2)}\!Z_{HH}+\frac{1}{4}\left(\delta^{(1)}\!Z_{HH}\right)^2\right]\nonumber\\
&+\delta^{(1)}\!Z_{HH}\,\delta^{(1)}m_H^2+\delta^{(1)}\!Z_{hH}\,\delta^{(1)}m_{hH}^2 + \delta^{(2)}m_H^2 ,
\end{align}
\end{subequations}
involving field renormalization constants and mass counterterms of one-and two-loop order. The two-loop mass counterterms are given by
\begin{subequations}
\begin{align}
\delta^{(2)}m_h^2 =& M_A^2 \cbe^4 \left(\delta^{(1)}\tbe\right)^2 - \frac{e}{2 M_W s_W} \cbe^2\; \delta^{(1)}\tbe\; \delta^{(1)}T_H \nonumber\\
&- \frac{e}{2 M_W s_W}\left[\delta^{(2)}T_h+\delta^{(1)}T_h\,\delta^{(1)}\!Z_W \right],\\[0.1cm]
\delta^{(2)}m_{hH}^2 =& M_A^2\cbb\, \delta^{(2)}\tbe +\cbb\,\delta^{(1)}\!M_A^2\,\delta^{(1)}\tbe-M_A^2\cbe^3\sbe \left(\delta^{(1)}\tbe\right)^2 \nonumber\\
&- \frac{e}{2 M_W s_W}\left[\delta^{(2)}T_H+\delta^{(1)}T_H\,\delta^{(1)}\!Z_W \right],\\[0.1cm]
\delta^{(2)}m_H^2 =& \delta^{(2)}\!M_A^2.
\end{align}
\end{subequations}
They involve the tadpole counterterms, the counterms for $\tan\beta$, as well as the renormalization constants of the electric charge, of the $W$-boson mass, and of $\sin\theta_W$ in the combination
\begin{align}
\delta^{(1)}Z_W = \frac{\delta^{(1)}e}{e}-\frac{\delta^{(1)}M_W}{M_W}-\frac{\delta^{(1)}s_W}{s_W}.
\end{align}
Also required are the one-loop mass counterterms,
\begin{subequations}
\begin{align}
\delta^{(1)}m_h^2 =& - \frac{e}{2 M_W s_W}\delta^{(1)}T_h ,\\
\delta^{(1)}m_{hH}^2 =& M_A^2\cbb\,\delta^{(1)}\tbe - \frac{e}{2 M_W s_W}\delta^{(1)}T_H ,\\
\delta^{(1)}m_H^2 =& \delta^{(1)}\!M_A^2 ,\\
\delta^{(1)}m_{AG}^2 =& -M_A^2\cbb\,\delta^{(1)}\tbe + \frac{e}{2 M_W s_W}\delta^{(1)}T_H.
\end{align}
\end{subequations}
At the one-loop level, the renormalization of $\tan\beta$ is given by the counterterm
\begin{align}
\delta^{(1)}\tbe ={}& \frac{1}{2}\tbe\left(\delta^{(1)}\!Z_{22}-\delta^{(1)}\!Z_{11}\right) +\frac{1}{2}\left(1-\tbb\right)\delta^{(1)}\!Z_{12} .
\end{align}
In the gaugeless limit and with the top Yukawa couplings only,
the corresponding two-loop counterterm for $\tan\beta$
reads as follows,
\begin{align}
\delta^{(2)}\tbe ={}& \frac{1}{2}\tbe\left(\delta^{(2)}\!Z_{22}-\delta^{(2)}\!Z_{11}\right) +\frac{1}{2}\left(1-\tbb\right)\delta^{(2)}\!Z_{12} \nonumber\\
& +\frac{1}{8}\tbe\left[3\left(\delta^{(1)}\!Z_{11}\right)^2-\left(\delta^{(1)}\!Z_{22}\right)^2\right] -\frac{1}{8}\left(1+2\tbe-\tbb-2\tbe^3\right)\left(\delta^{(1)}\!Z_{12}\right)^2\nonumber \\
& -\frac{1}{4}\tbe\,\delta^{(1)}\!Z_{11}\,\delta^{(1)}\!Z_{22}-\frac{1}{4}\left(1-2\tbb\right)\,\delta^{(1)}\!Z_{11}\,\delta^{(1)}\!Z_{12}-\frac{1}{4}\tbb\,\delta^{(1)}\!Z_{12}\,\delta^{(1)}\!Z_{22}.
\end{align}
Since we work in the real MSSM, the $A$-boson mass is used as a renormalized input parameter, with the counterterms determined by mass renormalization at one- and two-loop order,
\begin{align}
\delta^{(1)}\!M_A^2 =& \Sigma^{(1)}_{AA}(0)-M_A^2\,\delta^{(1)}\!Z_{AA},\\
\delta^{(2)}\!M_A^2 =& \Sigma^{(2)}_{AA}(0)-M_A^2\left[\delta^{(2)}\!Z_{AA}+\frac{1}{4}\left(\delta^{(1)}\!Z_{AA}\right)^2\right] \nonumber\\
&-\delta^{(1)}\!Z_{AA}\,\delta^{(1)}\!M_A^2-\delta^{(1)}\!Z_{AG}\,\delta^{(1)}m_{AG}^2,
\end{align}
where the external momentum in the $A$-boson self-energy is set to zero according to our approximation.

The tadpole counterterms are fixed by the requirement that the renormalized tadpoles vanish at the one- and two-loop level,
\begin{align}
T_{h,H}^{(1)} + \delta^{(1)}T_{h,H} &= 0, \\
T_{h,H}^{(2)}  +  \delta^{(2)}T_{h,H}^{\mathbf{Z}} &= 0,
\end{align}
where $T_{h,H}^{(i)}$ are the $i$-loop unrenormalized tadpoles of the $h$ and $H$ fields. The  two-loop counterterms  include field renormalization and are given by
\begin{subequations}
\begin{align}
\delta^{(2)}T_h^{\mathbf{Z}} &= \frac{1}{2}\left(\delta^{(1)}\!Z_{hh}\,\delta^{(1)}T_h+\delta^{(1)}\!Z_{hH}\,\delta^{(1)}T_H\right) + \delta^{(2)}T_h, \\
\delta^{(2)}T_H^{\mathbf{Z}} &= \frac{1}{2}\left(\delta^{(1)}\!Z_{HH}\,\delta^{(1)}T_H+\delta^{(1)}\!Z_{hH}\,\delta^{(1)}T_h\right) + \delta^{(2)}T_H.
\end{align}
\end{subequations}
With the conditions above all renormalization constants entering the renormalized self-energies in \Eq{renselftwoloop} are determined.

The two-loop  field renormalization constants appear exclusively in the $Z$-dependent two-loop counterterms of~\Eq{renselftwoloop}, either directly or through the two-loop mass, tadpole and $\tan\beta$ counterterms. In the combinations of \Eq{mzCT} they completely drop out and hence are not needed for the renormalized self-energies~(\ref{renselftwoloop}). This was already noted for the diagonal field counterterms in~\cite{Hollik:2014bua} for the \order{\alt^2} corrections.

The one-loop field renormalization constants $\delta^{(1)}\!Z_{ij}$ enter the two-loop renormalized self-energies \Eq{renselftwoloop} both through the counterterms and through the unrenormalized self-energies via one-loop subrenormalization. We extract the following dependence on $\delta^{(1)} \!Z_{ij}$,
\begin{subequations}
\begin{align}
\hat\Sigma_{hh}^{(2)}(0)\Big|_{\delta Z} &= \Sigma_{hh}^{(2)}(0)\Big|_{\delta Z} - \frac{e}{2 s_W M_W}\left(T_h^{(2)}\Big|_{\delta Z}+\frac{1}{2} \sbb T_h^{(1)} \delta^{(1)}\!Z_{hh}\right),\\
\hat\Sigma_{hH}^{(2)}(0)\Big|_{\delta Z} &= \Sigma_{hH}^{(2)}(0)\Big|_{\delta Z} - \frac{e}{2 s_W M_W}\left(T_H^{(2)}\Big|_{\delta Z}+\frac{1}{2} \sbb T_H^{(1)} \delta^{(1)}\!Z_{hh}\right),\\[0,1cm]
\hat\Sigma_{HH}^{(2)}(0)\Big|_{\delta Z} &= \Sigma_{HH}^{(2)}(0)\Big|_{\delta Z} - \Sigma_{AA}^{(2)}(0)\Big|_{\delta Z}\; \; .
\end{align}
\end{subequations}
The subscript $\delta Z$ indicates that only terms proportional to any of the field renormalization constants are kept. As a cross-check, we verified that adding a finite part to any $\delta^{(1)}\!Z_{ij}$ does not lead to additional divergencies. This is important for our method in \Sec{Combination_Section} to incorporate the different normalization  of the THDM fields as  a finite shift in the one-loop field renormalization constants of the MSSM.

%%%%%%%%%%%%%%%%%%%%%%%%%%%%%%%%%%%%%%%%%%%%%%%%%%%%%%%%%%%%%%%%%%%%%%%%%%%%%%
%%%%%%%%%%%%%%%%%%%%%%%%%%%%%%%%%%%%%%%%%%%%%%%%%%%%%%%%%%%%%%%%%%%%%%%%%%%%%%

\section{Scheme conversion for low \texorpdfstring{$M_A$}{MA}}\label{ConversionApp}

In this Appendix, we list the formulas necessary to convert the parameters of the stop sector from the OS to the $\DR$ scheme. Building upon the expressions given in \cite{Espinosa:1999zm,Draper:2013oza}, we extend those to the case of $M_A\neq M_S$.

First, we give the expression for calculating the $\DR$ top quark mass
of the MSSM in terms of  the OS top quark mass,
\begin{align}
\left(m_t^\DR\right)^2(Q)=& M_t^2\Bigg\{1-\frac{8}{3}k g_3^2\bigg[5+3\ln\frac{Q^2}{m_t^2}+\ln\frac{M_S^2}{Q^2}-\xf\bigg] \nonumber\\
&\hspace{0.8cm}+\frac{3}{2}k\frac{y_t^2}{\sbb}\bigg[\cbb\bigg(\frac{1}{2}-\ln\frac{M_A^2}{Q^2}\bigg)+\sbb\bigg(\frac{8}{3}+\ln\frac{Q^2}{m_t^2}\bigg)-\ln\frac{M_S^2}{Q^2}+\frac{1}{2}-\mf^2 f_2(\mf)\bigg]\Bigg\}.
\end{align}
with $M_S$ being the stop mass scale ($M_S^2\equiv m_{\tilde t_1}m_{\tilde t_2}$ with $m_{\tilde t_i}$ being the stop masses). For the conversion of this stop mass scale, we get
\begin{align}
\left(M_S^\DR\right)^2(Q) =& \left(M_S^\text{OS}\right)^2\Bigg\{1-\frac{16}{3}k g_3^2\Bigg[2-\ln\frac{M_S^2}{Q^2}\Bigg]\nonumber\\
&\hspace{1.6cm}+\frac{3}{4}k y_t^2\Bigg[\frac{2}{\tb^2}\ma^2\ln\frac{M_S^2}{M_A^2}+\frac{2}{\tb^2}\ma^2\bigg(1-\ln\frac{M_S^2}{Q^2}\bigg)\nonumber\\
&\hspace{1.6cm}+\frac{1}{\tb^2}\yf^2\bigg(\ma^2\ln\frac{M_S^2}{M_A^2}+(4-\ma^2)f_A(\ma)+4-2\ln\frac{M_S^2}{Q^2}\bigg)\nonumber\\
&\hspace{1.6cm}+4\xf^2\bigg(1-\frac{1}{2}\ln\frac{M_S^2}{Q^2}\bigg)\nonumber\\
&\hspace{1.6cm}+\frac{2}{\sbb}\bigg(\mf^4\ln\mf^2+(1-\mf^2)\Big(3-2\ln\frac{M_S^2}{Q^2}\Big)-(1-\mf^2)^2\ln(1-\mf^2)\bigg)\Bigg]\Bigg\},
\end{align}
and for the conversion of the stop mixing parameter,
\begin{align}
X_t^\DR(Q) =& M_S^{\text{OS}}\Bigg\{\xf^{\text{OS}}+\frac{4}{3}k g_3^2\Bigg[8+5\xf-\xf^2 +3\xf L\Bigg]\nonumber\\
&\hspace{1.cm}+\frac{1}{4}k y_t^2\Bigg[\frac{6}{\tb^2}\yf\bigg(\ma^2\ln\frac{M_S^2}{M_A^2}+(4-\ma^2)f_A(\ma)+2\ln\frac{M_S^2}{Q^2}-4\bigg) \nonumber\\
&\hspace{1.cm}-\frac{3}{\tb^2}\xf\ln\frac{M_S^2}{M_A^2}+\frac{1}{2}\xf\bigg(35-6\ln\frac{M_S^2}{m_t^2}-24\ln\frac{M_S^2}{Q^2}+\frac{24}{\sbb}\Big(1-\ln\frac{M_S^2}{Q^2}\Big)\bigg)\nonumber\\
&\hspace{1.cm}-\frac{6}{\sbb}\xf\bigg(1-\mf^2+\frac{1}{2}f_2(\mf)+\mf^4\ln\mf^2+(1-\mf^4)\ln(1-\mf^2)\bigg)\nonumber\\
&\hspace{1.cm}+\frac{3}{\tb^2}\xf\yf^2\bigg((1-\ma^2)\ln\frac{M_S^2}{M_A^2}-(3-\ma^2)f_A(\ma)-2\bigg)\nonumber\\
&\hspace{1.cm}+\xf^3\bigg(3\ln\frac{M_S^2}{m_t^2}-4\ln 2-6\ln|\xf|\bigg)\Bigg]\Bigg\}.
\end{align}
The appearing loop function $f_A$ depending on $\ma\equiv M_A/M_S$ is defined by
\begin{align}
f_A(\ma) &= \frac{\ma}{\sqrt{4-\ma^2}}\Bigg[\arctan\Bigg(\frac{\ma\sqrt{4-\ma^2}}{2-\ma^2}\Bigg)-\pi\Bigg]
\end{align}
with the limiting values
\begin{align}
f_A(0) &= 0, \qquad f_A(1) = -\frac{2}{3\sqrt{3}}\pi \, .
\end{align}

%%%%%%%%%%%%%%%%%%%%%%%%%%%%%%%%%%%%%%%%%%%%%%%%%%%%%%%%%%%%%%%%%%%%%%%%%%%%%%
%%%%%%%%%%%%%%%%%%%%%%%%%%%%%%%%%%%%%%%%%%%%%%%%%%%%%%%%%%%%%%%%%%%%%%%%%%%%%%

% \section{Renormalization group equations}\label{RGE_App}
%
% The RGEs have been derived using the {\tt Mathematica} package {\tt SARAH} \cite{Staub:2013tta}. We use the following notations,
% \begin{align}
% t &\equiv \ln(Q^2),\\
% k &\equiv \frac{1}{16\pi^2}.
% \end{align}
% $k$ is used to keep track of the loop order. For the convention used for the normalization of $\lambda_i$ and $v_i$, see Eq.~(\ref{HiggsPotential})-(\ref{HiggsFields}). All RGEs are given at the two-loop order in terms of the corresponding beta functions by
% \begin{align}
% \frac{d g_i}{dt}= k \beta^{(1)}_{g_i} + k^2 \beta^{(2)}_{g_i},
% \end{align}
% with $g_i$ being a generic coupling. The notation $\langle a; b \rangle$ indicates that $a$ is used for $Q<M_{\tilde g}$ and $b$ for $Q>M_{\tilde g}$.
%
% \subsection{Two Higgs doublet model}\label{RGE_THDM_App}
%
% \input{RGEs_THDM}
%
% \subsection{Two Higgs doublet model with Electroweakinos}\label{RGE_THDMsplit_App}
%
% \input{RGEs_THDMsplit}

%%%%%%%%%%%%%%%%%%%%%%%%%%%%%%%%%%%%%%%%%%%%%%%%%%%%%%%%%%%%%%%%%%%%%%%%%%%%%%
%%%%%%%%%%%%%%%%%%%%%%%%%%%%%%%%%%%%%%%%%%%%%%%%%%%%%%%%%%%%%%%%%%%%%%%%%%%%%%
%%%%%%%%%%%%%%%%%%%%%%%%%%%%%%%%%%%%%%%%%%%%%%%%%%%%%%%%%%%%%%%%%%%%%%%%%%%%%%

\newpage

\bibliographystyle{JHEP}
\bibliography{bibliography}{}

\end{document}